\documentclass[12pt]{article}
\usepackage[authoryear,round]{natbib}
\usepackage[usenames,dvipsnames]{color}
\usepackage{pdflscape}
\usepackage[colorlinks=true,linkcolor=blue,citecolor=blue,linkcolor=blue,anchorcolor=red,urlcolor=blue]{hyperref}
\usepackage{amsmath, amssymb, amsthm,authblk}
\usepackage{epsfig,euscript}
\usepackage{mathabx}

\usepackage{colortbl}

\usepackage[title]{appendix}  % for auto-sectioning Appendix
\usepackage{multirow} % for table
\usepackage[table,xcdraw]{xcolor}
\usepackage{longtable}

\usepackage{graphicx}
\usepackage{subcaption} % for 2 x 2 subfigure
\usepackage{mwe}
\usepackage{float}   % for putting table HERE

\usepackage{authblk} % for multiple author

\newcommand{\cM}{{\cal M}}

\newcommand{\cF}{\mbox{$\mathcal{F}$}}
\newcommand{\cD}{\mbox{$\mathcal{D}$}}
\newcommand{\rE}{\mbox{$\rm{E}$}}
\newcommand{\rvec}{\mbox{$\rm{vec}$}}

\newtheorem{theorem}{Theorem}
\newtheorem{lemma}{Lemma}
\newtheorem{corollary}{Corollary}
\newtheorem{proposition}{Proposition}

\renewcommand{\hat}{\widehat}
\def\singlespace{\def\baselinestretch{1}\@normalsize}

\def\endp{{\hfill \vrule width 5pt height 5pt\par}}

\newcommand{\cov}{{\rm Cov}}

\newcommand{\tr}{\mbox{tr}}
\newcommand{\var}{\mbox{Var}}

\newcommand{\bB}{{\mathbf B}}
\newcommand{\bF}{{\mathbf F}}
\newcommand{\bE}{{\mathbf E}}

\newcommand{\bH}{{\mathbf H}}
\newcommand{\bI}{{\mathbf I}}

\newcommand{\bM}{{\mathbf M}}
\newcommand{\bO}{{\mathbf O}}
\newcommand{\bQ}{{\mathbf Q}}

\newcommand{\bR}{{\mathbf R}}
\newcommand{\bS}{{\mathbf S}}
\newcommand{\bU}{{\mathbf U}}
\newcommand{\bV}{{\mathbf V}}
\newcommand{\bW}{{\mathbf W}}
\newcommand{\bX}{{\mathbf X}}

\newcommand{\bd}{{\mathbf d}}
\newcommand{\bc}{{\mathbf c}}
\newcommand{\be}{{\mathbf e}}
\newcommand{\bff}{{\mathbf f}}

\newcommand{\bh}{{\mathbf h}}

\newcommand{\bq}{{\mathbf q}}
\newcommand{\br}{{\mathbf r}}

\newcommand{\bx}{{\mathbf x}}
\newcommand{\by}{{\mathbf y}}

\newcommand{\bOmega}{\boldsymbol{\Omega}}

\newcommand{\bSigma}{\boldsymbol{\Sigma}}

\newcommand{\bGamma} {\boldsymbol{\Gamma}}

\newcommand{\bC}{{\mathbf C}}
\newcommand{\bD}{{\mathbf D}}
\newcommand{\bzero}{{\mathbf 0}}

\newcommand{\calD}{{\mathcal D}}

\newcommand{\calN}{{\mathcal N}}

\def\6bullets{\bullet\bullet\bullet\bullet\bullet\bullet}

\bibpunct{(}{)}{;}{a}{,}{,}

\graphicspath{ {figs/} }

\newcommand{\blind}{0}

\begin{document}

\def\spacingset#1{\renewcommand{\baselinestretch}%
{#1}\small\normalsize} \spacingset{1}
	
\if0\blind
{
	\title{\bf Helping Effects Against Curse of Dimensionality in Threshold Factor Models for Matrix Time Series }
	\author[1]{Xialu Liu \thanks{Xialu Liu is Assistant Professor, Management Information Systems Department, San Diego State University, San Diego, CA 92182. Email: xialu.liu@sdsu.edu. Xialu Liu is the corresponding author.}}
	\author[2]{Elynn Y. Chen \thanks{Elynn Chen is supported in part by NSF Grants DMS-1803241.} }
	\affil[1]{San Diego State University}
	\affil[2]{ORFE, Princeton University}
	\date{\vspace{-5ex}}
	% \date{Last modified on \today}
	% \thankstext{t1}{Supported in part by NSF Grants DMS-1206464 and DMS-1406266 and NIH grant R01-GM072611-9.}
	% \thankstext{t2}{Supported in part by NSF grants DMS-1317308, CAREER-DMS-1053987 and by the Howard B. Wentz Jr. Junior Faculty award.}
	\maketitle
} \fi

\if1\blind
{
	\bigskip
	\bigskip
	\bigskip
	\title{\bf Helping Effects Against Curse of Dimensionality in Threshold Factor Models for Matrix Time Series}
	\author{}
	\date{\vspace{-5ex}}
	\maketitle
	\medskip
} \fi

\bigskip	
	
\maketitle

\date{}

\begin{abstract}
As known, factor analysis is a popular method to reduce dimension for high-dimensional data. For matrix data, the dimension reduction can be more effectively achieved through both row and column directions \citep{wang2019}. In this paper, we introduce a threshold factor models to analyze matrix-valued high-dimensional time series data. The factor loadings are allowed to switch between regimes, controlling by a threshold variable. The estimation methods for loading spaces, threshold value, and the number of factors are proposed. The asymptotic properties of these estimators are investigated. Not only the strengths of thresholding and factors, but also their interactions from different directions and different regimes play an important role on the estimation performance. When the thresholding and factors are all strong across regimes, the estimation is immune to the impact that the increase of dimension brings, which breaks the curse of dimensionality. When the thresholding in two directions and factors across regimes have different levels of strength, we show that estimators for loadings and threshold value experience 'helping' effects against the curse of dimensionality. We also discover that even when the numbers of factors are overestimated, the estimators are still consistent. The proposed methods are illustrated with both simulated and real examples.
\end{abstract}

\vfill

\newpage
\spacingset{1.45} % DON'T change the spacing!

\section{Introduction}
Traditional multivariate time series models face computational challenges and loses efficiency when the dimension grows. Factor analysis is considered as an effective way to alleviate these problems by dimension reduction and to model the dynamics of high-dimensional time series \citep{geweke1977,chamberlain1983,pena1987,forni1998,forni2000,bai2002,stock2002a,stock2002b,pena2006,hallin2007,fan2016}. \cite{wang2019} further extended factor models to matrix-valued time series, achieving greater dimension reduction by utilizing the matrix structure of data and taking both row and column dimension reduction. Let $\bX_t$ ($t=1,\ldots,T$) be a matrix-valued time series.
\begin{equation*}
\label{eqn:2dmodel}
\bX_t=\bR \bF_{t} \bC' + \bE_t, \quad t=1,2,\ldots,T,
\end{equation*}
The dynamics of $\bX_t$ is driven by $\bF_t$ is a $k_1\times k_2$ unobserved matrix-valued
time series of common fundamental factors.
$\bR$ is a $p_1 \times k_1$ row loading matrix, and $\bC$ is a $p_2 \times k_2$
column loading matrix. $\bR\bF_t \bC'$ is the common component, and $\bE_t$ is a $p_1
\times p_2$ error matrix. $\bC$ and $\bR$ reflect the importance of common factors and
their interactions. 

Matrix-valued time series has many applications in economy, finance, and engineering. Here we consider an example with values of four economic indicators, GDP growth(GDP), unemployment(unem), risk-free rate(risk) and inflation rate(inf) from three countries (US, UK, and Japan). 
$\bX_t$ is a $4 \times 3$ matrix,
\[
\bX_t=\left( \begin{array}{ccc}
  \mbox{GDP-US} & \mbox{GDP-UK}  & \mbox{GDP-Japan} \\
  \mbox{unem-US} & \mbox{unem-UK}  & \mbox{unem-Japan} \\ 
  \mbox{risk-US} & \mbox{risk-UK}  & \mbox{risk-Japan} \\ 
  \mbox{inf-US} & \mbox{inf-UK}  & \mbox{inf-Japan} \\   
  \end{array}\right).
\]

Let the $\ell$-th columns of $\bX_t$, $\bR$, $\bC$, and $\bE_t$ be
$\bx_{t,\ell}$, $\br_{\ell}$, $\bc_\ell$, and $\be_{t,\ell}$,
respectively. Let $\by_{t,m\cdot}'$,$\be_{t,m\cdot}'$,$\br_{m\cdot}'$, and $\bc_{m\cdot}'$  be the row vectors that denote the $m$-th row of $\bX_t$, $\bE_t$, $\bR$, and $\bC$,
respectively. If we ignore other columns and only look at the $\ell$-th column of $\bX_t$, it can be expressed as
\begin{align}\label{row}
\bx_{t,\ell}=\bR (\bF_t \bc_{\ell\cdot})+\be_{t,\ell}.
\end{align}
This is a classic $p_1$-dimensional vector-valued factor model for economic indicators of $\ell$-th country. The dimension reduction is achieved in the sense that $p_1$-dimensional $\bx_{t,\ell}$ is driven by a $k_1$-dimensional process $(\bF_t \bc_{\ell \cdot})$. Since data in $p_1$ rows are reduced into $k_1$ rows, this is called {row dimension reduction}.

Let us say that the $\ell$-th country is U.S., (\ref{row}) can be written as

%\vspace{0.3cm}
\centerline{
\ \ \ U.S. indicators $\bx_{t,\ell}$ \hspace{0.8in} Loading matrix $\bR$ \hspace{0.5in}
 \hspace{0.5in} U.S. factors  $(\bF_t\bc_{\ell \cdot})$\ \ \ \ \  \ \ \ \ \ \ \
}
\vspace{-0.15in}
\begin{align}\label{row1}
\left(\begin{array}{c}
\mbox{GDP-US} \\ \mbox{unem-US} \\ \mbox{risk-US} \\ \mbox{inf-US} 
\end{array}\right)_t
\!\!=\!\left(\begin{array}{ccc}
\mbox{(loading)}_{1,1} &\ldots &\mbox{(loading)}_{k_1,1} \\
%[\mbox{loading-Unem}]_1, &\ldots, &[\mbox{loading-Unem}]_{k_1} \\ 
\mbox{(loading)}_{2,1}	&\ldots	&\mbox{(loading)}_{k_1,2}\\
\mbox{(loading)}_{3,1} &\ldots &\mbox{(loading)}_{k_1,4}\\
\mbox{(loading)}_{4,1} &\ldots &\mbox{(loading)}_{k_1,4}
\end{array}
\right)
\left(
\begin{array}{c}
\mbox{(factor)}_{1,\mbox{US}}\\
\vdots\\
\mbox{(factor)}_{k_1,\mbox{US}}\\
\end{array}
\right)_t
+\be_{t,\ell},
\end{align}
where
\\
\centerline{
\ \ \ U.S. factors  $(\bF_t\bc_{\ell\cdot})$ \hspace{0.8in} Factors $\bF_t$ \hspace{0.5in}
 \hspace{0.5in} U.S. loadings  $(\bc_{\ell \cdot})$\ \ \ \ \  \ \ \ \ \ \ \
}
\vspace{-0.15in}
\begin{align}\label{row2}
\left(
\begin{array}{c}
\mbox{(factor)}_{1,\mbox{US}}\\
\vdots\\
\mbox{(factor)}_{k_1,\mbox{US}}\\
\end{array}
\right)
\,=\!\left(\begin{array}{ccc}
\mbox{(factor)}_{1,1}, &\ldots, &\mbox{(factor)}_{1,k_2} \\
%[\mbox{loading-Unem}]_1, &\ldots, &[\mbox{loading-Unem}]_{k_1} \\ 
\vdots	&\vdots	&\vdots\\
\mbox{(factor)}_{k_1,1}, &\ldots, &\mbox{(factor)}_{k_1,k_2}
\end{array}
\right)_{t}
\left(
\begin{array}{c}
\mbox{(loading)}_{\mbox{US},1}\\
\vdots\\
\mbox{(loading)}_{\mbox{US},k_2}\\
\end{array}
\right).
\end{align}
In (\ref{row1}), the vector $(\bF_t\bc_{j\cdot})$ contains $k_1$ common {row} factors for the economic indicators of the $\ell$-th country. $\bR$ is its {row} loading matrix, in which the $m$-th column shows how much of an impact the $m$-th common row factor $(\bF_t\bc_{\ell\cdot})_m$ makes on the economic indicators. $\bF_t$ contains $k_1k_2$ common {fundamental factors}, which are building blocks of row factors. In (\ref{row2}), we can see that $m$-th row factor $\mbox{(factor)}_{m,\mbox{US}}$ on the left-hand side is constructed by the fundamental factors in $m$-th row of $\bF_t$, and $\bc_{\ell \cdot}$ reflects the interactions between fundamental factors for country $\ell$ when building the row factors with $\bF_t$. If only looking at $\ell$-th row of $\bX_t$, we could see the reason that $\bC$ is called the {column} loading matrix in a similar way.

The dimension reduction is achieved in model (\ref{eqn:2dmodel}) in the sense that  a $p_1\times p_2$ matrix-valued process $\bX_t$ is driven by a $k_1\times k_2$ matrix-valued process $\bF_t$. The $m$-th row in $\bF_t$ is used to construct the $m$-th row common factors, and the $\ell$-th column is used to construct the $\ell$-th column common factors. The $(m,\ell)$-th entry in $\bR_{p_1 \times k_1}$ shows how much of an impact the $\ell$-th row common factor makes on the $m$-th economic indicator. The $(m,\ell)$-th entry in $\bC_{p_2 \times k_2}$ quantifies the impact the $\ell$-th column common factor makes on the $m$-th country.

When dealing with matrix-valued time series, the classical factor analysis convert them to vectors by stacking the columns of matrix data on top of each other and ignore the interactions between columns. Matrix factor model (\ref{eqn:2dmodel}) overcomes the limitations, by fully utilizing the matrix structure, and achieve greater dimension reductions among two directions, columns and rows \citep{wang2019}.

Nonlinear dynamics have been a popular topic in factor models \citep{yalcin2001, cunha2010,fan2016,fan2017} during the past a few decades. Structural breaks \citep{breitung2011,chenl2014,han2015,ma2018,bai2017,su2017,barigozzi2018}, thresholding \citep{massacci2017,liu2019}, and Markov chain \citep{liu2016} were introduced to interpret the nonlinear behaviors observed in vector-valued time series data. However, nonlinear factor models for matrix-valued time series analysis is a new area and has not be exploited. In this paper, we introduce a threshold matrix factor model for high-dimensional matrix-valued time series, where the dynamic of $p_1\times p_2$ dimensional matrix time series is driven by a $k_1\times k_2$ dimensional matrix-valued factor process. In the model, loadings of the factor process vary across regimes, and there exists a threshold variable controlling the regime-switching mechanism. 

We propose estimation methods for loading spaces and threshold value, and investigate the asymptotic properties of the proposed estimators. When studying high-dimensional data, people often consider the cases when the dimension and sample size both go to infinity, and the increases in dimension may cause efficiency loss in estimation, which is called the curse of dimensionality. In this paper, we discuss the influence of the curse of dimensionality through the strength of the factor and the strength of the thresholding. It has been found that when the row factors, column factors, and thresholding are all strong, the estimation is immune to the curse of dimensionality. When two regimes have different level of factor strength, interactions between regimes are similar to vector-valued threshold factor models in \cite{liu2019}, and there exists a 'helping' effect, which means, comparing to one-regime factor models the estimation for strong regime does not hurt asymptotically due to the presence of a weak regime, while the convergence rate for the weak regime improves by introducing a strong regime. However, for matrix-valued threshold factor models the growth rates of $p_1$ and $p_2$ both need to be considered when defining the strength of regimes, which is different from vector-valued threshold factor models (See details in Section 3). Comparing to vector threshold factor models \citep{liu2019} which achieve dimension reduction only through one direction, matrix threshold factor models accomplish it through two directions (row and column). The efficiency of threshold value estimation is shown to be determined by the direction with stronger thresholding, and is also exposed to the 'helping' effect. The estimation of threshold value does not hurt by introducing a direction with weaker thresholding, and the estimator of threshold value gains some efficiency by the existence of a direction with stronger thresholding. 

To estimate the number of factors, we follow the idea in \cite{lam2012} and extend their ratio-based estimators for matrix-valued time series. \cite{bai2002,hallin2007} studied the methods to estimate the number of factors for factor models, but their methods perform poorly when there exists strong cross-sectional correlation in noise process \citep{lam2012}. The ratio-based estimator is first introduced by \cite{lam2012} and applied to difference scenarios in \cite{chang2015,liu2016,lee2018,wang2019,liu2019}. It has been shown that the probability of underestimation for the ratio-based estimator goes to zero asymptotically. Though its consistency has not been confirmed theoretically, the estimator performs well in numerical experiments. In this paper we show that even when the numbers of factors are overestimated, the asymptotic properties of estimated loading spaces and threshold value are the same with those when the numbers of factors are correctly estimated.

The rest of the paper is organized as follows. Section 2 introduces the threshold factor model for high-dimensional matrix-valued time series. In Section 3, the estimations for loading spaces, threshold value, and the number of factors are developed and the theoretical properties are also investigated. We apply our methods to simulated and real data, and present the results in Sections 4 and 5, respectively. All the regularity conditions, lemmas, detailed mathematical proofs are in the Appendix.

\section{Model}
Let $\bX_t$ be a $p_1 \times p_2$ observed matrix-valued time series, where 
\begin{align}\label{model}
\bX_t= \left\{
\begin{array}{cc}
\bR_1 \bF_t \bC_1'+\bE_{t}  &z_t <r_0, \\
\bR_2 \bF_t \bC_2'+\bE_{t} &z_t \geq r_0,\\
\end{array}\right. \quad t=1,\ldots, T.  
\end{align}
$\bF_t$ is a $k_1 \times k_2$ latent matrix-valued time series which consists of fundamental factors. $\bR_i$ is a $p_1 \times k_1$ row loading matrix, and $\bC_i$ is a $p_2\times k_2$ column loading matrix, for $i=1,2$. $\bE_{t}$ is a $p_1 \times p_2$ matrix which is the noise process and has no serial dependence. $z_t$ is an observed threshold variable, controlling the switchings between two regimes. Loading matrices $\{\bR_i, \bC_i\}$ are different across regimes.

\noindent{\bf Remark 1.} Only one threshold variable $z_t$ is incorporated into model (\ref{model}). We could introduce two threshold variables to control the regime-switching for row and column loading matrices separately, and the methods described in Section 3 also applies. For simplicity, in the rest of the paper, we focus on model (\ref{model}). 

\noindent{\bf Remark 2.} Model (\ref{model}) with $(k_1k_2)$ factors is a special case of matrix factor model \citep{wang2019}, since it can be written as 
\begin{align}
\bX_t=\widetilde{\bR} \widetilde{\bF}_t \widetilde{\bC}'+\bE_t,
\end{align}
where 
\begin{align*}
\widetilde{\bR}=\left(
\begin{array}{cc}
\bR_1& \bR_2
\end{array}
\right), \quad
\widetilde{\bF}_t=\left(
\begin{array}{cc}
\bF_t I_{t,1} &\bzero\\
\bzero	&\bF_t I_{t,2}
\end{array}
\right),\quad
\widetilde{\bC}=\left(
\begin{array}{c}
\bC_1\\
\bC_2
\end{array}\right),
\end{align*}
which is a one-regime matrix factor model with constraints and $(4k_1k_2)$ factors. The threshold factor model (\ref{model}) uses fewer factors and achieves greater dimension reduction by introducing regimes. Proposition 1 in \cite{liu2019} shows that ignoring the threshold variable and using the one-regime model may lead to a mis-specified model and inconsistent estimators. As is known, latent factors are challenging to interpret. By reducing the number of factors, model (\ref{model}) provides a more interpretable structure. 

\noindent{\bf Remark 3.} Most of papers on factor models in econometrics build their models based on the assumption that the factors have impact on most of the series. The noise process has weak serial dependence and weak cross-sectional dependence \citep{forni2000,bai2002,stock2002a,stock2002b,fan2017}. However, under this condition, the common component and noise process are not identifiable when the dimension is finite. \cite{pan2008}, \cite{lam2011}, \cite{lam2012},\cite{chang2015}, and \cite{liu2016} take another assumption that the factors capture all dynamics of the observed process, which means the noise is white, and can accommodate strong cross-sectional dependence. To make the common component and noise process separable, in this paper, we follow their settings, and further relax their assumption by allowing heteroscedasticity for the noise process.

As is known, the factor models have ambiguity issues, and $\bR_i$ and $\bC_i$ are not uniquely defined \citep{lam2011,lam2012,chang2015,liu2016,wang2019,liu2019}. Specifically, the model (\ref{model}) can be re-written as,
\begin{align*}
\bX_t= \sum_{i=1}^2
\left[ \bR_i \bU \left( \bU^{-1} \bF_t  \bV^{-1} \right) \bV \bC_i'\right]I_{t,i}+\bE_{t}, \quad t=1,\ldots, T,
\end{align*}
where $I_{t,1}=I(z_t < r_0)$ and $I_{t,2}=I(z_t \geq r_0)$. The row loading matrix, column loading matrix and factor process can be replaced by $\bR_i \bU$, $\bC_i\bV'$, and $\bU^{-1}\bF_t \bV^{-1}$, for $i=1,2$ and any non-singular matrices $\bU$ and $\bV$. However, the column spaces spanned by $\bR_i$ and $\bC_i$, ${\cal M}(\bR_i)$ and ${\cal M}(\bC_i)$, called row loading space and column loading space for regime $i$ respectively, are identifiable. Our aim is to estimate the row and column loading spaces, ${\cal M}(\bR_i)$ and ${\cal M}(\bC_i)$, instead of loading matrices, $\bR_i$ and $\bC_i$ for $i=1,2$. We can further decompose $\bR_i$ and $\bC_i$ as follows
\begin{align*}
\bR_i=\bQ_{1,i} \bW_{1,i}, \mbox{ and } \bC_i=\bQ_{2,i} \bW_{2,i},
\end{align*}
where $\bQ_{s,i}$ is $p_s \times k_s$ orthogonal matrices, and $\bW_{s,i}$ is $k_s \times k_s$ non-singular matrix, for $s,i=1,2$. By the definition, we have ${\cal M}(\bR_i)={\cal M}(\bQ_{1,i})$ and ${\cal M}(\bC_i)={\cal M}(\bQ_{2,i})$. In the following, we will estimate the orthonormal representatives of ${\cal M}(\bR_i)$ and ${\cal M}(\bC_i)$, $\bQ_{1,i}$ and $\bQ_{2,i}$, for $i=1,2$. Let $\bS_t=\sum_{i=1}^2 \bW_{1,i}\bF_t \bW_{2,i}I_{t,i}$, and the model (\ref{model}) can be re-expressed as
\[
\bX_t=\sum_{i=1}^2 \bQ_{1,i} \bS_t \bQ_{2,i}' I_{t,i} +\bE_t, \quad t=1,\ldots, T.
\]

\noindent{\bf Remark 4.} The stationarity of the factor process can be ruined by any nonsingular transformation using ($\bR_i \bU_i,\bC_i\bV_i',\bU_i^{-1}\bF_t \bV_i^{-1}I_{t,i}$) to replace ($\bR_i,\bC_i, \bF_t$) with $\bU_1\neq \bU_2$ and $\bV_1\neq \bV_2$. One advantage of our setting is that we do not impose stationarity assumptions or any specific models for the factor process. Our estimation only require the latent process to satisfy some mixing conditions stated in Appendix A.1.

Here is some notation we will use throughout the paper.  Let $\rvec(\cdot)$ be the vectorization operator, which converts a matrix to a vector by stacking columns fo the matrix on top of each other. For any matrix $\bH$, let $\|\bH\|_{\rm F}$ and $\|\bH\|_2$ denote the Frobenius and L-2 norms of $\bH$, $\sigma_i(\bH)$ is the $i$-th largest singular value of $\bH$, ${\rm rank}(\bH)$ is the rank of $\bH$, and $\|\bH \|_{\min}$ is the square root of the minimum nonzero eigenvalue of $\bH' \bH$. We use $\bh_{m\cdot}$ and $\bh_{\ell}$ to represent the vectors with the entries in $m$-th row and the $\ell$-th column of $\bH$ respectively, and $h_{m\ell}$ to represent the $m \ell$-th entry of $\bH$. For a square matrix $\bH$, $\tr(\bH)$ denotes its trace.

Our methods apply when ${\rE}(\bF_t)$ is nonzero. However, for ease of presentation, in the rest of the paper, we assume that the process $\bF_t$ has mean $\mathbf{0}$.

\section{Estimation} 
\subsection{Estimation of loading spaces}
Define the indicator functions $I_{t,1}(r)=I(z_t<r)$ and $I_{t,2}(r)=I(z_t \geq r)$. Here we present a procedure with two tentative threshold values, estimating the loading spaces with a partition in the form of $I_{t,1}(r_1)=I(z_t<r_1)$ and $I_{t,2}(r_2)=I(z_t \geq r_2)$, which combines the methods proposed by \cite{wang2019} and \cite{liu2019}. It can be seen that when $r_1\leq r_0$ and $r_2 \geq r_0$, the estimators are consistent.

Let $\bq_{s,i,\ell}$ be the $\ell$-th columns of $\bQ_{s,i}$ for $s,i=1,2$. Let $\bc_{i,\ell \cdot}$ and $\bq_{s,i,\ell \cdot}$ be the vector with entries in $\ell $-th row in $\bC_i$ and $\bQ_{s,i}$ for $s,i=1,2$. By the model (\ref{model}) we have
\begin{align}
\bx_{t,\ell}=\left\{
\begin{array}{ll}
\bR_1 \bF_t \bc_{1,\ell \cdot} +\be_{t,\ell}= \bQ_{1,1} \bS_t \bq_{2,1,\ell \cdot}+\be_{t,\ell}  & z_t <r_0,\\
\bR_2 \bF_t \bc_{2,\ell \cdot} +\be_{t,\ell}= \bQ_{1,2} \bS_t \bq_{2,2,\ell \cdot}+\be_{t,\ell}  & z_t \geq r_0.
\end{array}\right.
\end{align}

Let $h$ be a positive integer, and define auto-cross-covariances of the factor process and observed process in different partition,
\begin{align*}
\bOmega_{sq,ij,m\ell}(h,r_1,r_2)&=\frac{1}{T}\sum_{t=1}^{T-h}{\rm E}(\bS_{t} \bq_{2,i,m\cdot}  \bq_{2,j,\ell\cdot}' \bS_{t}' I_{t,i}(r_i) I_{t+h,j}(r_j)),\\
\bOmega_{x,ij,m\ell}(h,r_1,r_2)&=\frac{1}{T} \sum_{t=1}^{T-h} {\rm E}(\bx_{t,m} \bx_{t+h,\ell}' I_{t,i}(r_i) I_{t+h,j}(r_j)),
\end{align*}
for $i,j=1,2$, and $m,\ell=1,\ldots, p_2$.

Since the noise process is independent over time, when $h>0$, $r_1 \leq r_0$ and $r_2 \geq r_0$, we have
\begin{align}\label{zx}
\bOmega_{x,ij,m\ell}(h,r_1,r_2)= \bQ_{1,i} \bOmega_{sq,ij,m\ell}(h,r_1,r_2) \bQ_{1,j}'.
\end{align}
For a pre-determined positive integer $h_0$, define
\begin{align}\label{def_M}
\bM_{1,i}(r_1,r_2)=\sum_{h=1}^{h_0} \sum_{j=1}^2 \sum_{m=1}^{p_2}\sum_{\ell=1}^{p_2} \bOmega_{x,ij,m\ell} (h,r_1,r_2)\bOmega_{x,ij,m\ell}(h,r_1,r_2)', \mbox{ for } i=1,2.
\end{align}
By equation (\ref{zx}), when $r_1\leq r_0$ and $r_2 \geq r_0$, it follows that
\begin{align}\label{sandwich}
\bM_{1,i}(r_1,r_2)=\bQ_{1,i}\left( \sum_{h=1}^{h_0} \sum_{j=1}^2 \sum_{m=1}^{p_2}\sum_{\ell=1}^{p_2} \bOmega_{sq,ij,m\ell}(h,r_1,r_2) \bOmega_{sq,ij,m\ell}'(h,r_1,r_2)' \right) \bQ_{1,i}'.
\end{align}
$\bM_{1,i}(r_1,r_2)$ is a symmetric non-negative definite matrix sandwiched by $\bQ_{1,i}$ and $\bQ_{1,i}'$. If $\bM_{1,i}(r_1,r_2)$ has rank of $k_1$, its eigenvectors corresponding to its nonzero eigenvalues span the row loading space, ${\cal M}(\bQ_{1,i})$. Hence, ${\cal M}(\bQ_{1,i})$ can be estimated by the eigen-decomposition of sample version of $\bM_{1,i}(r_1,r_2)$. Let $\bq_{1,i,k}(r_1,r_2)$ be the unit eigenvector of $\bM_{1,i}(r_1,r_2)$ corresponding to the $k$-th largest eigenvalue, and we can now uniquely define $\bQ_{1,i}(r_1,r_2)$ by
\[
\bQ_{1,i}(r_1,r_2)=(\bq_{1,i,1}(r_1,r_2),\ldots, \bq_{1,i,k_1}(r_1,r_2)).
\]

Now we define the sample version of the above statistics.
\begin{align}
\hat{\bOmega}_{x,ij,m\ell}(h,r_1,r_2)&=\frac{1}{T}\sum_{t=1}^{T-h} \bx_{t,m}\bx_{t+h,\ell}' I_{t,i}(r_i) I_{t,j}(r_j),\label{Omegahat}\\
\hat{\bM}_{1,i}(r_1,r_2)&=\sum_{h=1}^{h_0} \sum_{j=1}^2 \sum_{m=1}^{p_2}\sum_{\ell=1}^{p_2} \hat{\bOmega}_{x,ij,m\ell} (h,r_1,r_2) \hat{\bOmega}_{x,ij,m\ell}(h,r_1,r_2)',\label{mhat}
\end{align}
for $i=1,2$. Let $\hat{\bq}_{1,i,k}(r_1,r_2)$ be the unit eigenvector of $\hat{\bM}_{1,i}(r_1,r_2)$ corresponding to the $k$-th largest eigenvalue. Then the row loading space in regime $i$ can be estimated by
\[
\widehat{{\cal M}(\bR_i)}={\cal M}(\hat{\bQ}_{1,i}(r_1,r_2)),
\]
where $\hat{\bQ}_{1,i}(r_1,r_2)=(\bq_{1,i,1}(r_1,r_2),\ldots, \bq_{1,i,k_1}(r_1,r_2))$.

For the column loading spaces, they can be estimated by performing the same procedure on the transposes of $\bX_t's$ to construct $\bM_{2,i}(r_1,r_2)$, for $i=1,2$.

\noindent{\bf Remark 5.} The sums in (\ref{Omegahat}) and (\ref{mhat}) run over a restricted range because $\bx_{t+h}$ is not available for $t+h>T$. Here we follow the definitions in \cite{shumway2017}, using (\ref{Omegahat}) and (\ref{mhat}) as sample auto-cross-covariance matrices, instead of the ones that would be obtained by dividing by $T-h$. For scalar time series, defining this way makes its auto-covariance function a non-negative definite function \citep{shumway2017}.

\noindent{\bf Remark 6.} If $\bOmega_{x,ij,m\ell}(h,r_1,r_2)$ is full rank for some $i,j=1,2$ and $m,\ell=1,\ldots, p_2$, theoretically any $h>0$ can be used to construct $\bM_{1,i}$ for loading space estimation. By summing over $h$, we only require $\bOmega_{x,ij,m\ell}(h,r_1,r_2)$ to be full rank for one of the $h$'s, instead of finding a particular $h$. Since the autocorrelation is often at its strongest at small time lags, a relatively small $h_0$ is usually adopted.

When studying high-dimensional time series, people not only let the sample size go to infinity, but also consider the case when the dimension increases to infinity. It is common to assume the norm of the loading matrices grows with the dimension. \cite{lam2011}, \cite{lam2012}, \cite{chang2015}, \cite{liu2016}, \cite{wang2019}, and \cite{liu2019} used the strength of factors to measure the growth rate as follows, 
\begin{align*}
\|\bR_i\|_2^2\asymp \|\bR_i\|_{\min}^2 \asymp p_1^{1-\delta_{1i}}, \quad \|\bC_i\|_2^2 \asymp \|\bC_i\|_{\min}^2 \asymp p_2^{1-\delta_{2i}},  
\end{align*}
where $ \delta_{si} \in [0, 1]$, for $s,i=1,2$, $s$ is the index for dimension reduction directions(row/column), and $i$ is the index for regimes.  The strength of the factors reflects the relative growth rate of information about $\bF_t$ carried by $\bX_t$ as the dimensions increase, comparing to the growth rate of the noise process. For example, if $\delta_{11}=0$, the row factors are strong in regime 1, the row loading matrix in regime 1 is dense, and $\bX_t$ is fully loaded with signal. If $\delta_{11}$=1, the row factors are extremely weak in regime 1, the row loading matrix in regime 1 is sparse, and only noise is added to $\bX_t$ as $p_1$ increases. 

For matrix-valued threshold factor models, we also need to consider the growth rate of $p_1$ and $p_2$ when comparing the strength of two regimes. If $p_1^{\delta_{11}-\delta_{12}}p_2^{\delta_{21}-\delta_{22}}\to 0$ as $p_1$ and $p_2$ go to infinity, loading matrices in regime 1 are denser, and data in regime 1 carry more useful information about $\bF_t$. Hence, we say regime 1 is stronger than regime 2. If $p_1^{\delta_{11}-\delta_{12}}p_2^{\delta_{21}-\delta_{22}} \to \infty$ as $p_1$ and $p_2$ go to infinity, data in regime 2 carry more useful information about $\bF_t$, and regime 2 is stronger than regime 1. If $p_1^{\delta_{11}-\delta_{12}}p_2^{\delta_{21}-\delta_{22}} \asymp 1$, regime 1 are as strong as regime 2.

Before presenting the asymptotic properties of the proposed estimators, we introduce a measure to quantify the distance of two linear spaces, which is first proposed by \cite{liu2019}. Let $\bS_1$ be a $p\times q_1$ full-rank matrix, and $\bS_2$ be a $p \times q_2$ full-rank matrix, where $p \geq q_1,q_2$. Let $\bO_i$ be an orthogonal representative of $\cM(\bS_i)$, i.e., $\cM(\bO_i)=\cM(\bS_i)$ and $\bO_i' \bO_i=\bI_{q_i}$, for $i=1,2$. The distance of $\cM(\bS_1)$ and $\cM(\bS_2)$ is defined as
\begin{align*}
\cD(\cM(\bS_1),\cM(\bS_2))=\sqrt{1-\frac{\tr(\bO_1\bO_1' \bO_2 \bO_2')}{\min\{q_1,q_2\}}}.
\end{align*}
It is a quantity between 0 and 1. It is 1 if and only if $\cM(\bS_1) \perp \cM(\bS_2)$, and is 0 if and only if $\cM(\bS_1) \subseteq \cM(\bS_2)$ or $\cM(\bS_2) \subseteq \cM(\bS_1)$. 

\begin{theorem}  \label{thm:loading_spdist}
Under Conditions A1-A4 and B1-B3 in Appendix \ref{appendix:regular_condition}, when $r_1 \leq r_0$ and $r_2 \geq r_0$, if true $k_1$ and $k_2$ are known and $p_1^{\delta_{11}/2+\delta_{12}/2}p_2^{\delta_{21}/2+\delta_{22}/2} T^{-1/2}=o(1)$, as $p_1,p_2, T \to \infty$, it holds that
\[
\cD(\cM(\hat{\bQ}_{s,i}(r_1,r_2)), \cM(\bQ_{s,i}(r_1,r_2)))=O_p(p_1^{\delta_{1i}/2+\delta_{1\min}/2} p_2^{\delta_{2i}/2+\delta_{2\min}/2} T^{-1/2}), \mbox{ for } s,i=1,2,
\]
where $\delta_{1\min}=\delta_{11}$ and $\delta_{2\min}=\delta_{21}$ if $p_1^{\delta_{11}-\delta_{12}}p_2^{\delta_{21}-\delta_{22}}=o(1)$; otherwise, 
$\delta_{1\min}=\delta_{12}$ and $\delta_{2\min}=\delta_{22}$.
\end{theorem}

Theorem 1 shows that when all the factors are strong, $\delta_{si}=0$ for $s,i=1,2$, the estimation does not hurt from the increase of $p_1$ and $p_2$, and the curse of dimension is offset by the signal of $\bF_t$. The 'helping' effect exists for loading spaces estimation, if factors have different levels of strength. Let us say the regime $i$ is weaker. The estimation for the weaker regime with convergence rate $p_1^{\delta_{1i}/2+\delta_{1\min}/2} p_2^{\delta_{2i}/2+\delta_{2\min}/2} T^{-1/2}$ gains efficiency from the introduction of the stronger regime, comparing to that in one-regime models with strength $\delta_{1i}$ and $\delta_{2i}$ whose convergence rate is $p_1^{\delta_{1i}}p_2^{\delta_{2i}}T^{-1/2}$ \citep{wang2019}. On the other hand, the loading space estimation in strong regime with convergence rate $p_1^{\delta_{1\min}} p_2^{\delta_{2\min}} T^{-1/2}$ does not suffer asymptotically from the existence of a weaker regime.

\subsection{Estimation of threshold value}
For simplicity, in the rest of the paper, if $r_1=r_2$, we only keep one input when defining matrices. For example, $\bQ_{s,i}(r,r)$, $\bq_{s,i,k}(r,r)$ and $\bM_{s,i}(r,r)$ are simplified as $\bQ_{s,i}(r)$, $\bq_{s,i,k}(r)$ and $\bM_{s,i}(r)$. Furthermore, we use $\bQ_{s,i}$, $\bq_{s,i,k}$ and $\bM_i$ to denote $\bQ_{s,i}(r_0)$, $\bq_{s,i,k}(r_0)$ and $\bM_{s,i}(r_0)$, where $r_0$ is the true threshold value, for $s,i=1,2$.

If we use $r$ as the tentative threshold value, the data are classified into 2 subsets, $S_1=\{t: z_t<r\}$ and $S_2=\{t:z_t \geq r\}$. Let $\bB_{s,i}=(\bq_{s,i,k_i+1}, \ldots, \bq_{s,i,p_i})$ be a $p_s \times k_s$ matrix, where $\bq_{s,i,k}$ is the unit eigenvector of $\bM_{s,i}$ corresponding to the $k$-th largest eigenvalue. $\cM(\bB_{s,i})$ is the complement of loading space $\cM(\bQ_{s,i})$ and $\bQ_{s,i}' \bB_{s,i}=\mathbf{0}$, for $s,i=1,2$. Define the objective function
\begin{align}
G(r)=\sum_{s=1}^2 \sum_{i=1}^2 \big\| \bB_{s,i}' \bM_{s,i}(r) \bB_{s,i} \big\|_2.
\end{align}
By the definition of $\bM_{s,i}$ in (\ref{def_M}), we can tell that $G(r)$ measures the sum of the squared norm of the projections of $\bOmega_{s,ij,m \ell}(h,r)$ onto the complement of loading spaces, $\cM(\bB_{s,i})$, for $h=1,\ldots, h_0$, $m,\ell=1,\ldots, p_i$, and $s,i,j=1,2$.

If $r=r_0$, the observations in two regimes are correctly classified into different subsets. Then by (\ref{sandwich}), $\bM_{s,i}$ is sandwiched by $\bQ_{s,i}$ and $\bQ_{s,i}'$. Hence $G(r)=0$.

However, if $r\neq r_0$, the observations from one regime is misclassified into two subsets, and one of the two subsets is mixed. $\bM_{s,i}$ is not sandwiched by $\bQ_{s,i}$ and $\bQ_{s,i}$, and the projection is not zero. The following proposition formally states the conclusions.

\begin{proposition}
Under Conditions A1-A4 and C1-C7 in Appendix \ref{appendix:regular_condition}, if $r=r_0$, then $G(r)=0$; if $r \neq r_0$, then $G(r)>0$.
\end{proposition}

A standard assumption for threshold variable estimation is imposed which is that $r_0$ is in a known region of the support of $z_t$, $r_0 \in (\eta_1,\eta_2)$. We use data corresponding to $z_t \leq \eta_1$ and $z_t \geq \eta_2$ to estimate $\cM(\bB_{s,1})$ and $\cM(\bB_{s,2})$, respectively, for $s=1,2$. By Theorem 1, they are both consistent. Let $\hat{\bB}_{s,i}(\eta_1,\eta_2)$ be the estimate $\bB_{s,i}$. The sample version of $G(r)$ is defined as
\[
\hat{G}(r)=\sum_{s=1}^2 \sum_{i=1}^2 \big\| \hat{\bB}_{s,i}(\eta_1,\eta_2)' \, \hat{\bM}_{s,i}(r) \, \hat{\bB}_{s,i}(\eta_1,\eta_2) \big\|_2.
\]
We estimate $r_0$ by
\[
\hat{r}=\arg \min_{r \in \{z_1,\ldots,z_T\} \cap (\eta_1,\eta_2)} \hat{G}(r).
\]

\medskip
When the dimension of two linear spaces goes to infinity, it is possible that the distance of the spaces also changes. We use a positive number $\beta_s \in[0,1]$ to quantify the growth rate of $\bQ_{s,1}$ and $\bQ_{s,2}$ as $p_s$ increases, $s=1,2$.
\[
[\cD(\cM(\bR_1),\cM(\bR_2))]^2 \asymp p_1^{\beta_1-1}, \mbox{ and  } [\cD(\cM(\bC_1),\cM(\bC_2))]^2 \asymp p_2^{\beta_2-1}.
\]
$\beta_1$ and $\beta_2$ not only reflect the distance of loading spaces, also measures how strong the row and column thresholding makes an impact on $\{\bX_t\}$. For example, if $\beta_1=0$, only a finite number of elements in $\bR_i$ is different across regimes, and we say the row  threshold is extremely weak; if $\beta_1=1$, the number of elements which undergo a change is $O(p_i)$, for $i=1,2$, and we say the row thresholding is strong. If $\cD(\cM(\bR_1),\cM(\bR_2)) > \cD(\cM(\bC_1),\cM(\bC_2))$, we say thresholding in row factors is stronger than column factors. If $\cD(\cM(\bR_1),\cM(\bR_2)) < \cD(\cM(\bC_1),\cM(\bC_2))$, we say thresholding in column factors is stronger than row factors. 

\begin{theorem}  \label{thm:threshold_value}
Under Conditions A1-A4 and C1-C7 in Appendix \ref{appendix:regular_condition}, with true $k_1$ and $k_2$, if $p_1^{\delta_{11}/2+\delta_{12}/2}p_2^{\delta_{21}/2+\delta_{22}/2} T^{-1/2}d_{\max}^{-1}=o(1)$, as $p_1,p_2, T \to \infty$, it holds that
\[
P(\hat{r} < r_0 -\epsilon) \leq \frac{Cp_1^{\delta_{11}/2+\delta_{1\min}/2} p_2^{\delta_{21}/2+\delta_{2\min}/2}}{\epsilon d_{\max}T^{1/2}}, \quad  P(\hat{r} > r_0 +\epsilon) \leq \frac{Cp_1^{\delta_{12}/2+\delta_{1\min}/2} p_2^{\delta_{22}/2+\delta_{2\min}/2}}{\epsilon d_{\max} T^{1/2}},
\]
for $\epsilon>0$, where $C$ is a positive constant and $d_{\max}=\max \{p_1^{\beta_1/2-1/2}, p_2^{\beta_2/2-1/2}\}$.
\end{theorem}

Theorem 2 shows that the estimator $\hat{r}$ is consistent. When all factors and thresholding are strong ($\delta_{si}=0, \beta_s=1$ for $s,i=1,2$), the estimator is immune to the curse of dimensionality. But if at least one is weak, the estimator gets less efficient when $p_1$ and $p_2$ increase. When two regimes have different levels of strength, the probability that $\hat{r}$ falls in the stronger regime is smaller than that in the weak regime, which is in line with the conclusions for vector-valued threshold factor models \citep{liu2019}. For the impact of thresholding, Theorem 2 shows that the performance of $\hat{r}$ is defined by the direction (row/column) with stronger thresholding. Assuming that threshold in row factors is stronger, comparing to vector-valued threshold factor models \citep{liu2019}, introducing column (weaker) dimension reduction does not hurt the threshold value estimation, and the existence of row (stronger) dimension reduction makes the estimator more efficient. The 'helping' effect also applies. It is worth noting that the overall convergence rate is determined by factor strengths of both regimes and thresholding effect of the stronger direction. The thresholding in the weaker direction will not change the estimation asymptotically.

The final estimation of loading spaces is obtained using $\hat{r}$ as the threshold value and following the procedure in Section 3.1.

\begin{theorem}  \label{thm:loading_spdist_est_r}
Under Conditions A1-A4 and C1-C7 in Appendix \ref{appendix:regular_condition}, with true $k_1$ and $k_2$, if $p_1^{\delta_{11}/2+\delta_{12}/2}p_2^{\delta_{21}/2+\delta_{22}/2} T^{-1/2}d_{\max}^{-1}=o(1)$, as $p_1,p_2, T\to \infty$, it holds that
\[
\cD(\cM(\hat{\bQ}_{s,i}(\hat{r})), \cM(\bQ_{s,i}))=O_p(p_1^{\delta_{1i}/2+\delta_{1\min}/2} p_2^{\delta_{2i}/2+\delta_{2\min}/2} T^{-1/2}d_{\max}^{-1}), \mbox{ for } s,i=1,2.
\]
\end{theorem}

From Theorem 3 we can see that when thresholding is strong in either row or column factors, i.e., $d_{\max}=1$, the asymptotics of the loading space estimators are the same with those when the true threshold value is known shown in Theorem 1. However, if thresholding is not strong in both directions ($\beta_1\neq 1$ and $\beta_2\neq 1$), the estimation suffers more from the increase of $p_1$ and $p_2$ comparing the case when $r_0$ is known.

\subsection{When the numbers of factors are unknown}
Since both the factor process $\bF_t$ and loadings are unobserved,  the numbers of factors $k_1$ and $k_2$ need to be estimated. \cite{lam2012} proposed a ratio-based estimator, and \cite{liu2016,liu2019} applied it to factor models with multiple regimes. Here we extend it for matrix-valued time series. Assume that $r_0$ is in a known interval $(\eta_1,\eta_2)$, and let
\begin{equation} \label{eqn:eigval_ratio}
\hat{k}_{s,i}=\arg \min_{1\leq k\leq R} \frac{\hat{\lambda}_{s,i,k+1}(\eta_1,\eta_2)}{\hat{\lambda}_{s,i,k}(\eta_1,\eta_2)}, \mbox{ for } s,i=1,2,
\end{equation}
where $\hat{\lambda}_{s,i,k}$ is the $k$-th largest eigenvalue of $\hat{\bM}_{s,i}(\eta_1,\eta_2)$. Since the eigenvalues practically will go to zero, here we cannot search up to $p_1$ or $p_2$. We follow \cite{lam2012} and use $R=p_s/2$ for $s=1,2$. 
\begin{corollary}
Under Conditions A1-A4 and C1-C7 in Appendix \ref{appendix:regular_condition}, if $p_1^{\delta_{11}/2+\delta_{12}/2}p_2^{\delta_{21}/2+\delta_{22}/2} T^{-1/2}=o(1)$, as $p_1,p_2, T \to \infty$, it holds that
\begin{eqnarray*}
\hat{\lambda}_{s,i,k+1}(\eta_1,\eta_2)/\hat{\lambda}_{s,i,k}(\eta_1,\eta_2) &\asymp& 1, \mbox{ for } k=1,\ldots,k_s-1,\\
\hat{\lambda}_{s,i,k_s+1}(\eta_1,\eta_2)/\hat{\lambda}_{s,i,k_s}(\eta_1,\eta_2) &=& O_p(p_1^{\delta_{1i}+\delta_{1\min}}p_2^{\delta_{2i}+\delta_{2\min}}T^{-1}),
\end{eqnarray*}
for $s,i=1,2$.
\end{corollary}
Corollary 1 presents the convergence rates of the ratios of estimated eigenvalues of $\bM_{s,i}(\eta_1,\eta_2)$. The stronger the regime is, the faster the ratio converges. Hence, we choose the one estimated by the regime with a larger 'strength' reflected by $\|\hat{\bM}_{s,i}(\eta_1,\eta_2)\|_2$ \citep{liu2016,liu2019}. We use
\begin{equation}
\hat{k}_s=\hat{k}_{s,\hat{q}_s}, \mbox{ where } \hat{q}_s=\arg \max_{q=1,2} \|\hat{\bM}_{s,q}(\eta_1,\eta_2)\|_2.   \label{eqn:est_ks}
\end{equation}

The corollary does not guarantee the consistency of the ratio-based estimator, and shows that the number of factors cannot be underestimated but may be overestimated \citep{lam2012}. The following results will tell that our estimators for loading spaces and threshold values are still consistent even when $k_1$ and $k_2$ are overestimated.

Let 
\begin{equation} \label{eqn:G(x)}
\hat{G}_{k_1,k_2}(r)=\sum_{s=1}^2\sum_{i=1}^2 \|\hat{\bB}_{s,i,k_s}(\eta_1,\eta_2)' \hat{\bM}_{s,i}(r) \hat{\bB}_{s,i,k_s}(\eta_1,\eta_2)\|_2,
\end{equation}
where $\hat{\bB}_{s,i,k_s}(\eta_1,\eta_2)=(\hat{\bq}_{i,s,k_s+1}(\eta_1,\eta_2),\ldots,\hat{\bq}_{s,i,p_s}(\eta_1,\eta_2))$, for $s,i=1,2$. When $k_1$ and $k_2$ are unknown, we estimate $r_0$ by

\begin{equation} \label{eqn:est_r}
\widetilde{r}=\arg \min_{r\in \{z_1,\ldots,z_T\}\cap(\eta_1,\eta_2)} \hat{G}_{\hat{k}_1,\hat{k}_2}(r).
\end{equation}

\begin{theorem} \label{thm:threshold_value2}
Under Conditions A1-A4 and C1-C8 in Appendix \ref{appendix:regular_condition}, if $p_1^{\delta_{11}/2+\delta_{12}/2}p_2^{\delta_{21}/2+\delta_{22}/2}T^{-1/2}d_{\max}^{-1}=o(1)$, as $p_1,p_2, T\to \infty$, it holds that
\[
P(\widetilde{r}<r_0-\epsilon)\leq \frac{Cp_1^{\delta_{11}/2+\delta_{1\min}/2}p_2^{\delta_{21}/2+\delta_{2\min}/2}}{\epsilon d_{\max} T^{1/2}},\quad
P(\widetilde{r}>r_0+\epsilon)\leq \frac{Cp_1^{\delta_{12}/2+\delta_{1\min}/2}p_2^{\delta_{22}/2+\delta_{2\min}/2}}{\epsilon d_{\max} T^{1/2}},
\]
for $\epsilon>0$, where $C$ is a positive constant.
\end{theorem}

The loading spaces are estimated using $\hat{k}_1$ and $\hat{k}_2$ as the numbers of factors and $\widetilde{r}$ as the threshold value. 
\begin{equation} \label{eqn:est_loading}
\widetilde{\bQ}_{s,i}(\hat{k}_s,\widetilde{r})=(\hat{\bq}_{s,i,1}(\widetilde{r}),\ldots,\hat{\bq}_{s,i,\hat{k}_s}(\widetilde{r})), \mbox{ for } s,i=1,2,
\end{equation}
where $\hat{\bq}_{s,i,k}(\widetilde{r})$ is the unit eigenvector of $\hat{\bM}_{s,i}(\widetilde{r})$ corresponding to its $k$-th largest eigenvalue.

Define $\widetilde{\bQ}_{s,i}(k_s,\widetilde{r})$, which consists of the first $k_s$ columns of $\widetilde{\bQ}_{s,i}(\hat{k}_s,\widetilde{r})$. The following theorem shows that its spanned space converges to the true loading space as fast as $\cM(\hat{\bQ}_{s,i}(\hat{r}))$ shown in Theorem 3.
\[
\widetilde{\bQ}_{s,i}(\widetilde{r})=(\hat{\bq}_{s,i,1}(\widetilde{r}),\ldots,\hat{\bq}_{s,i,{k}_s}(\widetilde{r})), \mbox{ for } s,i=1,2.
\]

\begin{theorem}\label{thm:loading_spdist_est_r2}
Under Conditions A1-A4 and C1-C8 in Appendix \ref{appendix:regular_condition}, if $p_1^{\delta_{11}/2+\delta_{12}/2}p_2^{\delta_{21}/2+\delta_{22}/2}T^{-1/2}d_{\max}^{-1}=o(1)$, as $p_1,p_2, T\to \infty$, it holds that
\[
\cD(\cM(\widetilde{\bQ}_{s,i}(\widetilde{r})),\cM(\bQ_{s,i}))=O_p(p_1^{\delta_{1i}/2+\delta_{1\min}/2}p_2^{\delta_{2i}/2+\delta_{2\min}/2}T^{-1/2}d_{\max}^{-1}), \mbox{ for } s,i=1,2.
\]
\end{theorem}
Theorems 4 and 5 show that when the numbers of factors are overestimated, though losing some efficiency, our estimators perform asymptotically as good as those when $k_1$ and $k_2$ are correctly estimated.

\subsection{Threshold Variable Identification}

Threshold variable searching has been well studied for univariate time series analysis \citep{tong1980,tong1990,chan1993}, and a typical candidate pool for threshold variable is the lag variables \citep{tsay1989,tsay1998}. However, it is a challenging problem to choose the threshold variable from such a large pool for high-dimensional data. In this paper, we adopt the method proposed by \cite{liu2019} to select the threshold variable. We estimate the loading spaces and threshold value using data $\{\bX_1,\ldots, \bX_{t_0}\}$. With those estimates, calculate the residual sum of squares for the remaining data,
\[
E=\sum_{t=t_0+1}^T \sum_{\ell=1}^{p_2} \sum_{i=1}^2 \left(\hat{\bB}_{1,i}' \bx_{t,\ell} \right)' \left(\hat{\bB}_{1,i}' \bx_{t,\ell} \right)I_{t,i}(\hat{r})+\sum_{t=t_0+1}^T \sum_{\ell=1}^{p_1} \sum_{i=1}^2 \left(\hat{\bB}_{2,i}' \bx_{t,\ell \cdot } \right)' \left(\hat{\bB}_{2,i}' \bx_{t,\ell \cdot} \right)I_{t,i}(\hat{r}).
\]
If the threshold variable is correctly selected and $r_0$ is given, then $\bB_{1,i}\bx_{t,\ell}I_{t,i}(r_0)=\bB_{1,i}\be_{t,\ell}I_{t,i}(r_0)$ and  $\bB_{2,i}\bx_{t,\ell\cdot}I_{t,i}(r_0)=\bB_{2,i}\be_{t,\ell \cdot}I_{t,i}(r_0)$. $E$ measures the residual sum of squares after the common factors are extracted. Hence, the preferred model is the one with minimum $E$ \citep{liu2019}.

\section{Simulation} \label{sec:simulation}
In this section, we present the empirical performance of the proposed method for synthetic data sets. In all the examples, the observed matrix time series $\bX_t$'s are generated according to Model~(\ref{model}). 
\begin{align*}
\bX_t= \left\{
\begin{array}{cc}
\bR_1 \bF_t \bC_1'+\bE_{t} &z_t <r_0, \\
\bR_2 \bF_t \bC_2'+\bE_{t} &z_t \geq r_0,\\
\end{array}\right. \quad t=1,\ldots, T.  
\end{align*}

The dimension of the latent factor matrix $\bF_t$ is fixed at $3 \times 3$. We simulate ${\rm vec}(\bF_t)$ from a vector autoregressive (VAR) model of order one. The AR coefficient matrix has diagonal entries of -0.8, 0.8, 0.9, -0.7, -0.9, 0.8, 0.7, 0.8, 0.7 and off-diagonal entries of 0. The threshold process $z_t$ follows an independent Gaussian process $\calN(0, 1)$. The threshold value $r_0$ is set at 0. The error process $\bE_t$ is a white noise process with mean $\bzero$ and a Kronecker product covariance structure, that is, $\cov({\rm vec}(\bE_t)) = \bGamma_2 \otimes \bGamma_1$, where $\bGamma_1$ and $\bGamma_2$ are of sizes $p_1 \times p_1$ and $p_2 \times p_2$, respectively. Both $\bGamma_1$ and $\bGamma_2$ have diagonal entries of 1 and off-diagonal entries of 0.2. 

We only consider the setting where the dimensions of the observed matrix time series and the number of time point are fixed at $(p_1, p_2, T) = (40, 40, 2400)$. The performance under different settings of $(p_1, p_2, T)$ can be summarized as that the error is smaller when $\frac{T}{p_1 p_2}$ is larger, which is similar to those shown in \cite{wang2019}, \cite{chen2017constrained} and \cite{liu2019}. 

The main focuses of this paper are the phenomenons associated specially with the threshold matrix factor models, namely, the factor-strength effect and the threshold-strength effect with respect to different regimes. They are manifested in settings with different combinations of factor strengths $\delta$'s and threshold strength $\beta$'s in the following two subsections. Loading matrices $\bR_1$, $\bR_2$, $\bC_1$, and $\bC_2$ are generated accordingly. The details will explained separately in each section. For all simulations, the reported results are based on 200 simulation runs. 

\subsection{Factor Strength Helping Effect} \label{sec:example1}

The general phenomenon of the \textit{factor-strength helping effect} in the threshold matrix factor model can be summarized as the fact that the strong regime actually provide helpful information for the weak regime. Thus, when at least one regime is strong, the proposed method obtains estimator with high precision even for weak regime.  This is aligned with the theoretical results presented in Theorem \ref{thm:loading_spdist_est_r} and Theorem \ref{thm:loading_spdist_est_r2}, and is also proved empirically through synthetic data in this example. 

Here, we fix the column factor strengths $\delta_{21}$ and $\delta_{22}$ at 0 and threshold strength in both row and column dimension reduction $\beta_1$ and $\beta_2$ at 1 (i.e. the two regimes are separated well apart for both row and column spaces), and study the factor strength effects by varying row factor strength $\delta_{11}$ and $\delta_{12}$ in different settings. Specifically, we consider different combinations of factor strength summarized in Table~\ref{table:fac_strength_comb}. All entries in $\bR_i$ and $\bC_i$ were generated independently from the uniform distribution on $[-p_s^{-\delta_{si}/2},p_s^{-\delta_{si}/2}]$ for $s,i=1,2$.

% \usepackage{multirow}

\begin{table}[htbp!]
	\caption{Different combinations of factor strength considered in Example 1.}
	\label{table:fac_strength_comb}
	\resizebox{\textwidth}{!}{%
		\begin{tabular}{|c|c|c|c|c|c|}
			\hline
			\multirow{3}{*}{Setting \#} & \multicolumn{2}{c|}{Row factor strength} & \multicolumn{2}{c|}{Column factor strength} & \multirow{3}{*}{Interpretation} \\ \cline{2-5}
			& Regime 1 & Regime 2 & Regime 1 & Regime 2 &  \\ \cline{2-5}
			& $\delta_{11}$ & $\delta_{12}$ & $\delta_{21}$ & $\delta_{22}$ &  \\ \hline
			1 & 0 & 0 & 0 & 0 & All regimes have strong factors. \\ \hline
			2 & 0.5 & 0 & 0 & 0 & Row factors are weak in regime one. \\ \hline
			3 & 0 & 0.5 & 0 & 0 & Row factors are weak in regime two. \\ \hline
			4 & 0.5 & 0.5 & 0 & 0 & Row factors are weak in both regimes. \\ \hline
		\end{tabular}%
	}
\end{table}

%\begin{table}[htbp!]
%	\caption{Different combinations of factor strength considered in Example 1.}
%	\label{table:fac_strength_comb}
%	\resizebox{\textwidth}{!}{%
%		\begin{tabular}{|c|c|c|c|c|c|}
%			\hline
%			\multirow{3}{*}{Setting \#} & \multicolumn{2}{c|}{Regime 1} & \multicolumn{2}{c|}{Regime2} & \multirow{3}{*}{Interpretation} \\ \cline{2-5}
%			& Row & Column & Row & Column &  \\ \cline{2-5}
%			& $\delta_{11}$ & $\delta_{21}$ & $\delta_{12}$ & $\delta_{22}$ &  \\ \hline
%			1 & 0 & 0 & 0 & 0 & All regimes have strong factors. \\ \hline
%			2 & 0.5 & 0 & 0 & 0 & Row factors are weak in regime one. \\ \hline
%			3 & 0 & 0 & 0.5 & 0 & Row factors are weak in regime two. \\ \hline
%			4 & 0.5 & 0 & 0.5 & 0 & Row factors are weak in both regimes. \\ \hline
%		\end{tabular}%
%	}
%\end{table}

We first investigate the accuracy in estimating the number of factors $k_1$ and $k_2$ when true threshold value $r_0$ is unknown. Table \ref{table:k1_k2_accuracy_exmp1} shows the frequency of estimated pair $(\hat{k}_1,\hat{k}_2)$ obtained by (\ref{eqn:est_ks}). Note that the last column of the table corresponds to the situation when the estimated $(\hat{k}_1,\hat{k}_2)$ correctly recover the truth $(3,3)$. When true threshold value $r_0$ is unknown, the proposed method only uses the sample points that are certain to be in either one of the two regimes. Specifically, we use the sample points where $z_t$ is under 25-th percentile for regime 1 and those where $z_t$ is above 75-th percentile for regime 2. Thus for each dimension, the effective sample size is only a quarter of the whole sample size. The proposed method can estimate $(k_1,k_2)$ with high precision when at least one regime is strong -- the strong regime actually provide helpful information for the weak regime. We call this phenomenon \textit{the helping effect from the strong regime to the weak regime}. When both regimes are weak, the proposed method tends to overestimate the number of factors. The results are in line with conclusions in Corollary 1.  Note that latent dimensions are almost always overestimated in our specific setting of $(p_1, p_2, T, \beta_1, \beta_2)$. However, this is not always the case when we use different settings of $(p_1, p_2, T, \beta_1, \beta_2)$. As shown in \cite{wang2019} and \cite{liu2019}, $(k_1, k_2)$ can be underestimated, especially when $T$ is small. 

%Comparing between the two table, we can observe that pair $(k_1,k_2)$ is estimated with higher accuracy when true threshold value $r_0$ is known. This is because the efficient sample size is larger in this case. 
% \usepackage{graphicx}
%\begin{table}[hbtp!]
%	\centering
%	\caption{Accuracy in estimating the number of factors $k_1$ and $k_2$ when true threshold value $r_0$ is known.}
%	\label{table:k1_k2_accuracy_r0}
%	\resizebox{\textwidth}{!}{%
%		\begin{tabular}{llll|llllllllllll}
%			\hline
%			pr & pc & gamma & deltas & (4,3) & (4,2) & (4,1) & (3,3) & (3,2) & (3,1) & (2,3) & (2,2) & (2,1) & (1,3) & (1,2) & (1,1) \\ \hline
%			20 & 20 & 0.5 & (0.5,0,0,0) & 0 & 0 & 0 & 0.05 & 0.4 & 0 & 0 & 0.5 & 0 & 0 & 0 & 0.05 \\ \hline
%		\end{tabular}%
%	}
%\end{table}

% \usepackage{multirow}
% \usepackage{graphicx}
\begin{table}[htbp!]
	\centering
		\caption{Accuracy in estimating the number of factors $k_1$ and $k_2$ when true threshold value $r_0$ is unknown in cases considered in Section \ref{sec:example1}}
	\resizebox{0.5\textwidth}{!}{%
		\begin{tabular}{|ccc|c|c|c|c|c|}
			\hline
			$p_1$ & $p_2$ & $T$ & $\beta_1, \beta_2$ & $\delta_{11}, \delta_{12}$ & (4,4) & (4,3) & (3,3) \\ \hline
			&  &  &  & 0, 0 & 0 & 0 & 1.00 \\ \cline{5-8} 
			&  &  &  & 0.5, 0 & 0 & 0 & 1.00 \\ \cline{5-8} 
			&  &  &  & 0, 0.5 & 0 & 0 & 1.00 \\ \cline{5-8} 
			\multirow{-4}{*}{40} & \multirow{-4}{*}{40} & \multirow{-4}{*}{2400} & \multirow{-4}{*}{1,1} & 0.5, 0.5 & 1.00 & 0 & 0 \\ \hline
		\end{tabular}%
	}
	\label{table:k1_k2_accuracy_exmp1}
\end{table}

\begin{figure}[htbp!]
	\centering	
	\includegraphics[width=0.65\textwidth, keepaspectratio]{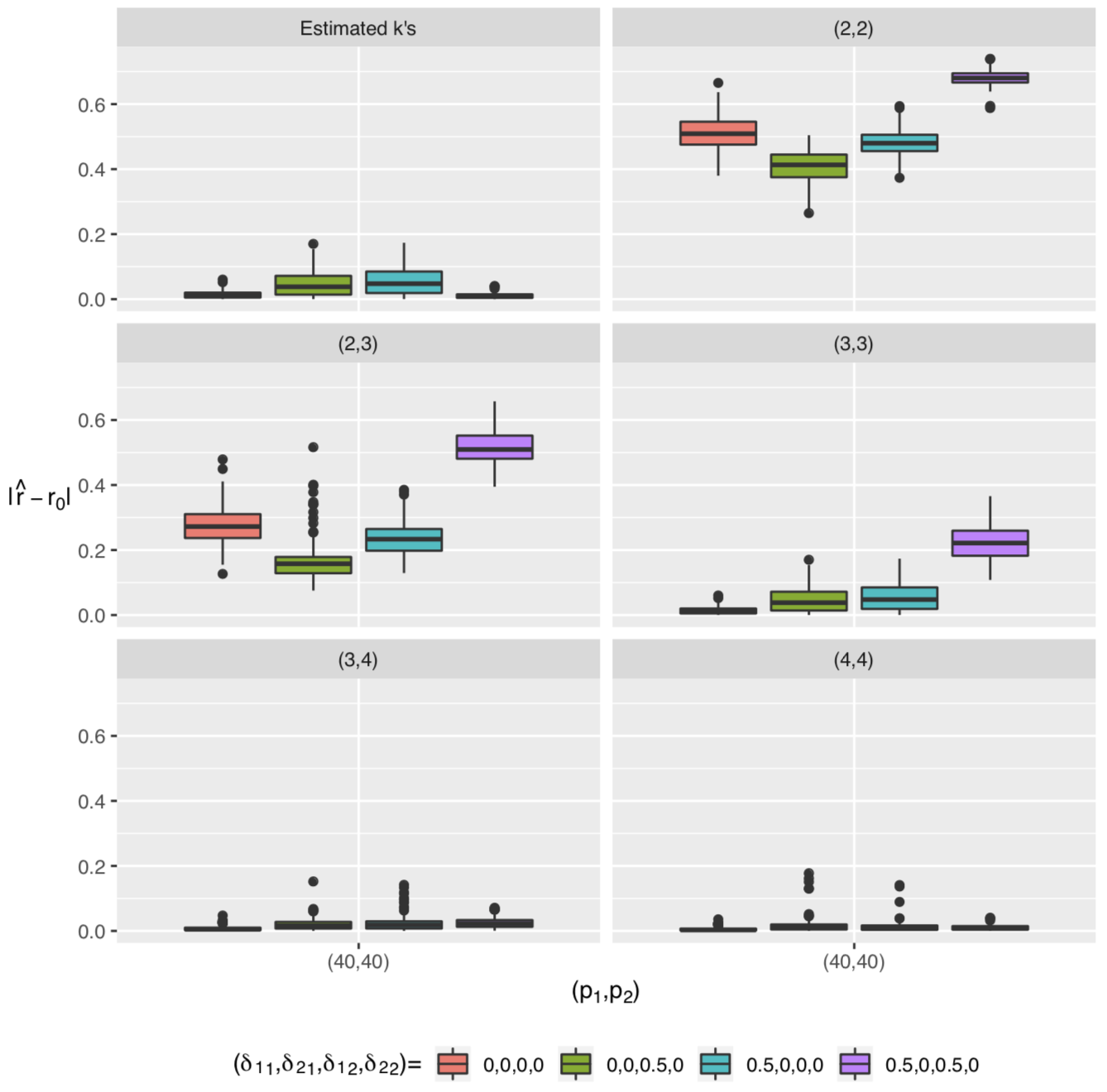}%{abs_err_rhat_betas_1_1}
	\caption{Box plots of the absolute error $|\hat{r} - r_0|$ when $(\delta_{21},\delta_{22})=(0,0)$ and $(\beta_1, \beta_2) = (1,1)$. Different panels correspond to different values of $(k_1, k_2)$ used in (\ref{eqn:G(x)}) and (\ref{eqn:est_r}). In each panel, four box plots show estimation results when $(\delta_{11},\delta_{12})=(0,0)$, $(\delta_{11},\delta_{12})=(0.5,0)$, $(\delta_{11},\delta_{12})=(0,0.5)$, and $(\delta_{11},\delta_{12})=(0.5,0.5)$, respectively.}
	\label{fig:abs_err_rhat_betas_1_1_exmp1}
\end{figure}

Secondly, we examine how well the proposed method can recover the threshold value without knowing true latent dimensions. Figure \ref{fig:abs_err_rhat_betas_1_1_exmp1} presents the box plots of the absolute error $|\hat{r} - r_0|$ when $(\delta_{21},\delta_{22})=(0,0)$ and $(\beta_1, \beta_2) = (1,1)$. Different panels correspond to different values of $(k_1, k_2)$ used in (\ref{eqn:G(x)}) and (\ref{eqn:est_r}). For example, the left panel in the first row corresponds to the case with latent dimension $(\hat{k}_1,\hat{k}_2)$ obtained by (\ref{eqn:est_ks}). The right panel on the first row with label $(2,2)$ correspond to the case with (underestimated) latent dimension $(2,2)$ to calculate $\hat{r}$. The two panels on the last row correspond to two overestimated case where the latent dimensions are $(3,4)$ and $(4,4)$, respectively. Table \ref{table:meansd_of_abs_err_r0_exmp1} shows the mean and standard deviation (in the parentheses) of the absolute error $|\hat{r} - r_0|$ with different combinations of $(k_1, k_2)$. We again observe \textit{the helping effect from the strong regime to the weak regime}. Consider the case when $(\delta_{11}, \delta_{21}, \delta_{12}, \delta_{22}) = (0, 0, 0.5, 0)$ or $(0.5, 0, 0, 0))$, the weak regime does not affect the performance of our algorithm. The results also show  that the threshold value $r_0$ is estimated with higher accuracy when the latent dimensions are correctly estimated or overestimated. When at least one of $(k_1, k_2)$ is underestimated, the errors of $\hat{r}$ are very large, as shown in the last five rows in Table \ref{table:meansd_of_abs_err_r0_exmp1}.

\begin{table}[htbp!]
	\centering
	
		\caption{Mean and standard deviation (in the parentheses) of the absolute error $|\hat{r} - r_0|$ with $\hat{r}$ estimated by (\ref{eqn:est_r}) with different combinations of $(k_1, k_2)$.}
	\resizebox{0.65\textwidth}{!}{%
		\begin{tabular}{|c|cccc|}
			\hline
			\multicolumn{5}{|c|}{$(p_1, p_2)=(40, 40)$, $T=2400$, $(\beta_1, \beta_2) = (1,1)$, $(\delta_{21},\delta_{22})=(0,0)$} \\ \hline
			$\delta_{11}, \delta_{12}$ & 0, 0 & 0, 0.5 & 0.5, 0 & 0.5, 0.5 \\ \hline 
			$(\hat{k}_1, \hat{k}_2)$ & 0.014 (0.011) & 0.047 (0.039) & 0.054 (0.039) & 0.011 (0.008) \\ \hline
			(4,4) & 0.005 (0.006) & 0.017 (0.024) & 0.013 (0.017) & 0.011 (0.008) \\ \hline
			(4,3) & 0.007 (0.007) & 0.023 (0.026) & 0.020 (0.022) & 0.039 (0.022) \\ 
			(3,4) & 0.008 (0.007) & 0.020 (0.017) & 0.024 (0.024) & 0.025 (0.016) \\ \hline
			(3,3) & 0.014 (0.011) & 0.047 (0.039) & 0.054 (0.039) & 0.223 (0.052) \\ \hline
			(4,2) & 0.314 (0.051) & 0.388 (0.048) & 0.440 (0.044) & 0.491 (0.053) \\ 
			(2,4) & 0.274 (0.057) & 0.163 (0.056) & 0.236 (0.053) & 0.486 (0.049) \\ \hline
			(3,2) & 0.312 (0.051) & 0.388 (0.049) & 0.433 (0.046) & 0.514 (0.051) \\ 
			(2,3) & 0.273 (0.057) & 0.164 (0.060) & 0.237 (0.052) & 0.517 (0.049) \\ \hline
			(2,2) & 0.509 (0.057) & 0.410 (0.047) & 0.481 (0.043) & 0.681 (0.024) \\ \hline
		\end{tabular}%
	}
	\label{table:meansd_of_abs_err_r0_exmp1}
\end{table}

\begin{figure}[htbp!]
	\centering	
	\includegraphics[width=0.6\textwidth, keepaspectratio]{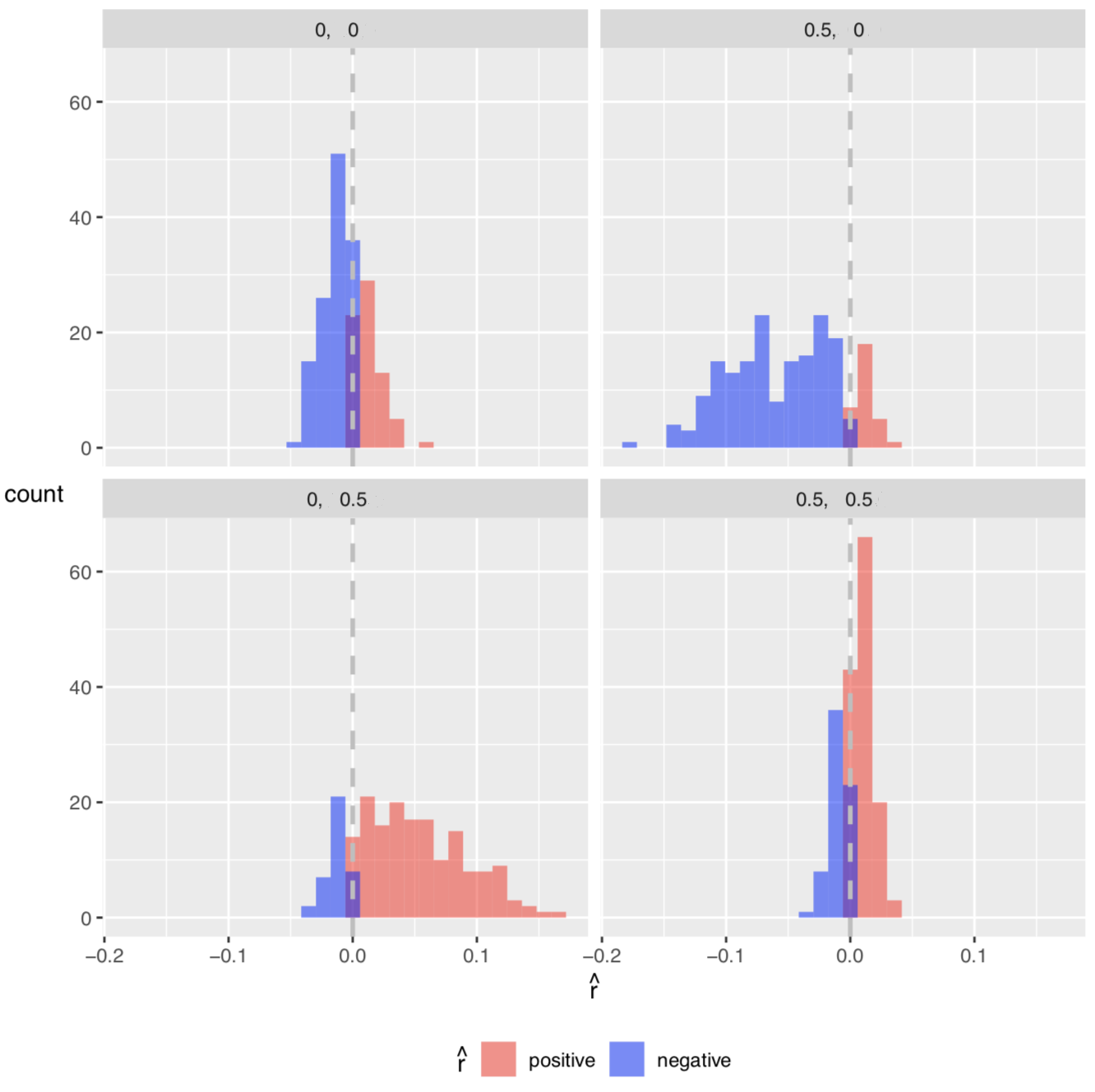}%{rhat_distn_betas_1_1}
	\caption{Histograms of $\hat{r}$ with $(\hat{k}_1, \hat{k}_2)$ estimated by (\ref{eqn:est_ks}) when $(\delta_{21},\delta_{22})=(0,0)$ and $(\beta_1, \beta_2) = (1,1)$. Panel titles show the values for $(\delta_{11},\delta_{12})$.} %Four panels show estimation results when $(\delta_{11},\delta_{12})=(0,0)$ (top left), $(\delta_{11},\delta_{12})=(0.5,0)$ (top right), $(\delta_{11},\delta_{12})=(0,0.5)$ (bottom left), and $(\delta_{11},\delta_{12})=(0.5,0.5)$ (bottom right), respectively.}
	\label{fig:rhat_distn_betas_1_1_exmp1}
\end{figure}

Figure \ref{fig:rhat_distn_betas_1_1_exmp1} presents the histograms of $\hat{r}$ with $(\hat{k}_1, \hat{k}_2)$ estimated by (\ref{eqn:est_ks}) when $(\delta_{21},\delta_{22})=(0,0)$ and $(\beta_1, \beta_2) = (1,1)$. Different panels correspond to different combination of row strengths $\delta_{11},  \delta_{12}$. From the figure, it is clear that $\hat{r}$ tends to be biased towards negative when regime 1 ($z_t < r_0$) has a weak factor (top right figure) and $\hat{r}$ tends to be biased towards positive when regime 2 ($z_t \ge r_0$) has a weak factor (bottom left), which confirms the conclusions in Theorem \ref{thm:threshold_value} and Theorem  \ref{thm:threshold_value2}. The last figure at the bottom right shows tight concentration around true threshold value zero. 

\begin{figure*}[htbp!]
	\centering
	\includegraphics[width=0.6\textwidth]{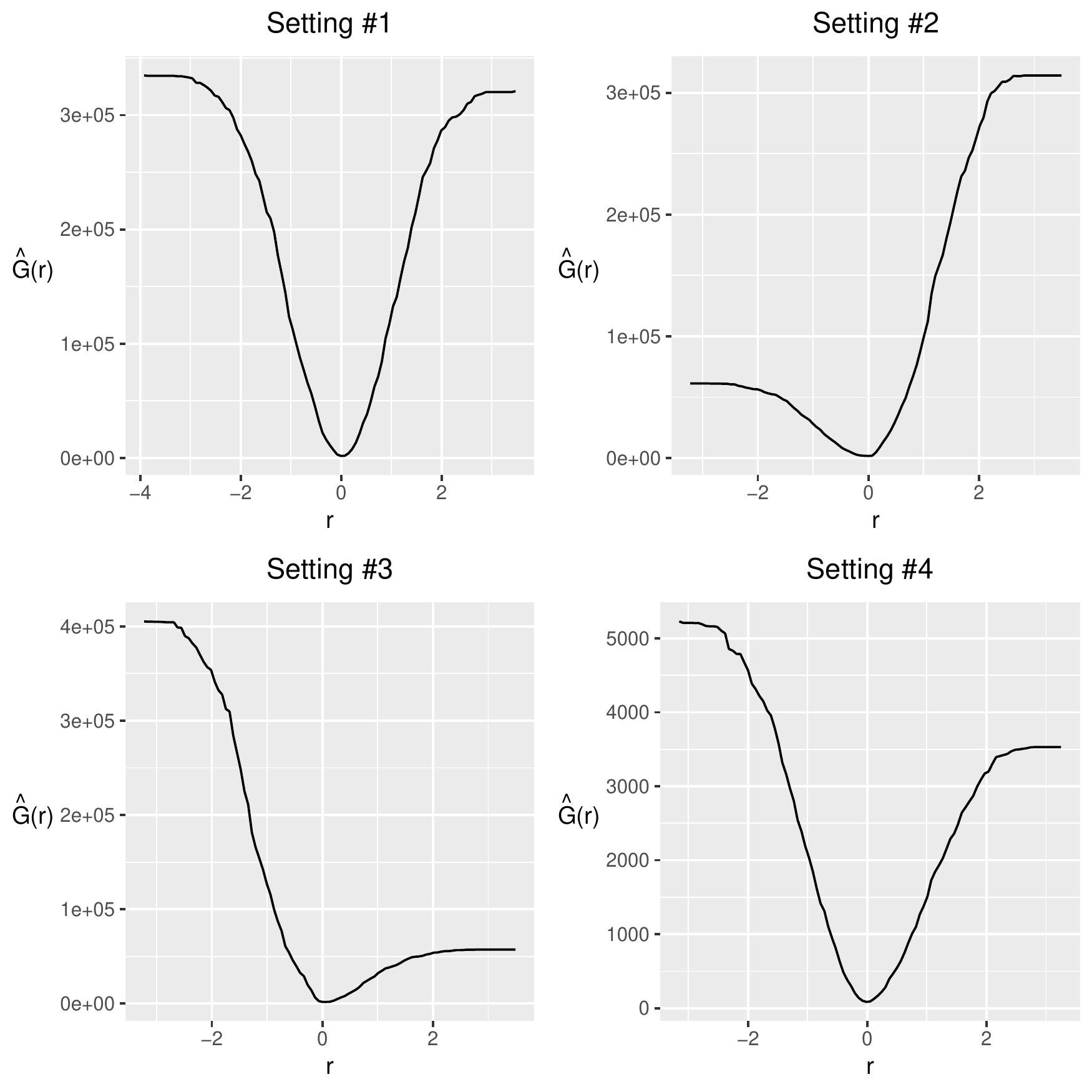}
    \caption[] {\small Plots of $\hat{G}(r)$'s under four settings in Table \ref{table:fac_strength_comb} for typical data sets of Example 1.}% We keep column factors strong in all both regimes under all settings and change the row factor strengths in different regimes. Specifically, under Setting \# 1 (\#4), row factors are strong (weak) in both regimes. Under Setting \#2 (\#3), row factors are strong (weak) in regime one (two) only. In all settings, $\hat{G}(r)$ is based on estimated $\hat{k}_1$ and $\hat{k}_2$. } 
    \label{fig:g(r)setting1234}
\end{figure*}

To investigate the bias in present of a single weak regime, we plot $\hat{G}(r)$ in Figure \ref{fig:g(r)setting1234} for typical data sets under the four different settings in Table \ref{table:fac_strength_comb}. %In all settings, the true $k_1$ and $k_2$ are unknown and $\hat{G}(r)$ is based on estimated $\hat{k}_1$ and $\hat{k}_2$. 
We can observe that $\hat{G}(r)$ approaches the theoretical minimum value 0 of $G(r)$ around $r=0$ when row factors are strong in at least one regime (Setting \#1, 2, 3). When row factors are weak in both regimes (Setting \#4), the range of $\hat{G}(r)$ is small, comparing with those in the other settings. Also, the minimum value of $\hat{G}(r)$ in Setting \#4 is larger than 0, but it still occurs around $r=0$. In Setting \#2 and \#3 where row factors have different levels of strength, $\hat{G}(r)$ is much large in the stronger regime. This also explains the existence of bias we observe in Figure \ref{fig:rhat_distn_betas_1_1_exmp1}. 

\begin{figure}[hbtp!]
	\centering
	\includegraphics[width=0.5\textwidth, keepaspectratio]{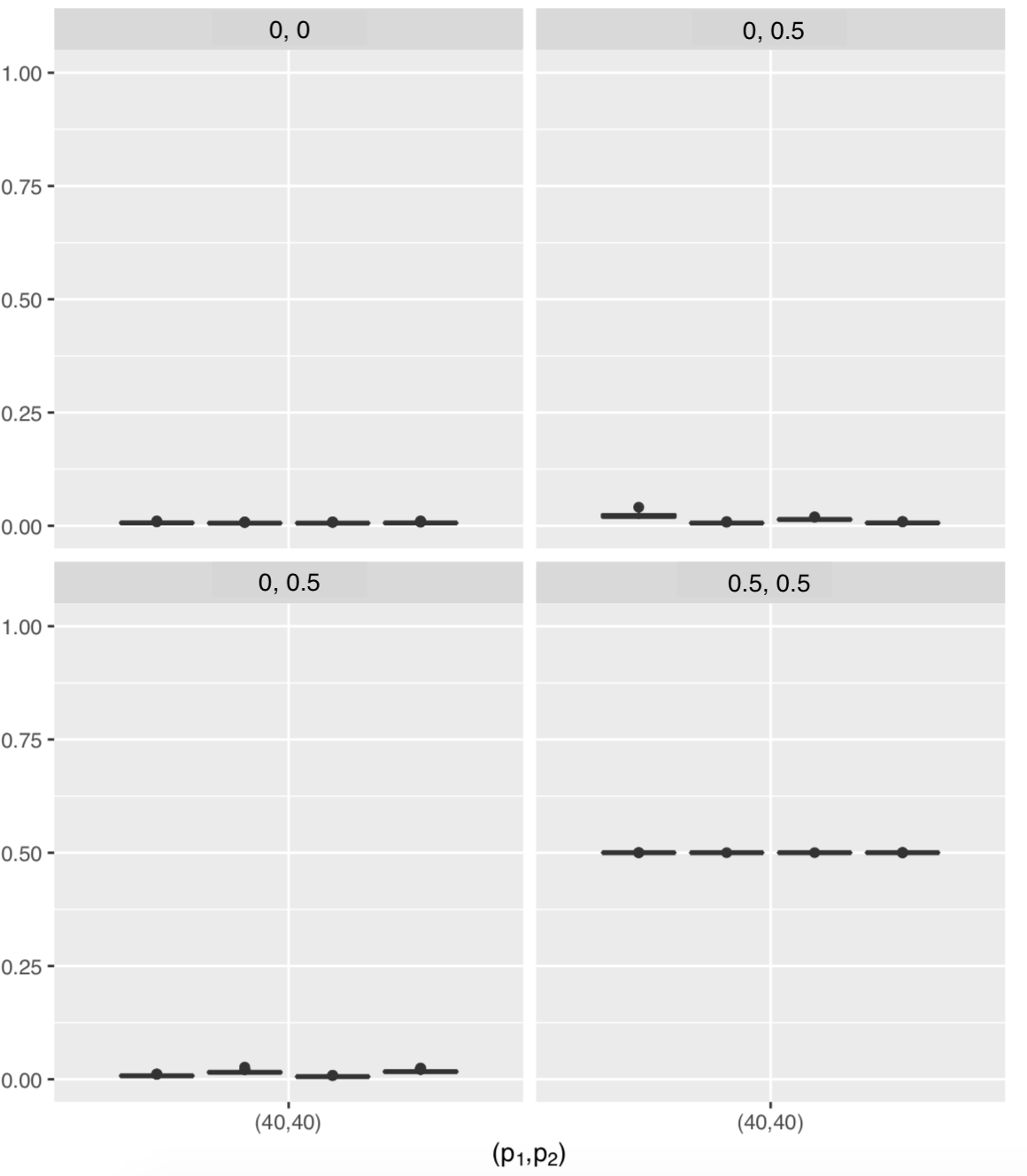}%{loading_spdist_1_1_y01}
	\caption{Boxplots of estimation errors for loading spaces under different row factor strength combinations when $(\delta_{21},\delta_{22})=(0,0)$ and $(\beta_1,\beta_2)=(1,1)$. Panel titles show the values of $\delta_{11}$ and $\delta_{12}$. Each panel includes four box plots, which are estimation results for $\hat{\bQ}_{1,1}$, $\hat{\bQ}_{2,1}$, $\hat{\bQ}_{1,2}$, and $\hat{\bQ}_{2,2}$, respectively. } 
	\label{fig:loading_spdist_1_1} 
\end{figure}

Thirdly, we investigate the accuracy in estimating the loading spaces when true $k_1$, $k_2$ and $r_0$ are unknown. The pair $(\hat{k}_1, \hat{k}_2)$ is estimated by (\ref{eqn:est_ks}) and $\hat{r}$ is estimated by (\ref{eqn:est_r}). Then a representative matrix of the row or column loading space is estimated according to (\ref{eqn:est_loading}) for each regime. Figure \ref{fig:loading_spdist_1_1} shows box plots of space distance for row and column loading matrices $\hat{\bQ}_{1,1}$, $\hat{\bQ}_{2,1}$, $\hat{\bQ}_{1,2}$, and $\hat{\bQ}_{2,2}$ under different factor strength combinations. As shown in the top-left figure $(\delta_{11},\delta_{12})=(0,0)$, when all factors are strong in both regimes, all loading spaces are estimated with high accuracy. As shown in the bottom-right figure $(\delta_{11},\delta_{12})=(0.5,0.5)$, when both regimes have weak row factors, the estimated row loading spaces are away from the truth. The estimation of column loading spaces is also affected by the weak row factors. As shown in the top-right figure $(\delta_{11},\delta_{12})=(0.5,0)$ and bottom-left figure $(\delta_{11},\delta_{12})=(0,0.5)$, when only one regime has a weak row factor, all loading spaces in the regime with the weak factor are estimated with lower accuracy than those in the regime with all strong factors. However, the discrepancies from the true loading spaces are smaller than those when both regimes contain strong factor.   

\subsection{Threshold Strength Helping Effect} \label{sec:example2}

The general phenomenon of the threshold strength helping effect in the threshold matrix factor model can be summarized as the fact that the row (column) loading with strong threshold strength can help with the column (row) loading with weak threshold strength. Thus at least one of row or column factors experiencing strong thresholding can guarantee an accurate estimation. This is aligned with the theoretical results presented in Theorem \ref{thm:threshold_value} and \ref{thm:loading_spdist_est_r} and is again verified empirical through the synthetic data set. 

In this example, we investigate the threshold strength effect as specified Condition 9 in Appendix \ref{appendix:regular_condition}. Specifically, we consider different combinations of threshold strength summarized in Table~\ref{table:threshold_strength_comb}. To ensure that the factors strengths and thresholding strengths are as we specify in Table~\ref{table:threshold_strength_comb}, by the fact that $\cD(\cM(\bQ_1),\cM(\bQ_2))$ is bounded by $\sqrt{2} \|\bQ_1-\bQ_2\|_2$ for any two $p \times q$ orthonormal matrices $\bQ_1$ and $\bQ_2$ shown in Theorem 3 of \cite{liu2016}, we do the following: first generate two matrices $\bR$ and $\bC$ with all entries independently generated from the uniform distribution on $[-1,1]$, and let $\bR_1=p_1^{-\delta_{11}/2}\bR$ and $\bC_1=p_2^{-\delta_{21}/2}\bC$; then sample $k_1p_1^{\beta_1}$ entries in $\bR$ and $k_2p_2^{\beta_2}$ entries in $\bC$, and replace them with new values independently generated from the uniform distribution on $[-1,1]$, where the new matrices are called $\bR^{(new)}$ and $\bC^{(new)}$;  let $\bR_2=p_s^{-\delta_{12}/2}\bR^{(new)}$ and $\bC_2=p_2^{-\delta_{22}/2}\bC^{(new)}$.

% \usepackage{multirow}
% \usepackage{graphicx}
\begin{table}[htpb!]
	\centering
	\caption{Different combinations of threshold strength considered in Example 2.}
	\label{table:threshold_strength_comb}
	\resizebox{\textwidth}{!}{%
		\begin{tabular}{|c|cc|cc|c|c|c|}
			\hline
			\multirow{3}{*}{Setting \#} & \multicolumn{2}{c|}{Row factor strength} & \multicolumn{2}{c|}{Column factor strength} & \multirow{2}{*}{Row threshold strength} & \multirow{2}{*}{Column threshold effect} & \multirow{3}{*}{Interpretation} \\ \cline{2-5}
			& Regime 1 & Regime 2 & Regime 1 & Regime 2 &  &  &  \\ \cline{2-7}
			& $\delta_{11}$ & $\delta_{12}$ & $\delta_{21}$ & $\delta_{22}$ & $\beta_1$ & $\beta_2$ &  \\ \hline
			1 & 0 & 0 & 0 & 0 & 1 & 1 & Both row and column regimes are orthogonal. \\ \hline
			2 & 0 & 0 & 0 & 0 & 0.5 & 1 & Row regimes are orthogonal. Column regimes are close. \\ \hline
			3 & 0 & 0 & 0 & 0 & 0.5 & 0.5 & Both row and column regimes are close. \\ \hline
			4 & 0.5 & 0.5 & 0 & 0 & 1 & 1 & Both row and column regimes are orthogonal. \\ \hline
			5 & 0.5 & 0.5 & 0 & 0 & 0.5 & 1 & Row regimes are orthogonal. Column regimes are close. \\ \hline
			6 & 0.5 & 0.5 & 0 & 0 & 0.5 & 0.5 & Both row and column regimes are close. \\ \hline
		\end{tabular}%
	}
\end{table}

First, we examine the accuracy in estimating the number of factors $k_1$ and $k_2$ when true threshold value $r_0$ is unknown under different threshold strength. Table \ref{table:k1_k2_accuracy_exmp2} shows the frequency of estimated pair $(\hat{k}_1,\hat{k}_2)$ obtained by (\ref{eqn:est_ks}). Here true threshold value $r_0$ is unknown, we use the sample points where $z_t$ is under 25-th percentile for regime 1 and those where $z_t$ is above 75-th percentile for regime 2 to estimate $(\hat{k}_1,\hat{k}_2)$. As shown in the table, threshold strength does not affect the accuracy of the estimators $(\hat{k}_1,\hat{k}_2)$. This is reasonable since the range of $z_t$ for separating regime 1 and regime 2 is correct. Thus $(\hat{k}_1,\hat{k}_2)$ is only affected by factor strength and efficient sample size -- a quarter of the whole sample size for each $k_i$, $i=1,2$.

\begin{table}[htbp!]
	\centering
		\caption{Accuracy in estimating the number of factors $k_1$ and $k_2$ when true threshold value $r_0$ is unknown in cases considered in Section \ref{sec:example2}.}
			\resizebox{0.5\textwidth}{!}{%
	\begin{tabular}{|c|c|c|c|c|c|c|c|}
		\hline
		$p_1$ & $p_2$ & $T$ & $\delta_{11}, \delta_{12}, \delta_{21}, \delta_{22}$ & $\beta_1, \beta_2$ & (4,4) & (4,3) & (3,3) \\ \hline
		\multirow{6}{*}{40} & \multirow{6}{*}{40} & \multirow{6}{*}{2400} & \multirow{3}{*}{0,0,0,0} & 1,1 & 0 & 0 & 1.00 \\ \cline{5-8} 
		&  &  &  & 0.5,1 & 0 & 0 & 1.00 \\ \cline{5-8} 
		&  &  &  & 0.5,0.5 & 0 & 0 & 1.00 \\ \cline{4-8} 
		&  &  & \multirow{3}{*}{0.5,0.5,0,0} & 1,1 & 1.00 & 0 & 0 \\ \cline{5-8} 
		&  &  &  & 0.5,1 & 1.00 & 0 & 0 \\ \cline{5-8} 
		&  &  &  & 0.5,0.5 & 0.99 & 0.01 & 0 \\ \hline
	\end{tabular}
	}
	\label{table:k1_k2_accuracy_exmp2}
\end{table}

\begin{figure}[htbp!]
	\centering	
	\includegraphics[width=0.6\textwidth, keepaspectratio]{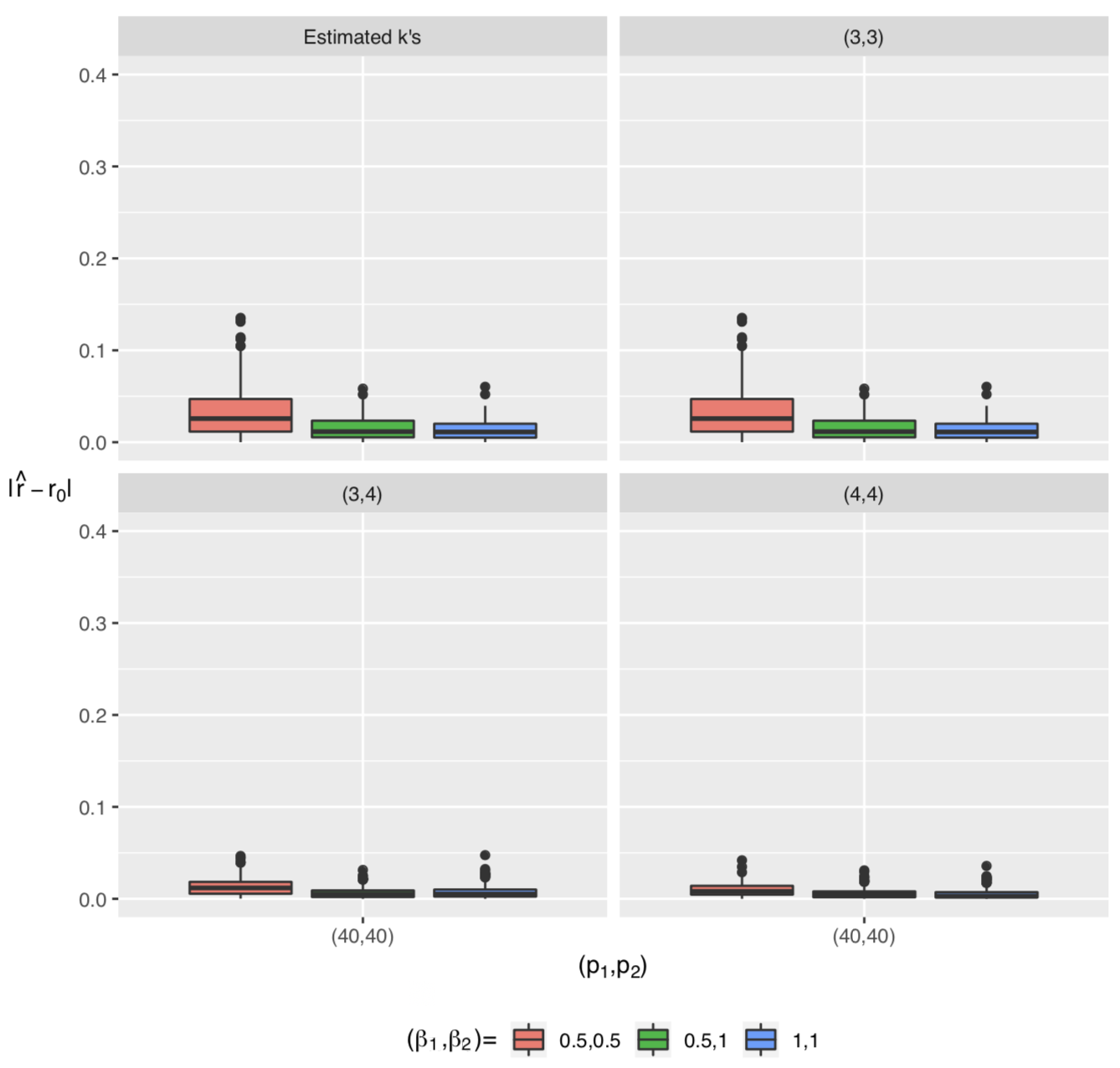}%{abs_err_rhat_deltas_0000}
	\caption{Box plots of the absolute error $|\hat{r} - r_0|$ with different thresholding strength $(\beta_1, \beta_2)$ with strong row and column factors $(\delta_{11}, \delta_{21}, \delta_{12}, \delta_{22}) = (0,0,0,0)$. Different panels correspond to different values of $(k_1, k_2)$ used in (\ref{eqn:G(x)}) and (\ref{eqn:est_r}). Each panel contains three box plots with different thresholding strengths, $(\beta_1,\beta_2)=(0.5,0.5)$, $(\beta_1,\beta_2)=(0.5,1)$, and $(\beta_1,\beta_2)=(1,1)$, respectively.}
	\label{fig:abs_err_rhat_betas_1_1_exmp2_0000}
\end{figure}

\begin{figure}[htbp!]
	\centering	
	\includegraphics[width=0.6\textwidth, keepaspectratio]{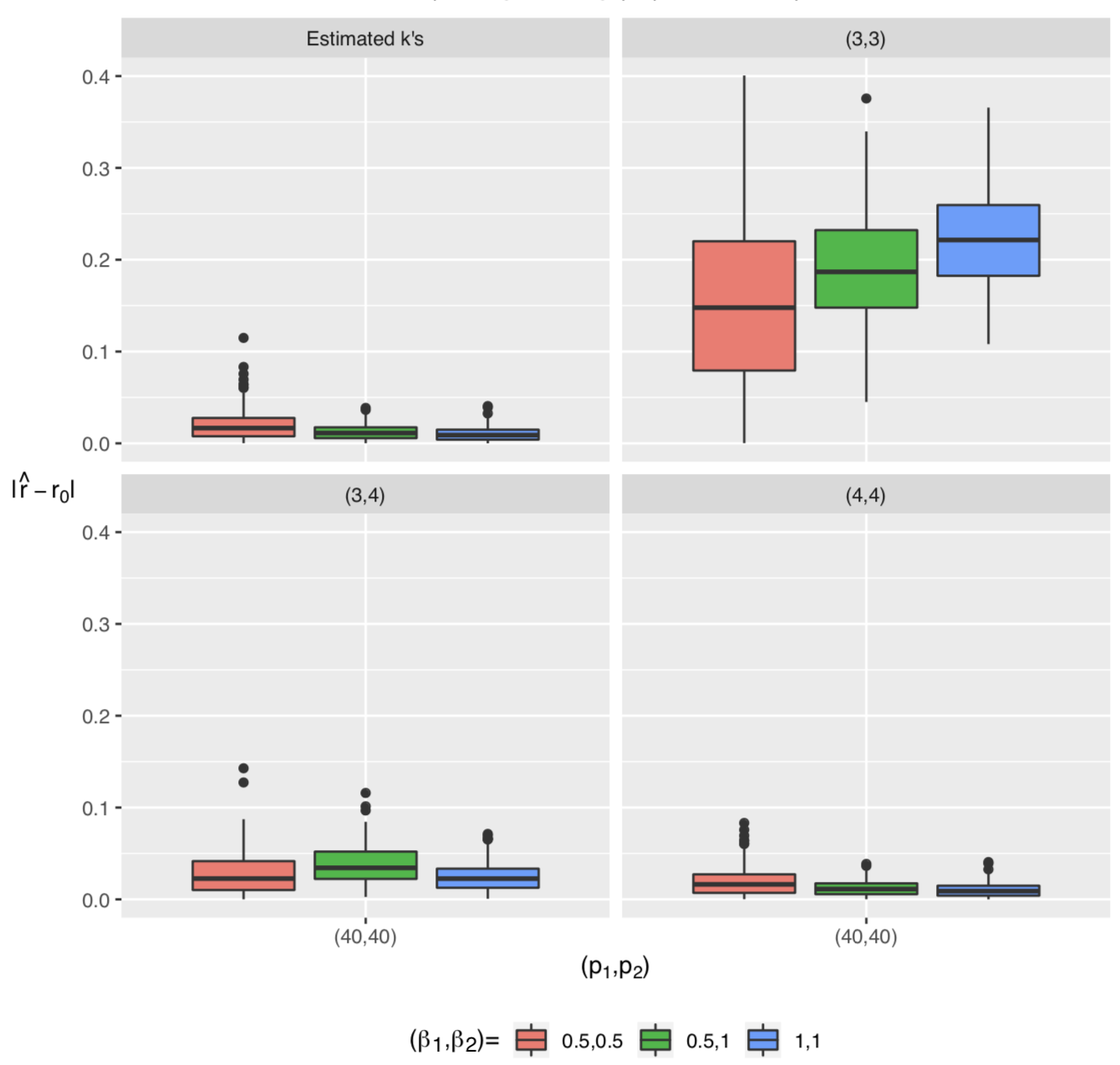}%{abs_err_rhat_deltas_5050}
	\caption{Box plots of the absolute error $|\hat{r} - r_0|$ with weak row and strong column factors $(\delta_{11}, \delta_{12}, \delta_{21}, \delta_{22}) = (0.5,0.5,0,0)$. Different panels correspond to different values of $(k_1, k_2)$ used in (\ref{eqn:G(x)}) and (\ref{eqn:est_r}), shown in the panel titles. Each panel contains three box plots with different thresholding strengths, $(\beta_1,\beta_2)=(0.5,0.5)$, $(\beta_1,\beta_2)=(0.5,1)$, and $(\beta_1,\beta_2)=(1,1)$, respectively.}
	\label{fig:abs_err_rhat_betas_1_1_exmp2_5050}
\end{figure}

Secondly, we examine how different threshold strengths affect the recovery of the threshold value without knowing true latent dimensions. Figures \ref{fig:abs_err_rhat_betas_1_1_exmp2_0000} and \ref{fig:abs_err_rhat_betas_1_1_exmp2_5050} present the box plots of the absolute error $|\hat{r} - r_0|$ under different threshold strength $(\beta_1, \beta_2)$ with $(\delta_{11}, \delta_{12}, \delta_{21}, \delta_{22}) = (0,0,0,0)$ and $(\delta_{11}, \delta_{12}, \delta_{21}, \delta_{22}) = (0.5,0.5,0,0)$, respectively. The absolute error $|\hat{r} - r_0|$ has smaller mean and variance under strong factors (Figure \ref{fig:abs_err_rhat_betas_1_1_exmp2_0000}) than those under weak row factor (Figure \ref{fig:abs_err_rhat_betas_1_1_exmp2_5050}). The performances for the cases where $(\beta_1, \beta_2) = (1,1)$ and $(\beta_1, \beta_2) = (0.5,1)$ are approximately the same, which is in line with conclusions in Theorem 2 and Theorem 4. Both are better than the case where $(\beta_1, \beta_2) = (0.5,0.5)$. We called this phenomena \textit{the helping effect from the strong column thresholding to the weak row thresholding}. 

\begin{table}[htbp!]
	\centering
	\resizebox{0.7\textwidth}{!}{%
		\begin{tabular}{|c|c|c|c|c|c|c|}
			\hline
			\multicolumn{7}{|c|}{$(p_1, p_2)=(40, 40)$, $T=2400$} \\ \hline
			$\delta_{11}, \delta_{12}, \delta_{21}, \delta_{22}$ & \multicolumn{3}{c|}{0,0,0,0} & \multicolumn{3}{c|}{0.5,0.5,0,0} \\ \hline
			$\beta_1, \beta_2$ & 1,1 & 0.5,1 & 0.5,0.5 & 1,1 & 0.5,1 & 0.5,0.5 \\ \hline
			$(\hat{k}_1, \hat{k}_2)$ & 0.014(0.011) & 0.015(0.013) & 0.034(0.029) & 0.011(0.008) & 0.013(0.009) & 0.020(0.018) \\ \hline
			(4,4) & 0.005(0.006) & 0.006(0.006) & 0.01(0.007) & 0.011(0.008) & 0.013(0.009) & 0.020(0.016) \\ \hline
			(4,3) & 0.007(0.007) & 0.013(0.010) & 0.015(0.011) & 0.039(0.022) & 0.095(0.053) & 0.121(0.078) \\ 
			(3,4) & 0.008(0.007) & 0.007(0.006) & 0.014(0.010) & 0.025(0.016) & 0.037(0.021) & 0.028(0.024) \\ \hline
			(3,3) & 0.014(0.011) & 0.015(0.013) & 0.034(0.029) & 0.223(0.052) & 0.191(0.058) & 0.160(0.103) \\ \hline
			(4,2) & 0.314(0.051) & 0.520(0.048) & 0.056(0.041) & 0.491(0.053) & 0.616(0.059) & 0.065(0.039) \\ 
			(2,4) & 0.274(0.057) & 0.358(0.037) & 0.092(0.046) & 0.486(0.049) & 0.468(0.048) & 0.093(0.033) \\ \hline			
			(3,2) & 0.312(0.051) & 0.517(0.048) & 0.061(0.040) & 0.514(0.051) & 0.623(0.057) & 0.064(0.039) \\ 
			(2,3) & 0.273(0.057) & 0.357(0.038) & 0.098(0.049) & 0.517(0.049) & 0.502(0.047) & 0.105(0.039) \\ \hline
			(2,2) & 0.509(0.057) & 0.589(0.044) & 0.080(0.040) & 0.681(0.024) & 0.630(0.043) & 0.079(0.037) \\ \hline			
		\end{tabular}%
	}
	\caption{Mean and standard deviation (in the parentheses) of the absolute error $|\hat{r} - r_0|$ with $\hat{r}$ estimated by (\ref{eqn:est_r}) with different combinations of $(k_1, k_2)$.}
	\label{table:meansd_of_abs_err_r0_exmp2}
\end{table}

\begin{figure}[htbp!]
	\centering	
	\includegraphics[width=0.6\textwidth, keepaspectratio]{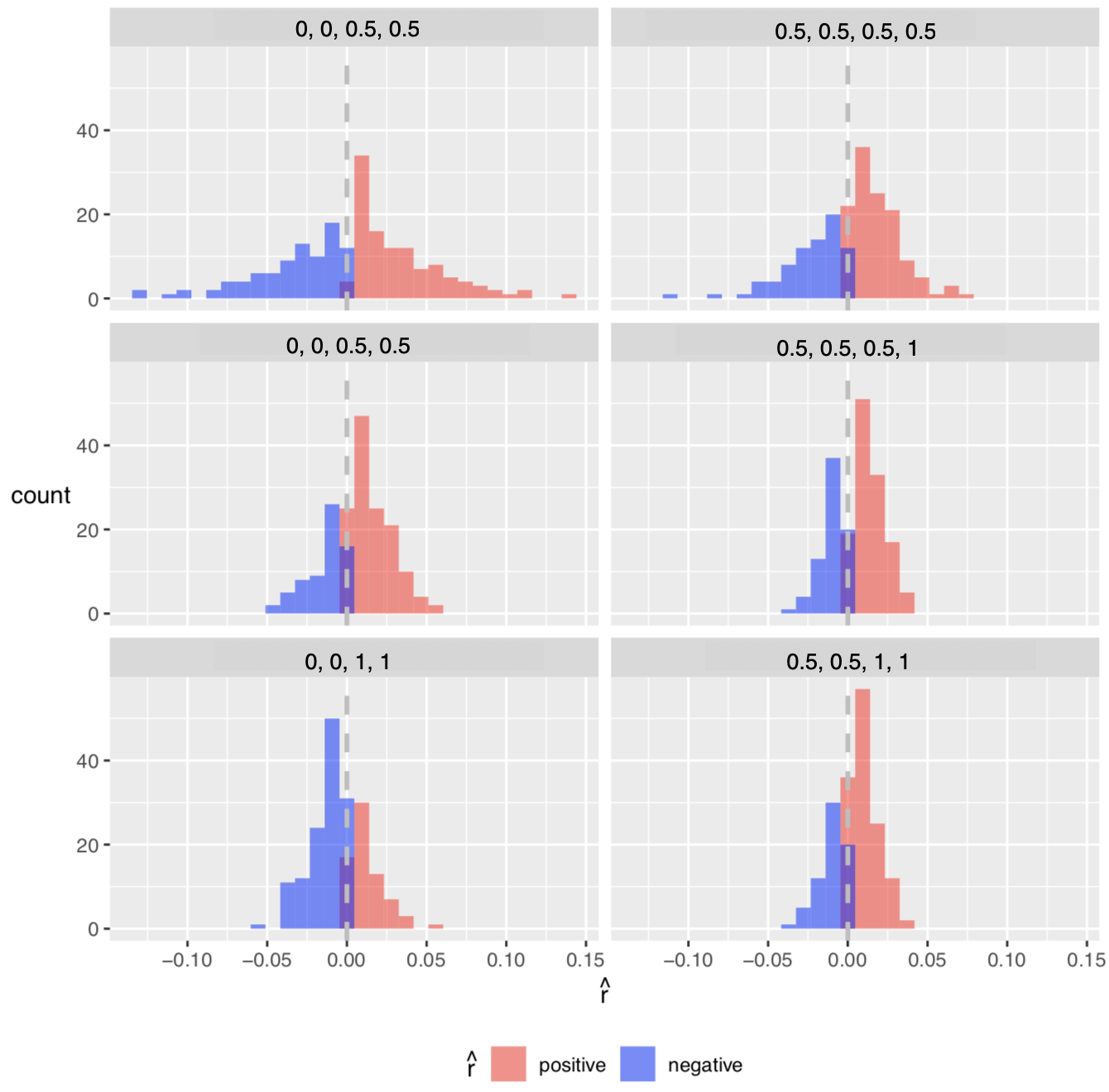}%{exmp2_rhat_distn_all}
	\caption{Histograms of $\hat{r}$ with $(\hat{k}_1, \hat{k}_2)$ estimated by (\ref{eqn:est_ks}) under different combinations of factor strength $(\delta_{11}, \delta_{21}, \delta_{12}, \delta_{22})$ and threshold strength $(\beta_1, \beta_2)$. Panel titles show the values of $\delta_{11}$, $ \delta_{21}$,  $\beta_1$, and $\beta_2$.}
	\label{fig:rhat_distn_betas_1_1_exmp2}
\end{figure}

Figure \ref{fig:rhat_distn_betas_1_1_exmp2} presents the histograms of $\hat{r}$ with $(\hat{k}_1, \hat{k}_2)$ estimated by (\ref{eqn:est_ks}) under different combinations of factor strength $(\delta_{11}, \delta_{12})$ and threshold strength $(\beta_1, \beta_2)$. The title of each panel shows the value of $\delta_{11}, \delta_{12}, \beta_1, \beta_2$. The three sub-figures in the left column all corresponds to the strong row and column factors case. The three sub-figures in the right column all corresponds to the weak row but strong column factors case. From the figure, it is clear that $\hat{r}$ is not biased. This is because the factor strengths are the same in both regime. However, when the thresholding effects are weak for both row and column, i.e. $(\beta_1, \beta_2) = (0.5, 0.5)$, the variances of $\hat{r}$ are larger. When the thresholding effects are strong for at least one of row or column loadings, i.e. $(\beta_1, \beta_2) = (0.5, 1)$ or $(1, 1)$, $\hat{r}$ concentrates around true threshold value zero. This again shows the \textit{the helping effect from the strong column thresholding to the weak row thresholding}.  

%Figure \ref{fig:g(r)setting157468} shows the plots of $\hat{G}(r)$ for typical data sets under the six different settings in Table \ref{table:threshold_strength_comb}. In all settings, the true $k_1$ and $k_2$ are unknown and $\hat{G}(r)$ is based on estimated $\hat{k}_1$ and $\hat{k}_2$. We can observe that $\hat{G}(r)$ has a unique minimal value around $r=0$ under all settings considered here. This is because the factor strengths in two regime are balanced. However, the valley of $\hat{G}(r)$ is flatter in settings weak strong factor strength and weak thresholding effect. This explains the existence of larger variance we observe in Figure \ref{fig:rhat_distn_betas_1_1_exmp2}. 

%\begin{figure*}[htbp!]
%	\centering
%	\includegraphics[width=0.7\textwidth]{gfunc-group2.pdf}
%	\caption[] {\small Plots of $\hat{G}(r)$ under three settings in Table \ref{table:threshold_strength_comb} for typical data sets of Example 2.} 
%	\label{fig:g(r)setting157468}
%\end{figure*}

\begin{figure}[hbtp!]
	\centering
	\includegraphics[width=0.7\textwidth, keepaspectratio]{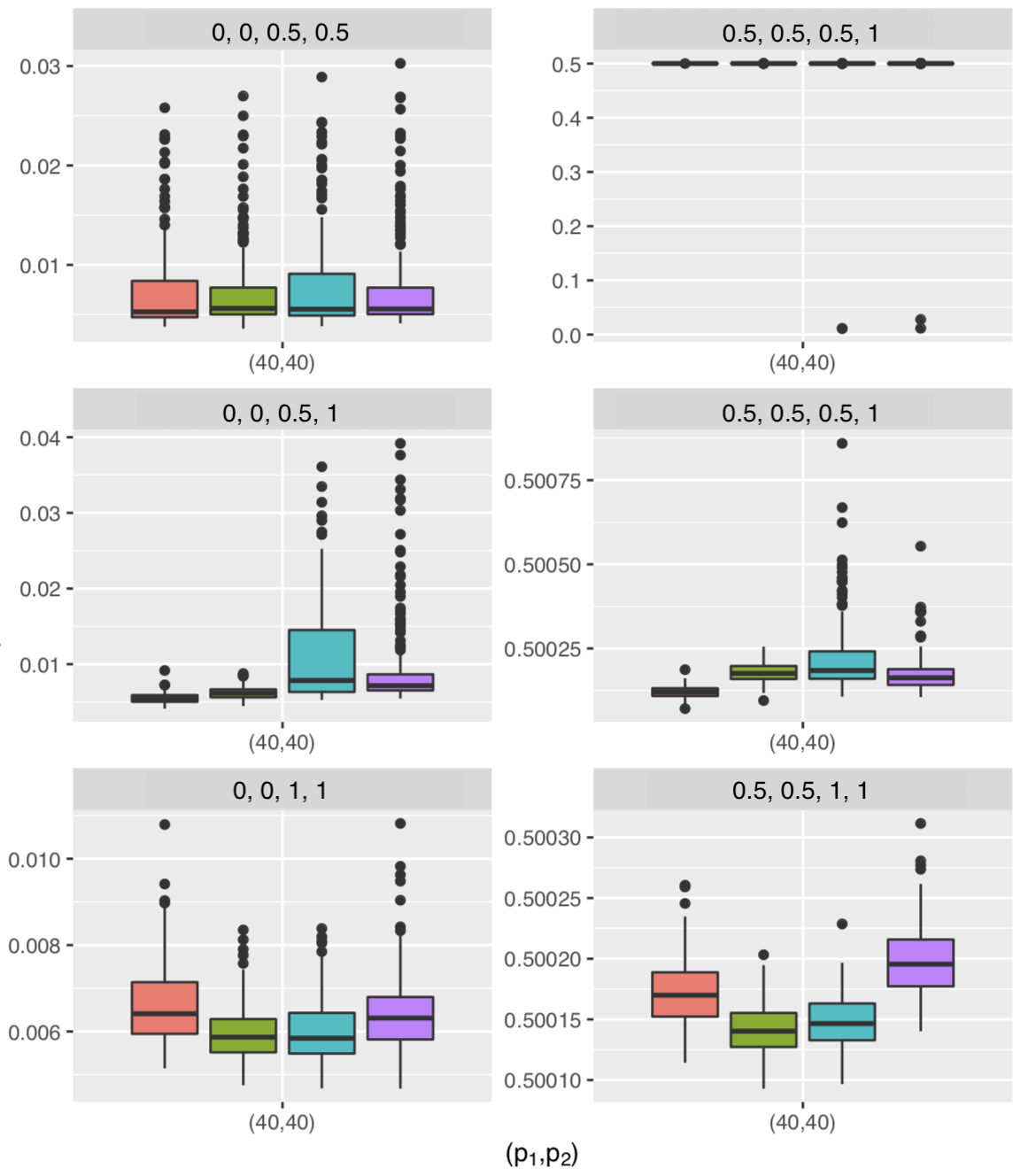}%{loading_spdist_exmp2_free}
	\caption{Boxplots of estimation errors for row and column loading spaces with different combinations of $\delta_{11}$, $\delta_{12}$, $\beta_{1}$, and $\beta_2$, whose values are shown in the panel titles. Each panel includes four box plots, which are estimation results for  $\hat{\bQ}_{1,1}$, $\hat{\bQ}_{2,1}$, $\hat{\bQ}_{1,2}$, and $\hat{\bQ}_{2,2}$, respectively. Note that the y-axis are under different scales. }. 
	\label{fig:loading_spdist_exmp2} 
\end{figure}

Thirdly, we investigate the accuracy in estimating the loading spaces when true $k_1$, $k_2$ and $r_0$ are unknown. The pair $(\hat{k}_1, \hat{k}_2)$ is estimated by (\ref{eqn:est_ks}) and $\hat{r}$ is estimated by (\ref{eqn:est_r}). Then a representative matrix of the row or column loading space is estimated according to (\ref{eqn:est_loading}) for each regime.
Figure \ref{fig:loading_spdist_exmp2} shows box plots of space distance for row and column loading matrices $\hat{\bQ}_{1,1}$, $\hat{\bQ}_{2,1}$, $\hat{\bQ}_{1,2}$, and $\hat{\bQ}_{2,2}$ under different combinations of factor strength $(\delta_{11}, \delta_{21})$ and thresholding strength $(\beta_1, \beta_2)$. Panel title shows the value of $\delta_{11}, \delta_{12}, \beta_1, \beta_2$. Note that to better present the variation of estimation errors in each setting, the sub-figures have different scales.
%For better comparison between sub-figures, we set the scales of y-axis the same for all sub-figures. For those who are interested in the details, Figure \ref{fig:loading_spdist_exmp2_free} in Appendix \ref{appendix:more_tableplot} is the same figure with different scales of y-axis among sub-figures.
As shown in the sub-figures in the left column with strong row/column factors in both regime, i.e. $(\delta_{11}, \delta_{21})=(0,0)$, all loading spaces are estimated with high accuracy. As shown in the the sub-figures in the right column with weak row but strong column factors in both regime, i.e. $(\delta_{11}, \delta_{12})=(0.5,0.5)$, the estimated row loading spaces are far from the truth. The thresholding strength affects the variance of the estimation. Estimation variances under strong thresholding $(\beta_1, \beta_2) = (1,1)$ are smaller than those under weak thresholding $(\beta_1, \beta_2) = (0.5,0.5)$. \textit{The helping effect from the strong column thresholding to the weak row thresholding} can be observed again by the case of $(\beta_1, \beta_2) = (1,1)$ with the cases of strong and weak threshold. 

\section{Application to Real Data} \label{sec:application}

\subsection{Example 1: Multinational Macroeconomic Indices} \label{sec:application_macro_indices}

\begin{figure}[ht!]
	\centering
	\includegraphics[width=\linewidth,height=\textheight,keepaspectratio=true]{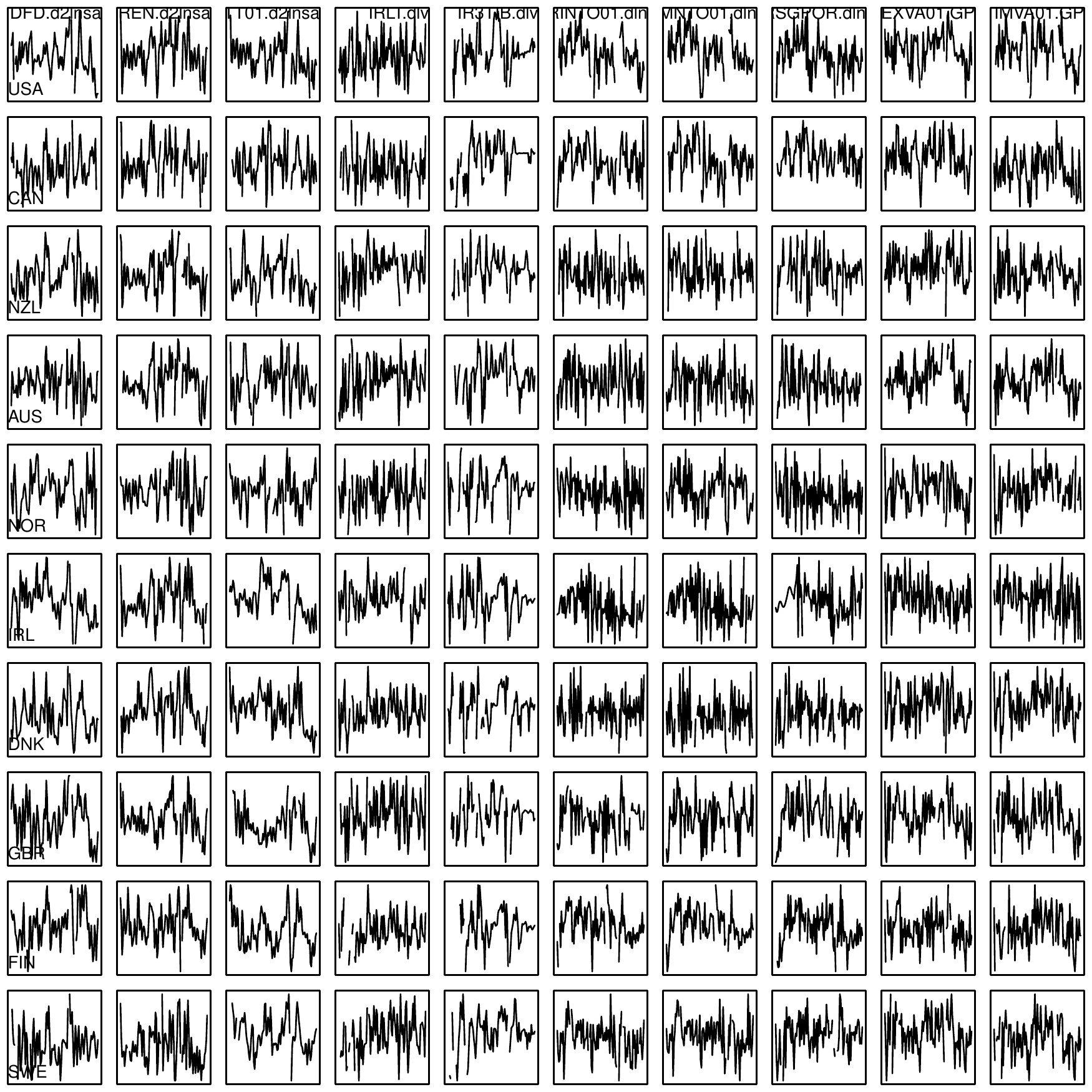}
	\caption{Time series plots of macroeconomic indicators of multiple countries (after data transformation). Only a subset of the countries and indicators is plotted due to the space limit.}
	\label{fig:oecd_mts_plot}
\end{figure}

We apply the threshold matrix factor model to the multinational macroeconomic indices dataset in \cite{chen2017constrained}. The dataset is collected from OECD. It contains 10 quarterly macroeconomic indices of 14 countries from 1990.Q2 to 2016.Q4  for 107 quarters. Thus, we have $T = 107$ and $p_1 \times p_2 = 14 \times 10$ matrix-valued time series. The countries include United States, Canada, New Zealand, Australia, Norway, Ireland, Denmark, United Kingdom, Finland, Sweden, France, Netherlands, Austria and Germany. The indexes cover four major groups, namely production (P:TIEC, P:TM, GDP), consumer price (CPI:Food, CPI:Ener, CPI:Tot), money market (IR:Long, IR:3-Mon), and international trade (IT:Ex, IT:Im).  Each original univariate time series is transformed by taking the first or second difference or logarithm to satisfy the mixing condition (Condition 1 in Appendix \ref{appendix:regular_condition}). See Table \ref{table:macro_index_data} in Appendix \ref{appendix:cross_country_macro_dataset} for detailed descriptions of the dataset and the transformation. Figure \ref{fig:oecd_mts_plot} shows the transformed time series of macroeconomic indicators of multiple countries.

It is well known that there exists the pattern of expansion and contraction in macroeconomy. Here we analyzed the data using a 2-regime threshold factor model, intending to capture the expansion and contraction economic behaviros. Macroeconomic research has identified different groups of business cycle indicators. For example, leading indicators include average weekly work hours in manufacturing, factory orders for goods, housing permits, stock prices, the index of consumer expectations, average weekly claims for unemployment insurance and the interest rate spread; Coincident indicators include the unemployment rate, personal income levels and industrial production; Lagging indicators include the average length of unemployment, labor cost per unit of manufacturing output, the average prime rate, the consumer price index and commercial lending activity. We consider the each one in the groups of leading, coincident and lagging business cycle indicator for the the threshold value (with both leads and lags $l=1, 2, 3$.) Namely, we use the OECD Composite Leading Indicator (CLI) from the leading indicator group, S\&P500 return from the coincident indicator group, and the Consumer Price Index (CPI) from the lagging indicator group. Note that CPI and CLI are available for each country while our model use only one scalar threshold process $z_t$. Since the countries in this data set are all developed countries from North American and European, the changes of business cycle are almost synchronous. We use the means of CLI and CPI across different countries. 

Table \ref{table:E_of_different_zzs} shows the value of $E$ for each candidate $\Delta \ln$ CLI, S\&P500 squared return and $\Delta \ln$ CPI in different leads. Here, $\Delta \ln$ denotes transformation of first difference after logarithm. The threshold processes using $\Delta \ln$CLI, S\&P500 squared return, and $\Delta \ln$CPI produce minimal $E$'s at lag 3, 1, and 3, respectively. In the following we use S\&P500 squared return at time $t-1$ as the threshold variable for time $t$, since it minimizes $E$ over all candidates.

\begin{table}[ht!] 
\centering
%\resizebox{\textwidth}{!}{%
\begin{tabular}{c|cccc}
\hline \hline
Threshold Process $z_t$ & $z_{t-4}$ & $z_{t-3}$ & $z_{t-2}$ & $z_{t-1}$  \\ \hline
$\Delta \ln CLI$ & 8768.02 & \textbf{8715.37}  & 8833.73  & 8824.84   \\ \hline
%$S\&P$ $return$ & 8622.03  & \textbf{8596.59} & 8850.86 & 8664.77  \\ \hline
$S\&P$ $return^2$ & 8710.79  & 8719.06 & 8734.01 & \textbf{8645.73}  \\ \hline
$\Delta \ln CPI$ & 8795.69 & \textbf{8743.94} & 8800.34 & 8848.71 \\ \hline \hline
\end{tabular}%
%}
\caption{Example 1: ESS for all threshold variable candidates in multinational macroeconomic. We use 25-th and 75-th percentiles of the threshold variable as $\eta_1$ and $\eta_2$ to boot start the estimation.} \label{table:E_of_different_zzs}
\end{table}

\begin{figure}[htbp!]
	\centering	
	\includegraphics[width=0.7\textwidth, keepaspectratio]{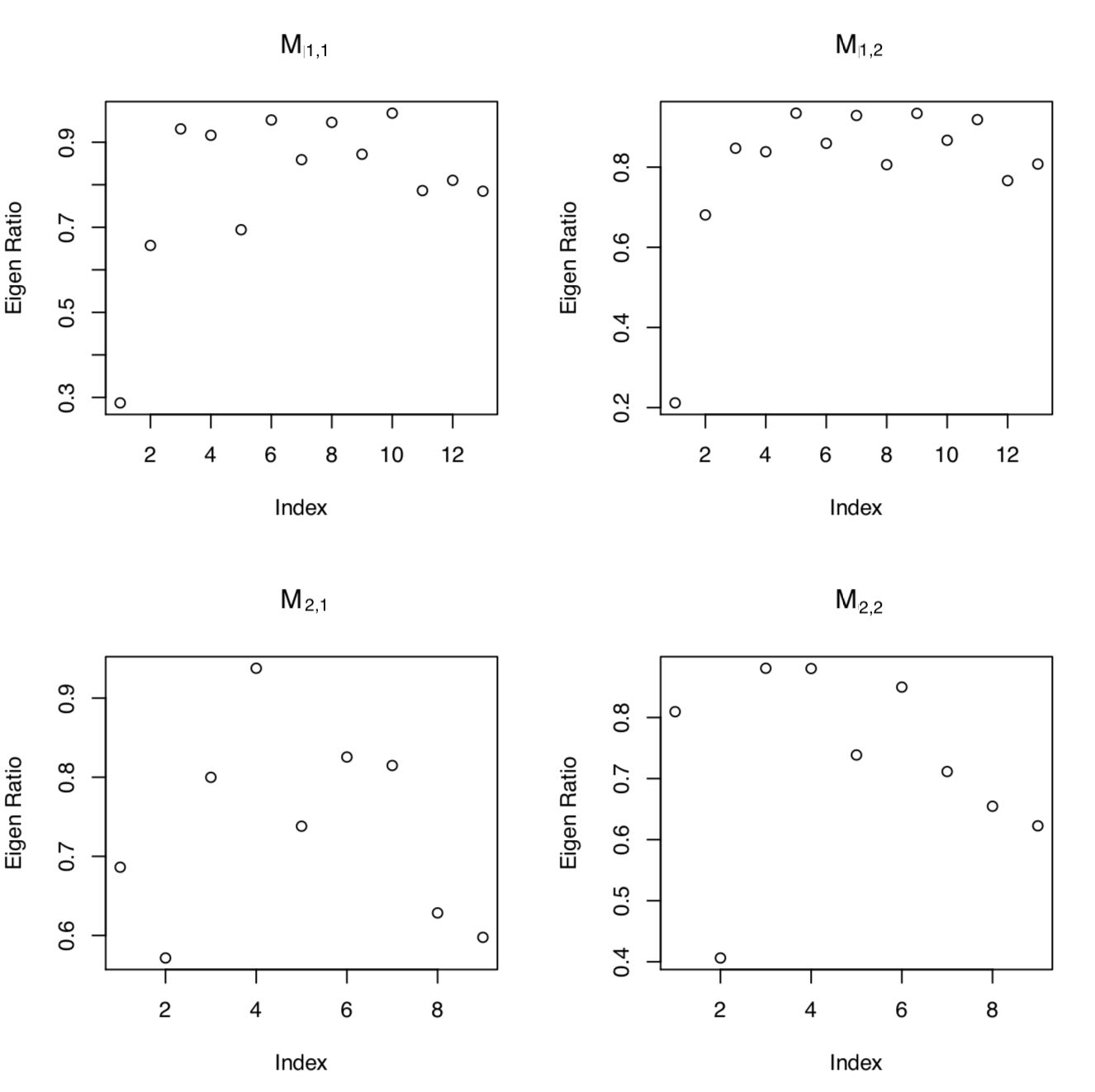}%{eigval_ratio_plot}
	\caption{Example 1: Ratios of eigenvalues in (\ref{eqn:eigval_ratio}) for $\hat{\bM}_{1,1}(\eta_1,\eta_2)$, $\hat{\bM}_{1,2}(\eta_1,\eta_2)$, $\hat{\bM}_{2,1}(\eta_1,\eta_2)$ and $\hat{\bM}_{2,2}(\eta_1,\eta_2)$, respectively.}
	\label{fig:eigval_ratio_plot}
\end{figure}

We use 25-th and 75-th percentiles of the threshold variable as $\eta_1$ and $\eta_2$ to estimate the number of factors. The four panels in Figure (\ref{fig:eigval_ratio_plot}) display the ratio of eigenvalues of $\hat{\bM}_{1,1}(\eta_1,\eta_2)$, $\hat{\bM}_{1,2}(\eta_1,\eta_2)$, $\hat{\bM}_{2,1}(\eta_1,\eta_2)$ and $\hat{\bM}_{2,2}(\eta_1,\eta_2)$, respectively. Ratio of the eigenvalues for row factors achieves its minimal at 1, while the ratio of eigenvalues for column factors achieves its minimal at 2. It yields that $\hat{k}_1=1$ and $\hat{k}_2=2$.

\begin{figure}[hbtp!]
	\centering
	\begin{subfigure}[b]{0.48\textwidth}
		\centering
		\includegraphics[width=\textwidth]{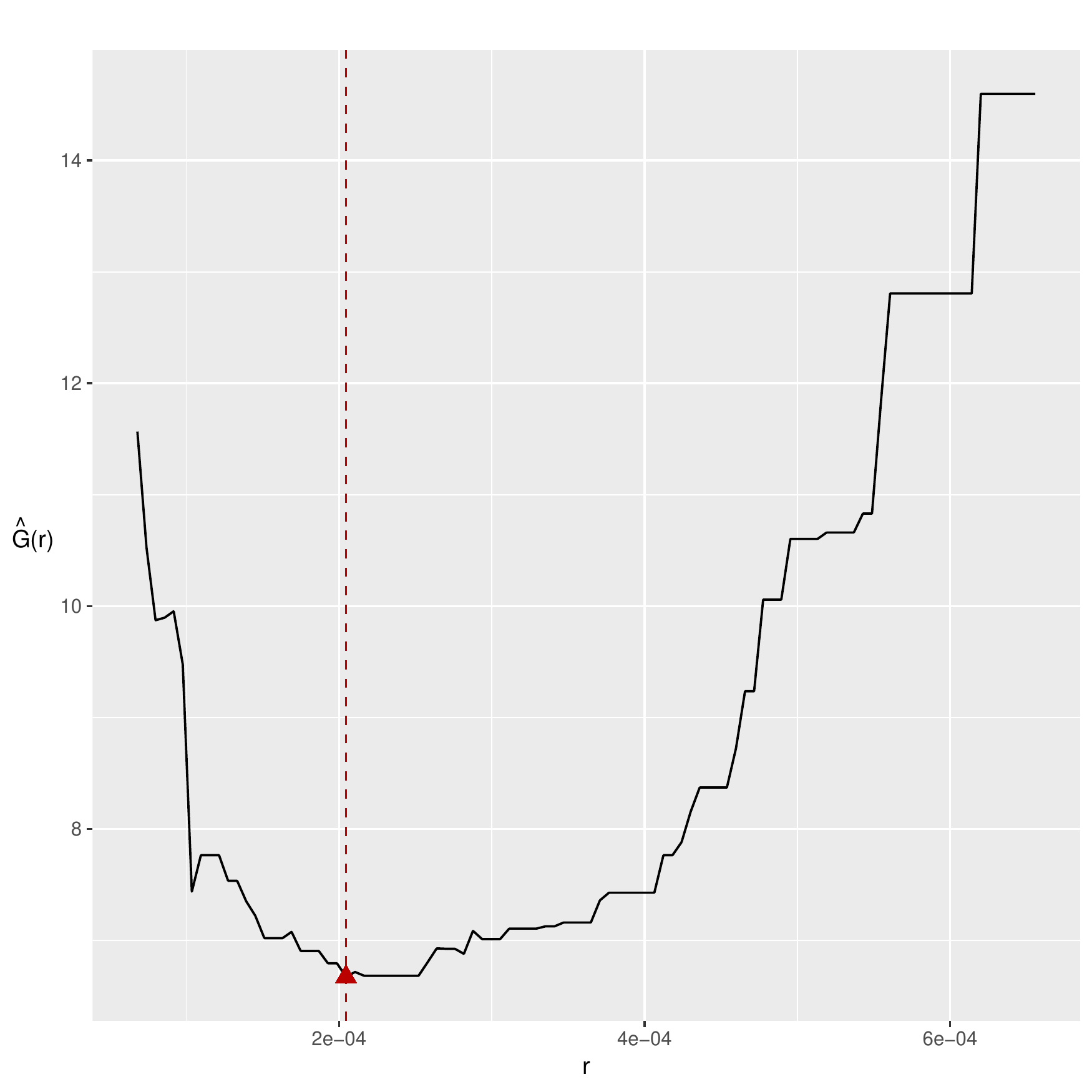}
		\caption{{\small $k_1 = 1$, $k_2=2$}}
		\label{fig:hat_gr_plot_1x2}
	\end{subfigure}
	\hfill
	\begin{subfigure}[b]{0.48\textwidth}
		\centering
		\includegraphics[width=\textwidth]{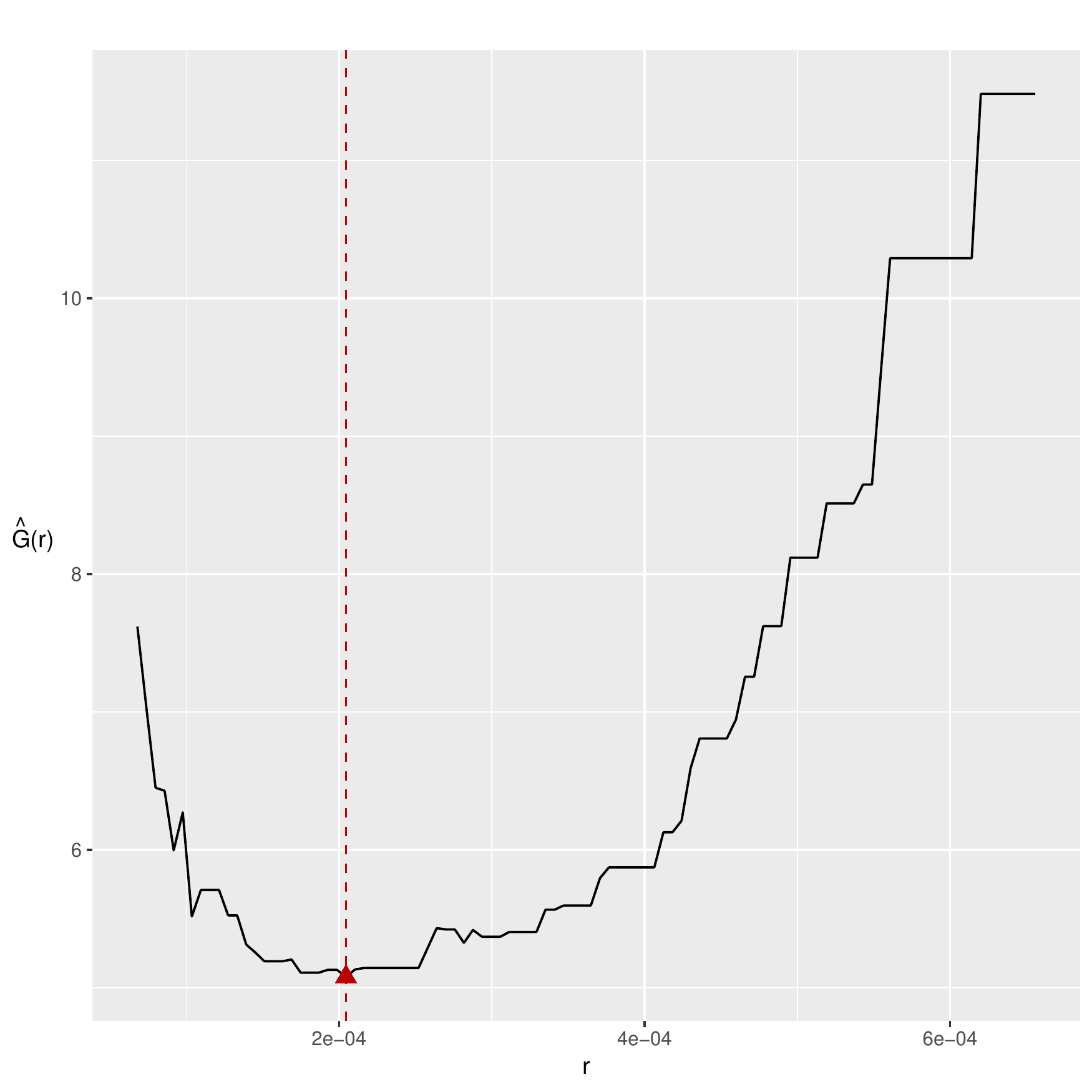}
		\caption{{\small $k_1 = 3$, $k_2=3$}}
		\label{fig:hat_gr_plot_3x3}
	\end{subfigure}
\caption{Example 1: Plot of $\hat{G}(r)$ for multinational macroeconomic indices data set.}  \label{fig:hat_gr_plot}
\end{figure}

To determine the threshold point $r_0$, we estimate $\hat{G}(r)$ based on (\ref{eqn:G(x)}) and find the minimizer of the function. Figure \ref{fig:hat_gr_plot_1x2} and Figure \ref{fig:hat_gr_plot_3x3} show the plots of  $\hat{G}(r)$ with estimated $(k_1, k_2)=(1,2)$ and overestimated $(k_1, k_2)=(3,3)$. Both show U-shape curves with a relatively flat bottom. By minimizing $\hat{G}(r)$, we have $\hat{r} =  2.0455 \cdot 10^{-4}$ for both $(k_1, k_2)=(1,2)$ and $(3,3)$. In this application, $\calD(\hat{\cM(\bR_{1})}, \hat{\cM(\bR_{2})})=0.5566$ and $\calD(\hat{\cM(\bC_{1})}, \hat{\cM(\bC_{2})})=0.4934$. The two regimes are well separated. 

\begin{figure}[hbtp!]
\centering
\includegraphics[width=\linewidth,height=\textheight,keepaspectratio=true]{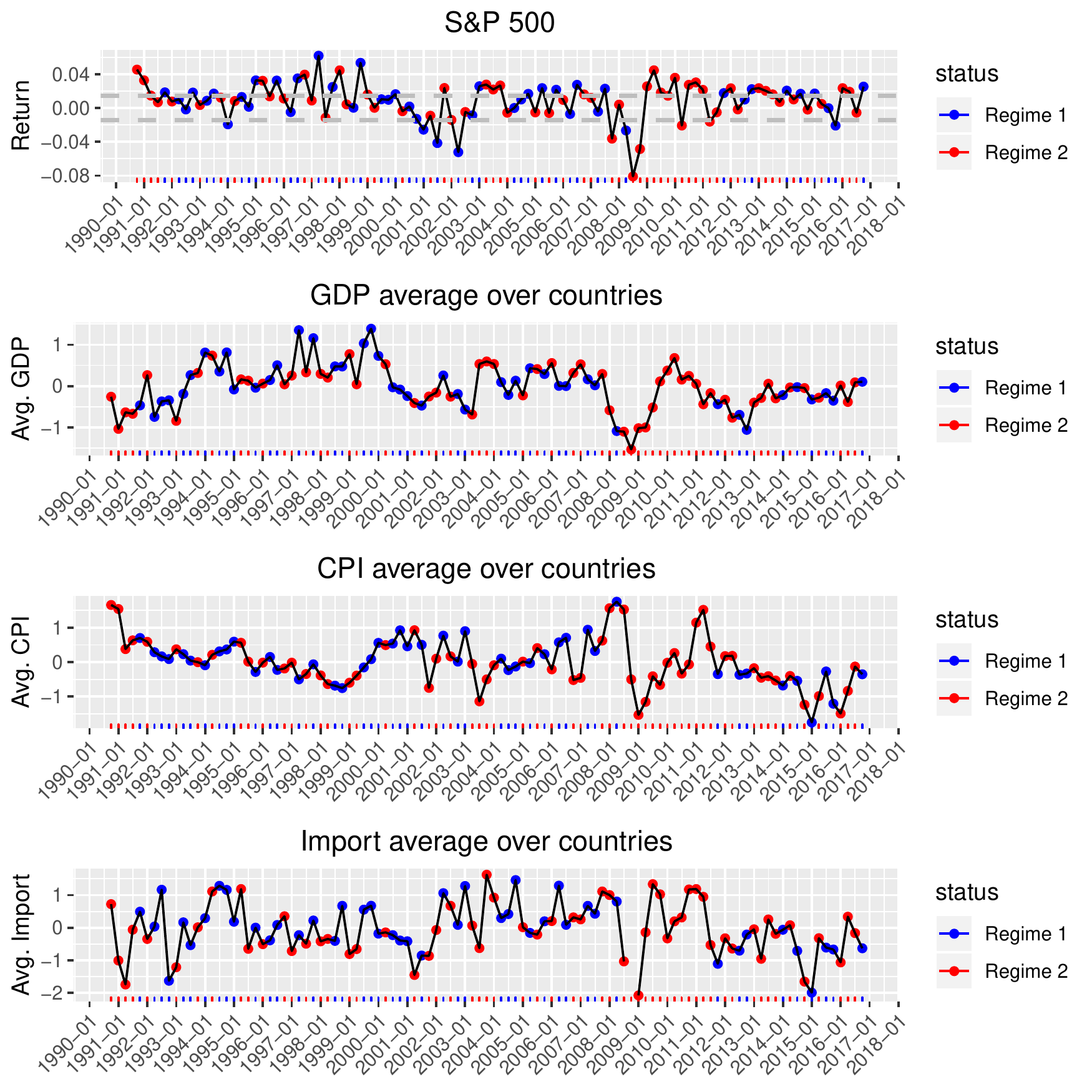}
\caption{Sample time series in the multinational macroeconomic indices data set in two regimes.}
\label{fig:regimes}
\end{figure}

Figure \ref{fig:regimes} shows sample time series in the multinational macroeconomic indices data set in two regimes. The proposed model gives very reasonable result. It shows clearly that Regime 1 corresponds to the high volatility periods and Regime 2 corresponds to low volatility periods. 

\subsection{Example 2: Fama-French 10 by 10 Series} \label{sec:application_Fama-French}

In this section, we investigate threshold matrix-variate factor models for the monthly market-adjusted return series of Fama-French $10 \times 10$ portfolios from January 1964 to December 2015 for 
624 months and overall $62,400$ observations. The portfolios are the intersections of 10 portfolios formed by size (market equity, ME) and 10 portfolios formed by the ratio of book equity to market equity (BE/ME). Thus, we have $T=624$ and $p_1 \times p_2 = 10 \times 10$ matrix-variate 
time series. The series are constructed by subtracting the monthly excess market returns from each of the original portfolio returns obtained from \cite{FFdata}, 
so they are free of the market impact. 

Table \ref{table:FF_E_of_different_zzs} shows the value of $E$ for each threshold variable candidate including S\&P500 return,S\&P500 squared return, and their lag variables. The minimal of $E$ is achieved at S\&P 500 return at lag 5. In the following we use it as the threshold variable.

\begin{table}[ht!] 
	\centering
%	\resizebox{\textwidth}{!}{%
		\begin{tabular}{c|cccccc}
			\hline \hline
			Threshold Process $z_t$ & $z_{t-6}$ & $z_{t-5}$ & $z_{t-4}$ & $z_{t-3}$ & $z_{t-2}$ & $z_{t-1}$  \\ \hline
			$S\&P 500$ $return$ & 10629.75 & \textbf{10607.12} & 10933.95 & 11234.59 & 11262.81 & 10673.53  \\ \hline
			$S\&P$ $return^2$ & 11727.39 & 11264.92 & 11834.70 & 11678.43 & 11263.96 & 10845.61 \\ \hline \hline 
		\end{tabular}%
%	}
	\caption{Example 2: ESS for all threshold variable candidates in monthly Fama French data set. We use 25-th and 75-th percentiles of the threshold variable as $\eta_1$ and $\eta_2$ to boot start the estimation.} \label{table:FF_E_of_different_zzs}
\end{table}

\begin{figure}[htbp!]
	\centering	
	\includegraphics[width=0.8\textwidth, keepaspectratio]{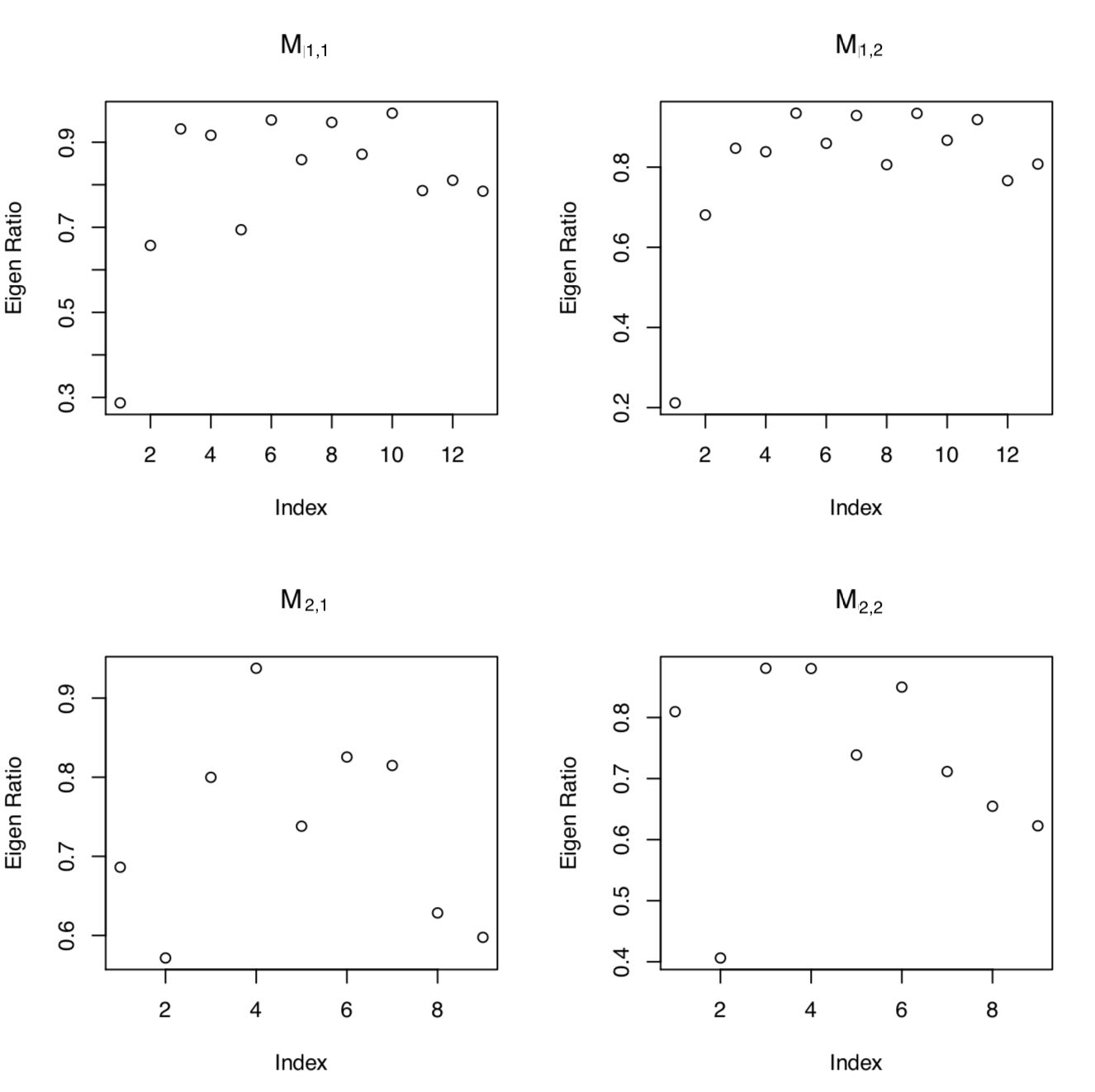}%{fama_eigval_ratio_plot}
	\caption{Example 2: Ratios of eigenvalues in (\ref{eqn:eigval_ratio}) for $\hat{\bM}_{1,1}(\eta_1,\eta_2)$, $\hat{\bM}_{1,2}(\eta_1,\eta_2)$, $\hat{\bM}_{2,1}(\eta_1,\eta_2)$ and $\hat{\bM}_{2,2}(\eta_1,\eta_2)$, respectively.}
	\label{fig:fama_eigval_ratio_plot}
\end{figure}

We use 25-th and 75-th percentiles of the threshold variable as $\eta_1$ and $\eta_2$ to estimate the number of factors. The four panels in Figure \ref{fig:fama_eigval_ratio_plot} display the ratio of eigenvalues of $\hat{\bM}_{1,1}(\eta_1,\eta_2)$, $\hat{\bM}_{1,2}(\eta_1,\eta_2)$, $\hat{\bM}_{2,1}(\eta_1,\eta_2)$ and $\hat{\bM}_{2,2}(\eta_1,\eta_2)$, respectively. Eigen-ratio for both row and column factors achieve minimal at 1. It yields that $\hat{k}_1=1$ and $\hat{k}_2=1$.

\begin{figure}[hbtp!]
	\centering
	\begin{subfigure}[b]{0.48\textwidth}
		\centering
		\includegraphics[width=\textwidth]{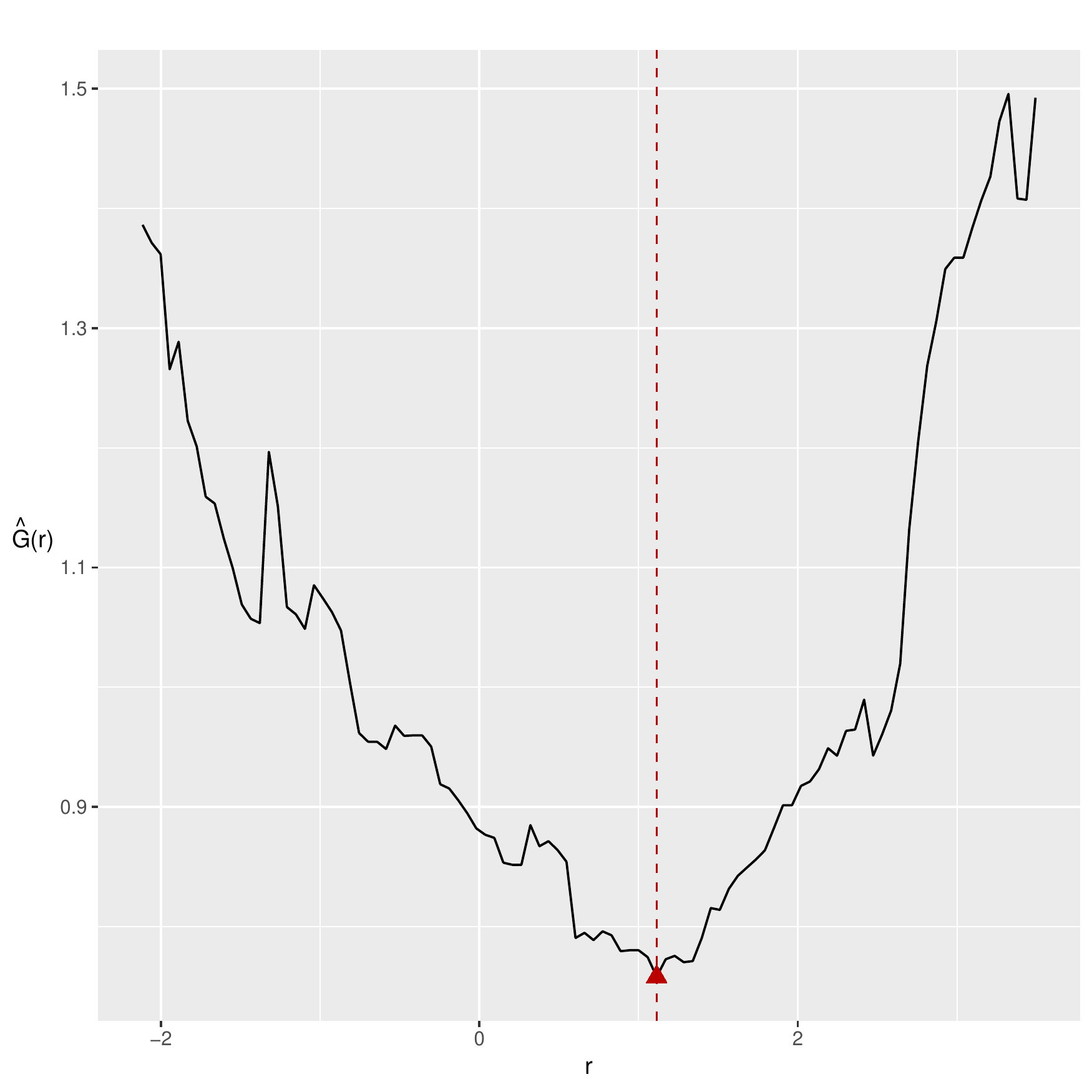}
		\caption{{\small $k_1 = 1$, $k_2=1$}}
		\label{fig:fama_hat_gr_plot_1x1}
	\end{subfigure}
	\hfill
	\begin{subfigure}[b]{0.48\textwidth}
		\centering
		\includegraphics[width=\textwidth]{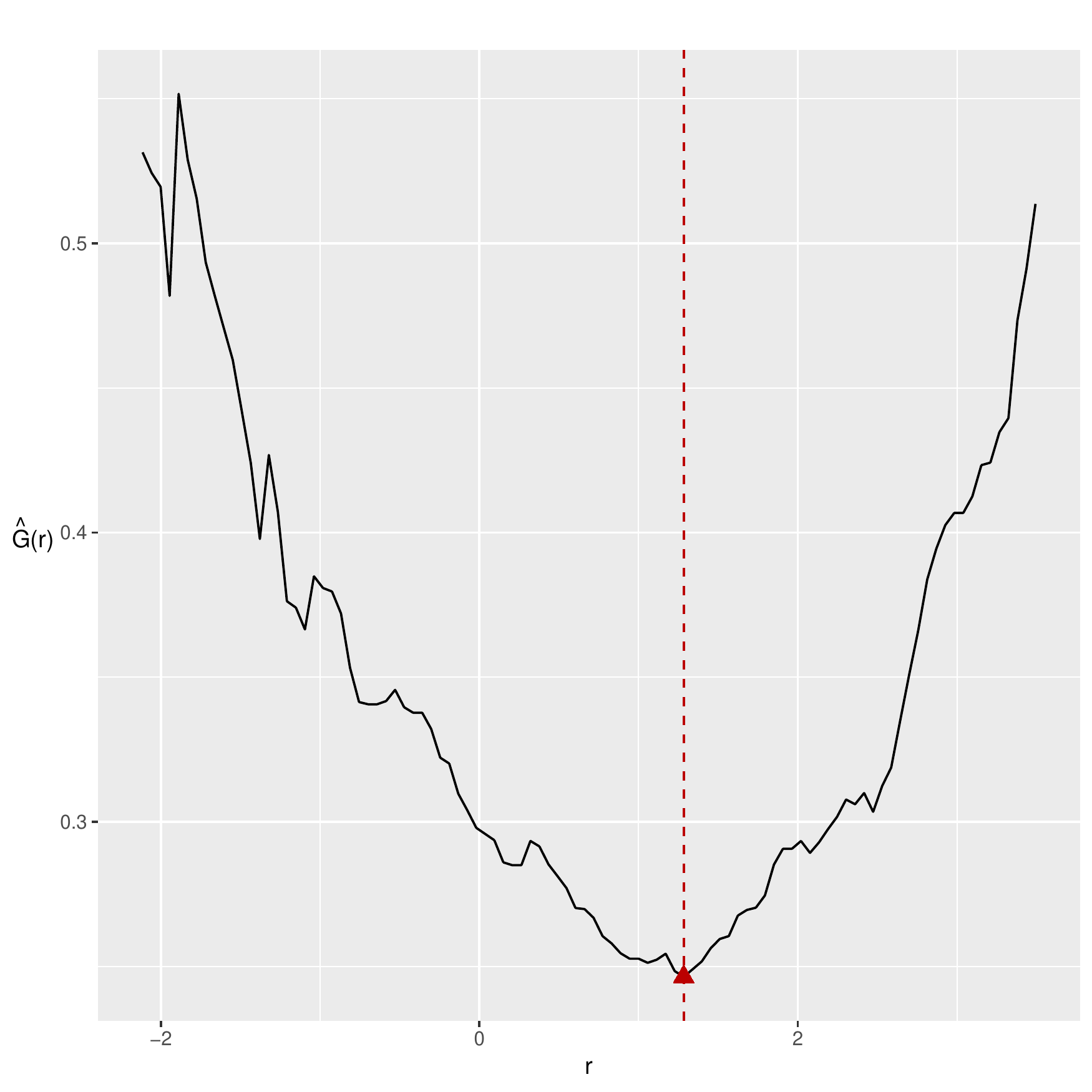}
		\caption{{\small $k_1 = 3$, $k_2=3$}}
		\label{fig:fama_hat_gr_plot_3x3}
	\end{subfigure}
	\caption{Example 2: Plot of $\hat{G}(r)$ for Fama French data set.}  \label{fig:fama_hat_gr_plot}
\end{figure}

To determine the threshold point $r_0$, we estimate $\hat{G}(r)$ based on (\ref{eqn:G(x)}) and find the $\hat{r}$ to minimize the function. Figure \ref{fig:fama_hat_gr_plot_1x1} and Figure \ref{fig:fama_hat_gr_plot_3x3} show the plots of  $\hat{G}(r)$ with estimated $(k_1, k_2)=(1,1)$ and overestimated $(k_1, k_2)=(3,3)$. Both show U-shape curves with a relatively flat bottom. By minimizing $\hat{G}(r)$, we have $\hat{r} =  1.2317$ for $(k_1, k_2)=(1,1)$ and  $\hat{r} = 1.1203$ for $(k_1, k_2)=(3,3)$, which are very close. In this application, $\calD(\hat{\cM(\bR_{1})}, \hat{\cM(\bR_{2})})=0.4647$ and $\calD(\hat{\cM(\bC_{1})}, \hat{\cM({\bC}_{2})})=0.5602$. Since the separation between these two regimes is moderate, we use overestimated dimension $(k_1, k_2)=(3,3)$ in the following steps. 

\begin{figure}[hbtp!]
	\centering
	\includegraphics[width=\linewidth,height=\textheight,keepaspectratio=true]{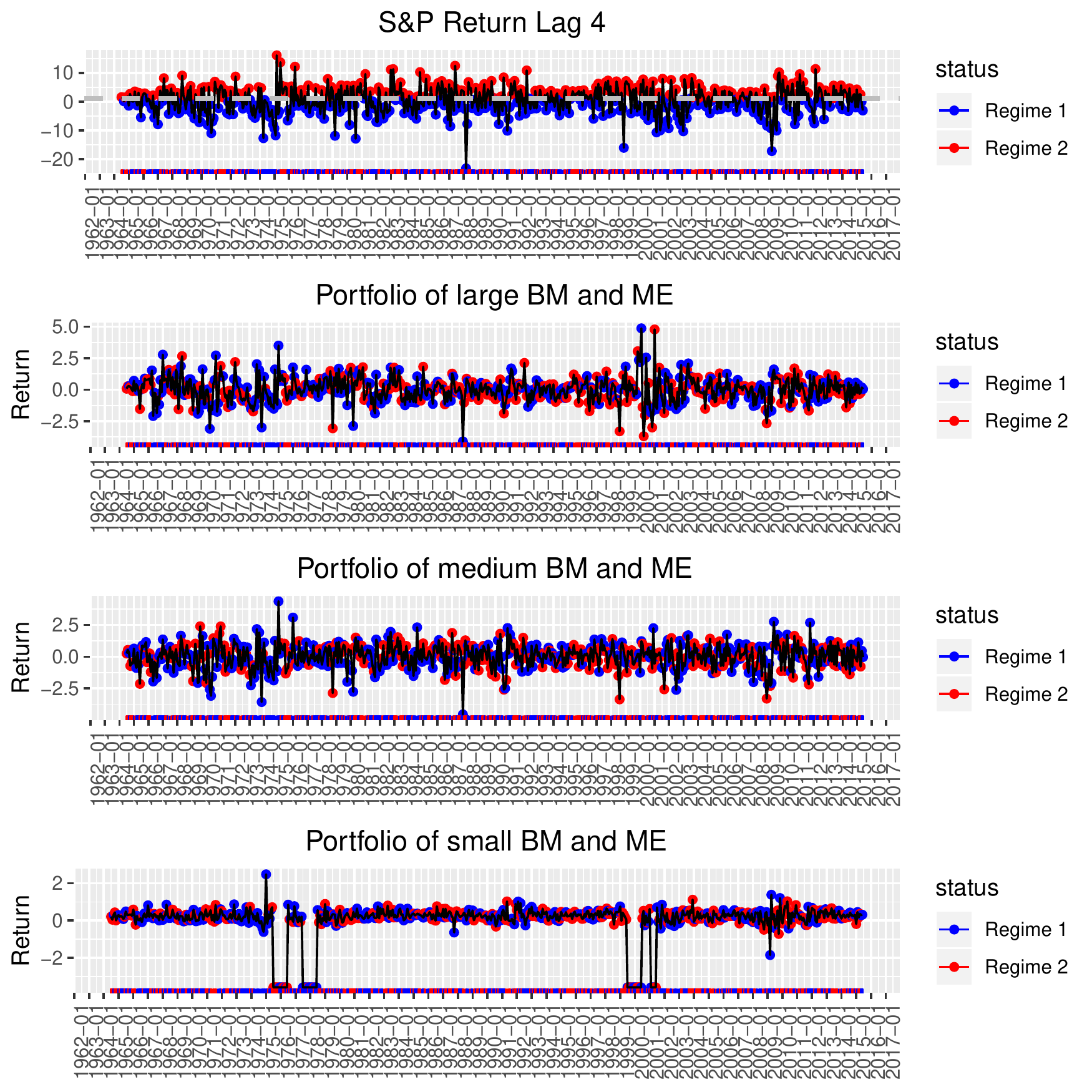}
	\caption{Sample time series in the Fama-French data set in two regimes.}
	\label{fig:fama_regimes}
\end{figure}

Figure \ref{fig:fama_regimes} shows sample time series in the monthly Fama-French data set in two regimes. It shows clearly that Regime 1 corresponds to the low return periods and Regime 2 corresponds to high return periods.

\section{Summary and Future Work}
In this article, we propose a threshold factor models for high-dimensional matrix-valued time series data. The loading spaces are allowed to change across regimes controlled by a threshold variable. The procedures to estimate threshold value, loading spaces in different regimes, and the number of factors are developed. The asymptotic properties of the estimators are investigated. We found that the strengths of factors in different regimes and the strength of thresholding play an important role in estimation results. When factors and thresholding are all strong, our estimation is immune to the curse of dimensionality. When the factors and thresholding are not all strong, the estimation of loading spaces and threshold value experience the 'helping' effect against the curse of dimensionality. The loading space estimation in the weaker regime gain efficiency and the impact of the curse of dimensionality is reduced due to the existence of a stronger regime. Comparing vector-valued factor models, the estimation of threshold value in matrix-valued models also benefits by introducing a direction with stronger thresholding. Determining the number of factor is a challenging issue for factor analysis. In this paper, we show that the proposed estimators are consistent even if the number of factors is overestimated.

Many potential research projects related to this topic have been left for future work. We only study two-regime models in this paper. It would be interesting to consider factor models with multiple regimes, and develop procedures for regime detection. Furthermore, we could also extend our approach for matrix time series analysis to tensor time series analysis, which is an interesting and new research direction.

%\section{Summary and Future Research}

\newpage
\bibliographystyle{apalike}
\bibliography{reference1}

\begin{thebibliography}{}

\bibitem[Bai et~al., 2017]{bai2017}
Bai, J., Han, X., and Shi, Y. (2017).
\newblock Estimation and inference of change points in high-dimensional factor
  models.
\newblock {\em Working paper}.

\bibitem[Bai and Ng, 2002]{bai2002}
Bai, J. and Ng, S. (2002).
\newblock Determining the number of factors in approximate factor models.
\newblock {\em Econometrica}, 70:191--221.

\bibitem[Barigozzi et~al., 2018]{barigozzi2018}
Barigozzi, M., Cho, H., and Fryzlewicz, P. (2018).
\newblock Simultaneous multiple change-point and factor analysis for
  high-dimensional time series.
\newblock {\em Journal of Econometrics}, 206:187--225.

\bibitem[Breitung and Eickmeier, 2011]{breitung2011}
Breitung, J. and Eickmeier, S. (2011).
\newblock Testing for structural breaks in dynamic factor models.
\newblock {\em Journal of Econometrics}, 163:71--84.

\bibitem[Chamberlain and Rothschild, 1983]{chamberlain1983}
Chamberlain, G. and Rothschild, M. (1983).
\newblock Arbitrage, factor structure and mean-variance analysis in large asset
  markets.
\newblock {\em Econometrica}, 70:191--221.

\bibitem[Chan, 1993]{chan1993}
Chan, K. (1993).
\newblock Consistency and limiting distribution of the least squares estimator
  of a threshold autoregressive model.
\newblock {\em The Annals of Statistics}, 21:520--533.

\bibitem[Chang et~al., 2015]{chang2015}
Chang, J., Guo, B., and Yao, Q. (2015).
\newblock High dimensional stochastic regression with latent factors,
  endogeneity and nonlinearity.
\newblock {\em Journal of Econometrics}, 189:297--312.

\bibitem[Chen et~al., 2017]{chen2017constrained}
Chen, E.~Y., Tsay, R.~S., and Chen, R. (2017).
\newblock Constrained factor models for high-dimensional matrix-variate time
  series.
\newblock {\em ArXiv e-prints}.

\bibitem[Chen et~al., 2014]{chenl2014}
Chen, L., Dolado, J., and Gonzalo, J. (2014).
\newblock Detecting big structural breaks in large factor models.
\newblock {\em Journal of Econometrics}, 180:30--48.

\bibitem[Cunha et~al., 2010]{cunha2010}
Cunha, F., Heckman, J., and Schennach, S. (2010).
\newblock Estimating the technology of cognitive and noncognitive skill
  information.
\newblock {\em Econometrica}, 78:883--931.

\bibitem[Fan et~al., 2016]{fan2016}
Fan, J., Liao, Y., and Wang, W. (2016).
\newblock Projected principal component analysis in factor models.
\newblock {\em The Annuals of Statistics}, 44:219--254.

\bibitem[Fan et~al., 2017]{fan2017}
Fan, J., Xue, L., and Yao, J. (2017).
\newblock Sufficient forecasting using factor models.
\newblock {\em Journal of Econometrics}, 201:292--306.

\bibitem[Fan and Yao, 2003]{fan2003}
Fan, J. and Yao, Q. (2003).
\newblock {\em Nonlinear Time Series: Nonparametric and Parametric Methods}.
\newblock Springer-Verlag, New York.

\bibitem[Forni et~al., 2000]{forni2000}
Forni, M., Hallin, M., Lippi, M., and Reichlin, L. (2000).
\newblock The generalized dynamic factor model: identification and estimation.
\newblock {\em Review of Economics and Statistics}, 82:540--554.

\bibitem[Forni and Reichlin, 1998]{forni1998}
Forni, M. and Reichlin, L. (1998).
\newblock Let's get real: a factor-analytical approach to disaggregated
  business cycle dynamics.
\newblock {\em Review of Economic Studies}, 65:453--473.

\bibitem[French, 2017]{FFdata}
French, K.~R. (2017).
\newblock {100 Portfolios Formed on Size and Book-to-Markete}.
\newblock
  \url{http://mba.tuck.dartmouth.edu/pages/faculty/ken.french/Data_Library/det_100_port_sz.html}.
\newblock [Online; Accessed 01-Jan-2017].

\bibitem[Geweke, 1977]{geweke1977}
Geweke, J. (1977).
\newblock The dynamic factor analysis of economic time series.
\newblock {\em Latent variable in socio-economic models}, ed. by D.J. Aigner
  and A.S. Goldberg, Amsterdam: North Holland.

\bibitem[Hallin and Liska, 2007]{hallin2007}
Hallin, M. and Liska, R. (2007).
\newblock Determining the number of factors in the general dynamic factor
  model.
\newblock {\em Journal of the American Statistical Association}, 102:603--617.

\bibitem[Han and Inoue, 2015]{han2015}
Han, X. and Inoue, A. (2015).
\newblock Tests for parameter instability in dynamic factor models.
\newblock {\em Econometric Theory}, 31:1117--1152.

\bibitem[Lam and Yao, 2012]{lam2012}
Lam, C. and Yao, Q. (2012).
\newblock Factor modeling for high-dimensional time series: inference for the
  number of factors.
\newblock {\em Annals of Statistics}, 40(2):694--726.

\bibitem[Lam et~al., 2011]{lam2011}
Lam, C., Yao, Q., and Bathia, N. (2011).
\newblock Estimation of latent factors for high-dimensional time series.
\newblock {\em Biometrika}, 98(4):901--918.

\bibitem[Lee and Shao, 2018]{lee2018}
Lee, C. and Shao, X. (2018).
\newblock Volatility martingale difference divergence matrix and its
  application to dimension reduction for multivariate volatility.
\newblock {\em Journal of Business \& Economic Statistics}, pages 1--13.

\bibitem[Liu and Chen, 2016]{liu2016}
Liu, X. and Chen, R. (2016).
\newblock Regime-switching factor models for high-dimensional time series.
\newblock {\em Statistica Sinica}, 26:1427--1451.

\bibitem[Liu and Chen, 2019]{liu2019}
Liu, X. and Chen, R. (2019).
\newblock Threshold factor models for high-dimensional time series.
\newblock {\em Manuscript}, Available at arXiv: 1809.03643.

\bibitem[Ma and Su, 2018]{ma2018}
Ma, S. and Su, L. (2018).
\newblock Estimation of large dimensional factor models with an unknown number
  of breaks.
\newblock {\em Journal of Econometrics}, 207:1--29.

\bibitem[Massacci, 2017]{massacci2017}
Massacci, D. (2017).
\newblock Least squares estimation of large dimensional threshold factor
  models.
\newblock {\em Journal of Econometrics}, 197:101--129.

\bibitem[Merikoski and Kumar, 2004]{merikoski2004}
Merikoski, J.~K. and Kumar, R. (2004).
\newblock Inequalities for spreads of matrix sums and products.
\newblock {\em Applied Mathematics E-Notes}, 4:150--159.

\bibitem[Pan and Yao, 2008]{pan2008}
Pan, J. and Yao, Q. (2008).
\newblock Modelling multiple time series via common factors.
\newblock {\em Biometrika}, 95:365--379.

\bibitem[Pe{\~n}a and Box, 1987]{pena1987}
Pe{\~n}a, D. and Box, G. E.~P. (1987).
\newblock Identifying a simplifying structure in time series.
\newblock {\em Journal of the American statistical Association}, 82:836--843.

\bibitem[Pe{\~n}a and Poncela, 2006]{pena2006}
Pe{\~n}a, D. and Poncela, P. (2006).
\newblock Nonstationary dynamic factor analysis.
\newblock {\em Journal of Statistical Planning and Inference}, 136:1237--1257.

\bibitem[Shumway and Stoffer, 2017]{shumway2017}
Shumway, R. and Stoffer, D. (2017).
\newblock {\em Time series analysis and its applications with R examples}.
\newblock Springer.

\bibitem[Stock and Watson, 2002a]{stock2002a}
Stock, J.~H. and Watson, M.~W. (2002a).
\newblock Forecasting using principal components from a large number of
  predictors.
\newblock {\em Journal of the American Statistical Association}, 97:1167--1179.

\bibitem[Stock and Watson, 2002b]{stock2002b}
Stock, J.~H. and Watson, M.~W. (2002b).
\newblock Macroeconomic forecasting using diffusion indices.
\newblock {\em Journal of Business \& Economic Statistics}, 20:147--162.

\bibitem[Su and Wang, 2017]{su2017}
Su, L. and Wang, X. (2017).
\newblock On time-varying factor models: estimation and testing.
\newblock {\em Journal of Econometrics}, 198:84--101.

\bibitem[Tong, 1990]{tong1990}
Tong, H. (1990).
\newblock {\em Nonlinear time series: a dynamic system approach}.
\newblock Clarendon Press, Oxford.

\bibitem[Tong and Lim, 1980]{tong1980}
Tong, H. and Lim, K. (1980).
\newblock Threshold autoregression, limit cycles and cyclical data.
\newblock {\em Journal of the Royal Statistical Society: Series B (Statistical
  Methodology)}, 42(3):245--292.

\bibitem[Tsay, 1989]{tsay1989}
Tsay, R. (1989).
\newblock Testing and modeling threshold autoregressive process.
\newblock {\em Journal of the American Statistical Association}, 84:231--240.

\bibitem[Tsay, 1998]{tsay1998}
Tsay, R. (1998).
\newblock Testing and modeling multivariate threshold models.
\newblock {\em Journal of the American Statistical Association}, 93:1188--1202.

\bibitem[Wang et~al., 2019]{wang2019}
Wang, D., Liu, X., and Chen, R. (2019).
\newblock Factor models for matrix-valued high-dimensional time series.
\newblock {\em Journal of Econometrics}, 208:231--248.

\bibitem[Yalcin and Amemiya, 2001]{yalcin2001}
Yalcin, I. and Amemiya, Y. (2001).
\newblock Nonlinear factor analysis as a statistical method.
\newblock {\em Statistical Science}, 16:275--294.

\end{thebibliography}

\newpage
%%%%%%%%%%%%%%%%%%%%%%%%%%%%%%%%%%%%%%%%%%%%%%%%%%%%%%%%%%%%%%%%%%%%%%
%%
%% Appendices 
%%
\begin{appendices}

\section{Regularity Conditions} \label{appendix:regular_condition}
%Let $\rvec(\cdot)$ be the vectorization operator, which converts a matrix to a vector by stacking columns fo the matrix on top of each other.
Define
\begin{align*}
\bSigma_f(h)&=\frac{1}{T} \sum_{t=1}^{T-h} \rE(\rvec(\bF_t) \rvec(\bF_{t+h})'), \\
I_t(h,c_1,c_2,c_3,c_4)&=I(c_1 <z_t<c_2, c_3<z_{t+h}<c_4),\\
\bSigma_{f}(h,c_1,c_2,c_3,c_4)&=\frac{\frac{1}{T}\sum_{t=1}^{T-h}\rE(\rvec(\bF_t) \rvec(\bF_{t+h})'I_t(h,c_1,c_2,c_3,c_4))}{\rE[I_t(h,c_1,c_2, c_3,c_4)]},\\
\bSigma_{f,i,j}(h,c_1,c_2)&=\frac{\frac{1}{T}\sum_{t=1}^{T-h}\rE(\rvec(\bF_t) \rvec(\bF_{t+h})'I_{t,i}(c_1)I_{t+h,j}(c_2))}{\rE[I_{t,i}(c_1)I_{t+h,j}(c_2)]},\\
\bOmega_{fc,ij,m\ell}(h,c_1,c_2,c_3,c_4)&=\frac{1}{T} \sum_{t=1}^{T-h} \rE(\bF_t \bc_{i,m} \bc_{j,\ell \cdot}' \bF_{t+h}'I_t(h,c_1,c_2,c_3,c_4)),\\
\hat{\bOmega}_{fc,ij,m \ell}(h,c_1,c_2,c_3,c_4)&=\frac{1}{T} \sum_{t=1}^{T-h} \bF_t \bc_{i,m\cdot} \bc_{j,\ell \cdot}' \bF_{t+h}'I_t(h,c_1,c_2,c_3,c_4),\\
\bSigma_{t,e}&=\var(\rvec(\bE_t)).
\end{align*}

\noindent {\bf Condition A1.} The process $(\bF_t,z_t)$
is $\alpha$-mixing. Specifically, for
some $\gamma > 2$, the mixing coefficients
satisfy the condition $\sum_{h=1}^\infty \alpha(h)^{1-2/\gamma} <
\infty$, where
\[
\alpha(h) = \sup_i \sup_{A \in {\cF}^i_{-\infty}, B \in
  {\cF}^{\infty}_{i+h}} | P(A \cap B) - P(A) P(B) |,
\]
and ${\cF}_i^j$ is the $\sigma$-field generated by $\{(\bF_t,z_t):i
\leq t \leq j\}$.
\vspace{.1in}

\noindent {\bf Condition A2.} Let $f_{t,m\ell}$ be the $(m \ell)$-th entry of
$\bF_t$. For any $m=1,\ldots, k_1$, $\ell=1, \ldots, k_2$, and
$t=1,\ldots, T$, $\rE (|f_{t,m\ell}|^{4\gamma})\leq \sigma_f^{4\gamma}$, where $\sigma_f$ is a positive constant and $\gamma$ is given in Condition
A1. There exists an $1 \leq h \leq h_0$ such that
${\rm rank}(\bSigma_f(h)) \geq k_{\max}$, and $\sigma_{k_{\max}}(
\bSigma_f(h))$ is uniformly bounded, where $k_{\max}=\max\{k_1, k_2\}$, as $p_1$ and $p_2$ go to infinity and $k_1$ and $k_2$ are
fixed. For $m=1,\ldots, k_1$ and $\ell=1,\ldots, k_2$, $\frac{1}{T-h}\sum_{t=1}^{T-h}\cov(\bff_{t,m}, \bff_{t+h,m}) \neq \mathbf{0}$, $\frac{1}{T-h}\sum_{t=1}^{T-h}\cov(\bff_{t,\ell\cdot}, \bff_{t+h,\ell \cdot}) \neq \mathbf{0}$.

\noindent {\bf Condition A3.}  The absolute value of each element in $\bSigma_{t,e}$ remains
bounded by $\sigma_e^2$ as $p_1$ and $p_2$ increase to infinity for $t=1,\ldots, T$. $\cov(\rvec(\bE_{t_1}), \rvec(\bF_{t_2}))=\bzero$ and $\cov(\rvec(\bE_{t_1}), \rvec(\bE_{t_2}))=\bzero$ for $t_1,t_2=1,\ldots, T$.

We do not impose stationarity assumptions for the factor process $\bF_t$ and the noise process $\bE_t$, but $\bF_t$ should satisfy the mixing condition as specified in Condition A1. Condition A2 ensures that the autocovariance matrices contain useful information from different fundamental factors and the model is not redundant. Condition A3 allows the noise process to accommodate strong cross-sectional dependence by only imposing entry-wise constraints on its covariance matrix. Under Condition A3, $\bE_t$ is an exogenous and independent process.

\medskip
\noindent {\bf Condition A4.} For $i=1,2$, there exist constants $\delta_{1i}$ and
$\delta_{2i} \in [0,1]$ such that $\|\bR_i\|_2^2 \asymp p_1^{1-\delta_{1i}}
\asymp \|\bR_i\|_{\min}^2$ and $\|\bC_i\|_2^2 \asymp p_2^{1-\delta_{2i}}
\asymp \|\bC_i\|_{\min}^2$, as $p_1$ and $p_2$ go to infinity and $k_1$
and $k_2$ are fixed.

\noindent{\bf Condition B1.} Assume $z_t$ is a continuous random variable and the process $\{z_t\}$ is stationary. The marginal probability of $z_t$ satisfies that $P(z_t \leq r_1)>0$ and $P(z_t \geq r_2)>0$. $P(z_{t+h}<r_1\mid z_t)>0$ and $P(z_{t+h}>r_2 \mid z_t)>0$ for $z_t\in (-\infty, r_1)\cup (r_2,+\infty)$ and $h=1,\ldots, h_0$. 

Conditions B1 guarantees that there is enough information in partial data with $\{z_t, z_{t+h} \in (-\infty, r_1)\cup (r_2,+\infty) \}$ for loading space estimation.

\medskip
\noindent {\bf Condition B2.} There exists a positive integer $\bar{h}_{ij}\in[1,h_0]$ such that ${\rm rank}(\bSigma_{f,i,j}(\bar{h}_{ij},r_1,r_2))\geq k_{\max}$ and $\|\bSigma_{f,i,j}(\bar{h}_{ij},r_1,r_2)\|_{\min}$ is uniformly bounded above 0, for $i,j=1,2$.

\medskip
\noindent{\bf Condition B3.} $\bM_{s,i}(r_1,r_2)$ has $k_s$ distinct positive
eigenvalues for $s,i=1,2$.

$\bR_i \bF_t \bC_i'$ can be viewed as
the signal part of the observation $\bX_t$, and $\bE_t$ as the
noise. The signal strength, or the strength of the factors, can be
measured by the $L_2$-norm of the loading matrices which is assumed
to grow with the dimensions. The constants $\{\delta_{si}\}$ in Condition A4 control the strength levels of factors. Condition B2 implies that autocovariance matrices in both regimes account for $\bM_{s,i}(r_1,r_2)$, and guarantees the 'helping' effect since the one corresponding to the strong regime carries more information about $\bF_t$ and improves the estimation efficiency. Under Condition B3, we can unique define $\bQ_{s,i}(r_1,r_2)$, where $\bQ_{s,i}=(\bq_{s,i,1}(r_1,r_2),\ldots, (\bq_{s,i,k_i}(r_1,r_2))$ and $\bq_{s,i,k}(r_1,r_2)$ is the unit eigenvector of $\bM_{s,i}(r_1,r_2)$ corresponding to the $k$-th largest eigenvalue.

\medskip
\noindent{\bf Condition C1.} Assume $r_0 \in (\eta_1,\eta_2)$. $z_t$ is a continuous random variable, and the process $\{z_t\}$ is stationary. The marginal probability of $z_t$ satisfies that $P(z_t \leq \eta_1)>0$ and $P(z_t \geq \eta_2)>0$. The densities of $z_t$, $f(z_t)$ is continuous, and there exist two positive constants $\tau_1$ and $\tau_2$ such that $\tau_2 \leq f(z_t)\leq \tau_1$ uniformly in $[\eta_1,\eta_2]$. The conditional probability of $z_{t+h}$ satisfies that $P(z_{t+h}<\eta_1\mid z_t)>\pi_1>0$ and $P(z_{t+h}>\eta_2 \mid z_t)>\pi_2>0$ for any $z_t\in (-\infty, \eta_1)$ or $(\eta_2,+\infty)$ and $h=1,\ldots, h_0$. %The density of $z_{t+h}$ given $z_t$, $f(z_{t+h}\mid z_t)$ is continuous and $\tau_4< f(z_{t+h}\mid z_t)<\tau_3$ uniformly when $z_t \in [\eta_1,\eta_2]$.

\medskip
\noindent {\bf Condition C2.} There exists a positive integer $\tilde{h}_{ij} \in[1,h_0]$ such that ${\rm rank}(\bSigma_{f,i,j}(h_{ij},\eta_1,\eta_2))\geq k_{\max}$ and $\|\bSigma_{f,i,j}(h_{ij},\eta_1,\eta_2)\|_{\min}$ is uniformly bounded above 0, for $i,j=1,2$.

\medskip
\noindent{\bf Condition C3.} $\bM_{s,i}(\eta_1,\eta_2)$ has $k_s$ distinct positive
eigenvalues for $s,i=1,2$. 

\medskip
\noindent {\bf Condition C4.} There exists a positive integer $h_{ij} \in[1,h_0]$ such that ${\rm rank}(\bSigma_{f,i,j}(h_{ij},r_0))\geq k_{\max}$ and $\|\bSigma_{f,i,j}(h_{ij},r_0)\|_{\min}$ is uniformly bounded above 0, for $i,j=1,2$.

\medskip
\noindent{\bf Condition C5.} $\bM_{s,i}(r_0)$ has $k_s$ distinct positive
eigenvalues for $s,i=1,2$.

\medskip
\noindent{\bf Condition C6.} For any $r \in(\eta_1,r_0)$, there exists an integer $h_1^*\in[1,h_0]$ such that ${\rm rank}(\bSigma_{f}(h_1^*, -\infty, r_0, -\infty, r_0))\geq k_{\max}$ and ${\rm rank}(\bSigma_{f}(h_1^*, -\infty, r_0, r,+\infty))\geq k_{\max}$. For any $r \in (r_0, \eta_2)$, there exists an integer $h_2^*\in[1,h_0]$ such that ${\rm rank}(\bSigma_{f}(h_2^*,r_0,r,r,+\infty))\geq k_{\max}$ and ${\rm rank}(\bSigma_{f}(h_2^*,r_0,r,-\infty,r_0))\geq k_{\max}$. The minimum nonzero singular values of these four matrices mentioned are all uniformly bounded above $\gamma_0$, where $\gamma_0>0$.

Condition C1-3 indicate that the estimators for loading spaces are consistent when only data with $\{z_t, z_{t+h} \in (-\infty, \eta_1)\cup (\eta_2,+\infty) \}$ are used. Condition C4 is a natural generalization of Condition B3 and avoid a redundant model. Condition C5 uniquely defines $\bQ_{s,i}$ for $s,i=1,2$. Condition C6 guarantees that the autocovariance matrices of $\rvec(\bF_t)$ with mixed data from two regimes carries all the information from the latent factor process.

\medskip
\noindent{\bf Condition C7.} {\bf Strength of thresholding.} $\cM(\bR_1)\neq \cM(\bR_2)$ and $\cM(\bC_1)\neq \cM(\bC_2)$. There exists two positive constants $\beta_1,\beta_2 \in [0,1]$ such that $[\cD(\cM(\bR_1),\cM(\bR_2))]^2 \asymp p_1^{\beta_1-1}$ and  $[\cD(\cM(\bC_1),\cM(\bC_2))]^2 \asymp p_2^{\beta_2-1}$.

$\beta_1$ and $\beta_2$ reflect the growth rate of the distances of loading spaces as $p_1$ and $p_2$ go to infinity. It also measures how much the thresholding effect $\{\bX_t\}$ experiences. If $\beta_1=\beta_2=0$, only a finite number of elements in $\bR_i$ and $\bC_i$ are different across regimes when $p_1$ and $p_2$ go to infinity; if $\beta_1=\beta_2=1$, the number of elements which undergo a change is $O(p_i)$, for $i=1,2$.

\medskip
\noindent{\bf Condition C8.} When $\hat{k}_s>k_s$ there exists two positive constants $d_1$ and $d_2$ such that $[\cD(\cM(\bQ_{s,1}^*),\cM(\bQ_{s,2}))]^2>d_sp_s^{\beta_s-1}$ and $[\cD(\cM(\bQ_{s,1}),\cM(\bQ_{s,2}^*)]^2>d_sp_s^{\beta_s-1}$, where $\bQ_{s,i}^*=(\bQ_{s,i},\bU_{s,i})$ is a $p_s\times \hat{k}_s$ matrix, for $s,i=1,2$ and any $p_s \times (\hat{k}_s-k_s)$ matrix $\bU_{s,i}$ such that $\dim(\cM(\bU_{s,i}),\cM(\bQ_{s,i}))=0$.

If the numbers of factors are overestimated, Condition C8 implies that the two augmented loading spaces $\cM(\bQ_{s,1}^*)$ and $\cM(\bQ_{s,2}^*)$  are still differentiable for $s=1,2$.

\newpage
\section{Lemmas and Proofs} \label{appendix:lemmas_proofs}
In the appendix, only the theoretical results for $s=1$ are demonstrated, since those for $s=2$ are similar. Moreover, we mainly focus on the proofs when $r>r_0$ and $\epsilon>0$ because those for $r\leq r_0$ or $\epsilon<0$ can be obtained in a similar fashion. We use $C$'s and $C_i$'s to denote generic uniformly positive constants which only depend on the parameters.

%\[
%\bOmega_{fc,ij,m\ell}(h,r)=\frac{1}{T-h} \sum_{t=1}^{T-h} \rE(\bF_t \bc_{m} \bc_{\ell \cdot}' \bF_{t+h}')I_{t,i}(r)I_{t,j}(r),
%\]
%\[
%\hat{\bOmega}_{fc,ij, m \ell}(h,r)=\frac{1}{T-h} \sum_{t=1}^{T-h} \bF_t \bc_{i,m\cdot} \bc_{j,\ell \cdot}' \bF_{t+h}'I_{t,i}(r)I_{t,j}(r),
%\]
Let $\bD_t$ be the common component of $\bX_t$, i.e., $\bD_t=\sum_{i=1}^2 \bR_i \bF_t \bC_iI_{t,i}(r_0)$. It can also be written as $\bD_t=\bX_t-\bE_t$. Define
\[
\bOmega_{d,ij,m\ell}(h,r)=\frac{1}{T} \sum_{t=1}^{T-h} \rE(\bd_{t,m}' \bd_{t+h,\ell}I_{t,i}(r)I_{t+h,j}(r)),
\]
\[
\hat{\bOmega}_{d,ij,m\ell}(h,r)=\frac{1}{T} \sum_{t=1}^{T-h} \bd_{t,m}' \bd_{t+h,\ell}I_{t,i}(r)I_{t+h,j}(r),
\]
\[
\hat{\bOmega}_{de,ij,m\ell}(h,r)=\frac{1}{T} \sum_{t=1}^{T-h} \bd_{t,m}' \be_{t+h,\ell}I_{t,i}(r)I_{t+h,j}(r),
\]
\[
\hat{\bOmega}_{ed,ij,m\ell}(h,r)=\frac{1}{T} \sum_{t=1}^{T-h} \be_{t,m}' \bd_{t+h,\ell}I_{t,i}(r)I_{t+h,j}(r).
\]
%\[
%{\bOmega}_{m\ell}(h,c_1,c_2,c_3,c_4)=\frac{1}{T-h}\sum_{t=1}^{T-h} \rE( \bR_i \bF_t \bc_{i,m\cdot} \bc_{j,\ell\cdot}' \bF_{t+h}' \bR_j' I_t(h,c_1,c_2,c_3,c_4),
%\]
%\[
%\hat{\bOmega}_{m\ell}(h,c_1,c_2,c_3,c_4)=\frac{1}{T-h}\sum_{t=1}^{T-h}  \bR_i \bF_t \bc_{i,m\cdot} \bc_{j,\ell\cdot}' \bF_{t+h}' \bR_j' I_t(h,c_1,c_2,c_3,c_4),
%\]
%\[
%\hat{\bOmega}_{e,ij,m\ell}=\frac{1}{T-h} \sum_{t=1}^{T-h} \be_{t,m} \be_{t+h,\ell}' I_{t,i}(r)I_{t,j}(r).
%\]

\begin{lemma}\label{F}
Let $f_{t,qv}$ denote the $(qv)$-th entry in $\bF_t$. Under Conditions A1-A2 and C1, for any $q,m=1,2,\ldots,k_1$, and $v, \ell=1,\ldots, k_2$, if holds that
%\[
%\|\hat{\bOmega}_{fc,ij,m\ell}(h,r_0+\epsilon)-\bOmega_{fc,ij,m\ell}(h,r_0+\epsilon)\|_2 =O_p(T^{-1/2}), 
%\]
\[
\rE \left\{ \frac{1}{T} \sum_{t=1}^{T-h} \left[f_{t,qv}f_{t+h,m\ell}I_t(h,c_1,c_2,c_3,c_4)-\rE (f_{t,qv}f_{t+h,m\ell } I_t(h,c_1,c_2,c_3,c_4)\right] \right\}^2 \leq \frac{(3h+8\alpha)\rho_{c_1,c_2}\rho_{c_3,c_4} \sigma_f^4}{T},
\]
and
\[
\Bigg|\frac{1}{T} \sum_{t=1}^{T-h} \rE( f_{t,qv}f_{t+h,m\ell}I_t(h,c_1,c_2,c_3,c_4)) \Bigg|=\rho_{c_1,c_2}\rho_{c_3,c_4}\sigma_f^2,
\]
where $\alpha=\sum_{u=1}^{\infty}\alpha(u)^{1-2/\gamma}$, and $c_1<c_2<c_3<c_4$ can be any real numbers in $(\eta_1,\eta_2)$, $-\infty$, or $+\infty$. $\rho_{c_1,c_2}=1$ if at least one of them is $-\infty$ or $+\infty$, and $\rho_{c_1,c_2}=\tau_1(c_2-c_1)$ if $c_1$ and $c_2$ are both real numbers, where $\tau_1$ is given in Condition C1.
\end{lemma}
\noindent{\it Proof:} Condition C1 indicates that 
$P(c_1<z_t<c_2)\leq \rho_{c_1,c_2}$ for $t=1,\ldots, T$. By Condition A2 and Jensen's inequality we can show that $\rE(f_{t,qv})^2<\sigma_f^2$, $\rE(f_{t,qv}^4)<\sigma_f^4$, and $\rE(f_{t,qv}^{2\gamma}|)<\sigma_f^{2\gamma}$ for $q=1,\ldots,k_1$, $v=1,\ldots,k_2$ and $t=1,\ldots,T$. Under Conditions A1 and A2, by Proposition 2.5 in \cite{fan2003}, we have
\begin{eqnarray*}
\lefteqn{\rE \left\{ \frac{1}{T} \sum_{t=1}^{T-h} \Big[ f_{t,qv}f_{t+h,m\ell}I_t(h,c_1,c_2,c_3,c_4)-\rE (f_{t,qv}f_{t+h,m\ell }I_{t}(h,c_1,c_2,c_3,c_4)) \Big]  \right\}^2}\\
&=& \frac{1}{T^2} \sum_{|t_1-t_2| \leq h} \rE [f_{t_1,qv}f_{t_1+h,m\ell }I_{t_1}(h,c_1,c_2,c_3,c_4)- \rE (f_{t_1,qv}f_{t_1+h,m\ell }I_{t_1}(h,c_1,c_2,c_3,c_4))] \\
&& \cdot [f_{t_2,qv}f_{t_2+h,m\ell }I_{t_2}(h,c_1,c_2,c_3,c_4)-\rE (f_{t_2,qv}f_{t_2+h,m\ell }I_{t_2}(h,c_1,c_2,c_3,c_4))]\\
&&+\frac{1}{T^2} \sum_{|t_1-t_2| >h} \rE [f_{t_1,qv}f_{t_1+h,m\ell }I_{t_1}(h,c_1,c_2,c_3,c_4)- \rE (f_{t_1,qv}f_{t_1+h,m\ell }I_{t_1}(h,c_1,c_2,c_3,c_4))] \\
&& \cdot [f_{t_2,qv}f_{t_2+h,m\ell }I_{t_2}(h,c_1,c_2,c_3,c_4)-\rE (f_{t_2,qv}f_{t_2+h,m\ell }I_{t_2}(h,c_1,c_2,c_3,c_4))]\\
& \leq& \frac{[(2h+1)T-h^2-h]\rho_{c_1,c_2}\rho_{c_3,c_4}\sigma_f^4}{T^2} +\frac{8(T-h)\rho_{c_1,c_2}\rho_{c_3,c_4}\sigma_f^4}{T^2}\sum_{u=1}^{T-2h-1}\alpha(u)^{1-2\gamma}\\
&\leq&  \frac{[(2h+1)T-h^2-h]\rho_{c_1,c_2}\rho_{c_3,c_4}\sigma_f^4}{T^2} +\frac{8\alpha\rho_{c_1,c_2}\rho_{c_3,c_4}\sigma_f^4}{T}\leq \frac{(3h+8\alpha)\rho_{c_1,c_2}\rho_{c_3,c_4} \sigma_f^4}{T},
\end{eqnarray*}
and
\begin{eqnarray*}
\lefteqn{ \Bigg| \frac{1}{T} \sum_{t=1}^{T-h} \rE (f_{t,qv}f_{t+h,m\ell}I_t(h,c_1,c_2,c_3,c_4) )\Bigg|}\\
&=& \frac{1}{T}\sum_{t=1}^{T-h} \rE |f_{t,qv}f_{t+h,m\ell }| \cdot E|I_{t}(h,c_1,c_2,c_3,c_4)| \leq \frac{(T-h)\rho_{c_1,c_2}\rho_{c_3,c_4}\sigma_f^2}{T}\leq \rho_{c_1,c_2}\rho_{c_3,c_4}\sigma_f^2.
\end{eqnarray*}
%where $C$ is a positive constant.
\endp

\begin{lemma}\label{f}
For $i,j=1,2$, $m,\ell=1,\ldots,p_2$, it holds that
\begin{eqnarray*}
\|\bOmega_{fc,ij,m\ell}(h,c_1,c_2,c_3,c_4)\|_2^2 \leq \Big\| \frac{1}{T} \sum_{t=1}^{T-h} \rE(\bF_{t+h} \otimes \bF_t I_t(h,c_1,c_2,c_3,c_4))]\Big\|_F^2 \| \bc_{i,m \cdot}\|_2^2 \cdot \| \bc_{j,\ell \cdot} \|_2^2.
\end{eqnarray*}
and
\begin{eqnarray*}
\lefteqn{\|\hat{\bOmega}_{fc,ij,m\ell}(h,c_1,c_2,c_3,c_4)-\bOmega_{fc,ij,m\ell}(h,c_1,c_2,c_3,c_4)\|_2^2}\\
& \leq& \Big\| \frac{1}{T} \sum_{t=1}^{T-h} [\bF_{t+h} \otimes \bF_t I_t(h,c_1,c_2,c_3,c_4)-\rE(\bF_{t+h} \otimes \bF_t I_t(h,c_1,c_2,c_3,c_4))]\Big\|_F^2 \| \bc_{i,m \cdot}\|_2^2 \cdot \| \bc_{j,\ell \cdot} \|_2^2,
\end{eqnarray*}
where $c_1<c_2<c_3<c_4$ can be real numbers, $-\infty$, or $+\infty$.
\end{lemma}
\noindent{\it Proof:} By the definition and properties of Kronecker product, we have
\begin{eqnarray}
\lefteqn{\|\bOmega_{fc,ij,m\ell}(h,c_1,c_2,c_3,c_4)\|_2^2 } \nonumber\\
&\leq& \|\bOmega_{fc,ij,m\ell}(h,c_1,c_2,c_3,c_4)\|_F^2
= \|{\rm vec} (\bOmega_{fc,ij,m\ell}(h,c_1,c_2,c_3,c_4)) \|_2^2  \nonumber \\
&=& \Big\| \frac{1}{T} \sum_{t=1}^{T-h} {\rm vec}(\rE (\bF_t \bc_{i,m \cdot} \bc_{j,\ell \cdot}' \bF_{t+h}'I_t(h,c_1,c_2,c_3,c_4)) )\Big\|_2^2 \nonumber \\
&=&\Big\| \frac{1}{T} \sum_{t=1}^{T-h} [ \rE(\bF_{t+h} \otimes \bF_t\cdot I_t(h,c_1,c_2,c_3,c_4))] {\rm vec}(\bc_{i,m \cdot} \bc_{j,\ell \cdot}') \Big\|_2^2  \nonumber \\
& \leq & \Big\| \frac{1}{T} \sum_{t=1}^{T-h} [\rE(\bF_{t+h} \otimes \bF_t\cdot I_t(h,c_1,c_2,c_3,c_4))]\Big\|_2^2 \| {\rm vec}(\bc_{i,m \cdot} \bc_{j,\ell \cdot}') \|_2^2  \nonumber \\
& \leq & \Big\| \frac{1}{T} \sum_{t=1}^{T-h} [ \rE(\bF_{t+h} \otimes \bF_tI_t(h,c_1,c_2,c_3,c_4)]\Big\|^2_F \| \bc_{i,m \cdot} \bc_{j,\ell \cdot}' \|_F^2  \nonumber \\
& \leq & \Big\| \frac{1}{T} \sum_{t=1}^{T-h} [\rE(\bF_{t+h} \otimes \bF_t I_t(h,c_1,c_2,c_3,c_4))]\Big\|^2_F \| \bc_{i,m \cdot}\|_2^2 \cdot \| \bc_{j,\ell \cdot} \|_2^2. \nonumber
\end{eqnarray}
The other inequality can be proven similarly.
\endp

\begin{lemma}\label{fc}
Under Conditions A1-A2, A4 and C1, for $i,j=1,2$, it holds that
\begin{eqnarray*}
\lefteqn{\sum_{m=1}^{p_2}\sum_{\ell=1}^{p_2} \rE \|\hat{\bOmega}_{fc,ij, m \ell}(h,c_1,c_2,c_3,c_4) -\bOmega_{fc,ij,m\ell}(h,c_1,c_2,c_3,c_4)\|_2^2}\\
&\leq&(3h+8\alpha)\rho_{c_1,c_2}\rho_{c_3,c_4}k_1^2k_2^4  a_0^4\sigma_f^4 p_2^{2-\delta_{2i}-\delta_{2j}} T^{-1},
\end{eqnarray*}
and
\begin{eqnarray*}
\sum_{m=1}^{p_2}\sum_{\ell=1}^{p_2} \| {\bOmega}_{fc,ij, m \ell}(h,c_1,c_2,c_3,c_4) \|_2^2\leq  \rho_{c_1,c_2}^2\rho_{c_3,c_4}^2 k_1^2 k_2^4a_0^4\sigma_f^4  p_2^{2-\delta_{2i}-\delta_{2j}},
\end{eqnarray*}
where $a_0$ satisfies $\|\bC_i\|_2 \leq a_0 p_i^{1/2-\delta_{2i}/2}$ for $i=1,2$, and $c_1<c_2<c_3<c_4$can be real numbers in $(\eta_1,\eta_2)$, $-\infty$ or $+\infty$.
\end{lemma}
\noindent{\it Proof:} Condition A4 implies that there exists a positive constant $a_0$ such that $\|\bC_i\|_2 \leq a_0p_2^{1/2-\delta_{2i}/2}$ for $i=1,2$. By Lemma \ref{F} and Lemma \ref{f}, it follows
\begin{eqnarray*}
\lefteqn{\sum_{m=1}^{p_2}\sum_{\ell=1}^{p_2}\rE \|\hat{\bOmega}_{fc,ij,m\ell}(h,c_1,c_2,c_3,c_4) -\bOmega_{fc,ij,m\ell}(h,c_1,c_2,c_3,c_4)\|_2^2}\\
&=& \left(\sum_{m=1}^{p_2} \|\bc_{i,m\cdot}\|_2^2 \right) \left(\sum_{\ell=1}^{p_2} \|\bc_{j,\ell \cdot}\|_2^2 \right)\\
&&\cdot \rE \Big\| \frac{1}{T} \sum_{t=1}^{T-h} [ \bF_{t+h} \otimes \bF_t I_t(h,c_1,c_2,c_3,c_4)-\rE(\bF_{t+h} \otimes \bF_t I_t(h,c_1,c_2,c_3,c_4))]\Big\|_F^2\\ 
&\leq & \|\bC_i\|_F^2\|\bC_j\|_F^2\cdot \rE \Big\| \frac{1}{T} \sum_{t=1}^{T-h} [ \bF_{t+h} \otimes \bF_tI_t(h,c_1,c_2,c_3,c_4) -\rE(\bF_{t+h} \otimes \bF_t I_t(h,c_1,c_2,c_3,c_4))]\Big\|_F^2\\
&\leq &k_2^2 \|\bC_i\|_2^2\|\bC_j\|_2^2 \cdot \rE \Big\| \frac{1}{T} \sum_{t=1}^{T-h} [ \bF_{t+h} \otimes \bF_tI_t(h,c_1,c_2,c_3,c_4) -\rE(\bF_{t+h} \otimes \bF_t I_t(h,c_1,c_2,c_3,c_4))]\Big\|_F^2\\
&\leq&(3h+8\alpha) \rho_{c_1,c_2}\rho_{c_3,c_4}k_1^2k_2^4 a_0^4\sigma_f^4 p_2^{2-\delta_{2i}-\delta_{2j}} T^{-1}.
\end{eqnarray*}
We can also obtain the bound of $\sum_{m=1}^{p_2}\sum_{\ell=1}^{p_2} \|{\bOmega}_{fc,ij, m \ell}(h,c_1,c_2,c_3,c_4) \|_2^2$ with Lemma \ref{F} and Lemma \ref{f}, 
\begin{eqnarray*}
\lefteqn{\sum_{m=1}^{p_2}\sum_{\ell=1}^{p_2} \|\bOmega_{fc,ij,m\ell}(h,c_1,c_2,c_3,c_4)\|_2^2}\\
&=& \left(\sum_{m=1}^{p_2} \|\bc_{i,m\cdot}\|_2^2 \right) \left(\sum_{\ell=1}^{p_2} \|\bc_{j,\ell \cdot}\|_2^2 \right) \Big\| \frac{1}{T} \sum_{t=1}^{T-h} \rE(\bF_{t+h} \otimes \bF_t I_t(h,c_1,c_2,c_3,c_4))\Big\|_F^2\\ 
&\leq & \|\bC_i\|_F^2\|\bC_j\|_F^2 \Big\| \frac{1}{T} \sum_{t=1}^{T-h} \rE(\bF_{t+h} \otimes \bF_t I_t(h,c_1,c_2,c_3,c_4))\Big\|_F^2\\
&\leq &k_2^2\|\bC_i\|_2^2\|\bC_j\|_2^2 \Big\| \frac{1}{T} \sum_{t=1}^{T-h} \rE(\bF_{t+h} \otimes \bF_t I_t(h,c_1,c_2,c_3,c_4))\Big\|_F^2\\
&\leq&\rho_{c_1,c_2}^2\rho_{c_3,c_4}^2 k_1^2 k_2^4 a_0^4 \sigma_f^4  p_2^{2-\delta_{2i}-\delta_{2j}}).
\end{eqnarray*}
\endp

\begin{lemma}\label{hatX}
Under Conditions A1-A4 and C1, for $i,j=1,2$ and any $\epsilon\in (\eta_1-r_0,\eta_2-r_0)$, it holds that
\[
\sum_{m=1}^{p_2}\sum_{\ell=1}^{p_2} \rE \|\hat{\bOmega}_{x,ij, m \ell}(h,r_0+\epsilon) -\bOmega_{x,ij,m\ell}(h,r_0+\epsilon)\|_2^2 \leq 100(3h+8\alpha)k_1^2k_2^4 a_1^8\sigma_0^4 \tau_0^2 \delta_{\eta}^2 p_1^2p_2^2T^{-1},
\]
where $\sigma_0=\max \{\sigma_f, \sigma_e\}$, $\tau_0=\max\{\tau_1,1\}$, $\delta_{\eta}=\max \{\eta_2-\eta_1,1\}$, and $a_1>1$ satisfies $\|\bR_i\|_2 \leq a_1 p_1^{1/2-\delta_{1i}/2}$ and $\|\bC_i\|_2 \leq a_1p_2^{1/2-\delta_{2i}/2}$ for $i=1,2$.
\end{lemma}
\noindent{\it Proof:} By Condition A4, we can find a positive constant $a_1>1$ such that $\|\bR_i\|_2 \leq a_1 p_1^{1/2-\delta_{1i}/2}$ and $\|\bC_i\|_2 \leq a_1 p_2^{1/2-\delta_{2i}/2}$ for $i=1,2$. By Lemmas \ref{F} and \ref{fc},  when $\epsilon>0$ and $i=j=1$,
\begin{eqnarray*}
\lefteqn{\sum_{m=1}^{p_2}\sum_{\ell=1}^{p_2} \rE \|\hat{\bOmega}_{d,11,m\ell}(h,r_0+\epsilon) -\bOmega_{d,11,m\ell}(h,r_0+\epsilon)\|_2^2}\\
%&\leq&\sum_{m=1}^{p_2}\sum_{\ell=1}^{p_2} 4 \rE \|\bR_1( \hat{\bOmega}_{fc,11,m\ell}(h,-\infty,r_0,-\infty,r_0) -\bOmega_{fc,11,m\ell}(h,-\infty,r_0,-\infty,r_0) )\bR_1' \|_2^2\\
%&&+\sum_{m=1}^{p_2}\sum_{\ell=1}^{p_2} 4\rE \|\bR_1( \hat{\bOmega}_{fc,12,m\ell}(h,-\infty,r_0,r_0,r_0+\epsilon) -\bOmega_{fc,12,m\ell}(h,-\infty,r_0,r_0,r_0+\epsilon) )\bR_2' \|_2^2\\
%&&+\sum_{m=1}^{p_2}\sum_{\ell=1}^{p_2} 4 \rE \|\bR_2( \hat{\bOmega}_{fc,21,m\ell}(h,r_0,r_0+\epsilon,-\infty,r_0) -\bOmega_{fc,21,m\ell}(h,r_0,r_0+\epsilon,-\infty,r_0) ) \bR_1' \|_2^2\\
%&&+\sum_{m=1}^{p_2}\sum_{\ell=1}^{p_2} 4\rE \|\bR_2( \hat{\bOmega}_{fc,22,m\ell}(h,r_0,r_0+\epsilon,r_0,r_0+\epsilon) -\bOmega_{fc,22,m\ell}(h,r_0,r_0+\epsilon,r_0,r_0+\epsilon) )\bR_2' \|_2^2\\
&\leq&\sum_{m=1}^{p_2}\sum_{\ell=1}^{p_2} 4 \|\bR_1\|_2^4 \cdot \rE\| ( \hat{\bOmega}_{fc,11,m\ell}(h,-\infty,r_0,-\infty,r_0) -\bOmega_{fc,11,m\ell}(h,-\infty,r_0,-\infty,r_0) ) \|_2^2\\
&&+\sum_{m=1}^{p_2}\sum_{\ell=1}^{p_2} 4\|\bR_1\|_2 \cdot  \rE\|  \hat{\bOmega}_{fc,12,m\ell}(h,-\infty,r_0,r_0,r_0+\epsilon) -\bOmega_{fc,12,m\ell}(h,-\infty,r_0,r_0,r_0+\epsilon) \|_2^2\cdot  \|\bR_2\|_2^2\\
&&+\sum_{m=1}^{p_2}\sum_{\ell=1}^{p_2} 4  \|\bR_2\|_2^2 \cdot  \rE\| \hat{\bOmega}_{fc,21,m\ell}(h,r_0,r_0+\epsilon,-\infty,r_0) -\bOmega_{fc,21,m\ell}(h,r_0,r_0+\epsilon,-\infty,r_0) \|_2^2 \cdot \|\bR_1\|_2^2\\
&&+\sum_{m=1}^{p_2}\sum_{\ell=1}^{p_2} 4 \|\bR_2\|_2^4 \cdot \rE\| \hat{\bOmega}_{fc,22,m\ell}(h,r_0,r_0+\epsilon,r_0,r_0+\epsilon) -\bOmega_{fc,22,m\ell}(h,r_0,r_0+\epsilon,r_0,r_0+\epsilon) \|_2^2 \\
%&=&4(3h+8\alpha)k_1^2k^4_2a_1^8 \sigma_f^4 p_1^{2-2\delta_{11}} p_2^{2-2\delta_{21}}T^{-1}+8(3h+8\alpha)k_1^2k^4_2 a_1^8\tau_1 \sigma_f^4  \epsilon p_1^{2-\delta_{11}-\delta_{12}} p_2^{2-2\delta_{21}-\delta_{22}}T^{-1}\\
%&&+4(3h+8\alpha)k_1^2k^4_2a_1^8 \tau_1^2 \sigma_f^4 C_2^4\epsilon^2 p_1^{2-2\delta_{12}} p_2^{2-2\delta_{22}}T^{-1}\\
&\leq& 4(3h+8\alpha)k_1^2k^4_2 a_1^8 \sigma_f^4 (p_1^{2-2\delta_{11}} p_2^{2-2\delta_{21}}+2\tau_1\epsilon p_1^{2-\delta_{11}-\delta_{12}} p_2^{2-2\delta_{21}-\delta_{22}}+ \tau_1^2 \epsilon^2 p_1^{2-2\delta_{12}} p_2^{2-2\delta_{22}})T^{-1}.
\end{eqnarray*}
For the interaction of the common component and noise,
\begin{eqnarray}
\lefteqn{\sum_{m=1}^{p_2} \sum_{\ell=1}^{p_2} \rE \|\hat{\bOmega}_{de,11,m\ell}(h,r_0+\epsilon)\|_2^2} \nonumber \\
& \leq&2 \sum_{m=1}^{p_2} \sum_{\ell=1}^{p_2} \|\bR_1\|_2^2 \cdot  \rE \Big\|  \frac{1}{T} \sum_{t=1}^{T-h}  \bF_t \bc_{1,m\cdot}  \be_{t+h,\ell}' I(z_t<r_0) I_{t+h,1}(r_0+\epsilon) \Big\|_2^2 \nonumber \\
&&+ 2\sum_{m=1}^{p_2} \sum_{\ell=1}^{p_2} \|\bR_2\|_2^2  \cdot  \rE \Big\|  \frac{1}{T} \sum_{t=1}^{T-h}  \bF_t \bc_{2,m\cdot}  \be_{t+h,\ell}' I(r_0<z_t<r_0+\epsilon) I_{t+h,1}(r_0+\epsilon) \Big\|_2^2  \nonumber \\
&\leq& 2 \|\bR_1\|_2^2   \left( \sum_{\ell=1}^{p_2} \rE \Big\|\frac{1}{T} \sum_{t=1}^{T-h} \be_{t+h,\ell} \otimes \bF_t  I(z_t<r_0)\Big\|_2^2 \right)\left( \sum_{m=1}^{p_2} \|\bc_{1,m\cdot}\|^2_2\right) \nonumber \\
&&+2 \|\bR_2\|_2^2   \left( \sum_{\ell=1}^{p_2} \rE \Big\| \frac{1}{T} \sum_{t=1}^{T-h} \be_{t+h,\ell} \otimes \bF_t  I(r_0<z_t<r_0+\epsilon)\Big\|_2^2 \right)\left( \sum_{m=1}^{p_2} \|\bc_{2,m\cdot}\|^2_2\right) \nonumber \\
&\leq& 2 \|\bR_1\|_2^2   \left( \sum_{\ell=1}^{p_2} \rE \Big\|\frac{1}{T} \sum_{t=1}^{T-h} \be_{t+h,\ell} \otimes \bF_t  I(z_t<r_0)\Big\|_F^2 \right) \|\bC_{2}\|^2_F \nonumber \\
&&+2 \|\bR_2\|_2^2   \left( \sum_{\ell=1}^{p_2} \rE \Big\| \frac{1}{T} \sum_{t=1}^{T-h} \be_{t+h,\ell} \otimes \bF_t  I(r_0<z_t<r_0+\epsilon)\Big\|_F^2 \right)  \|\bC_{2}\|^2_F \nonumber \\
%&\leq& 2k_2 \|\bR_1\|_2^2  \|\bC_{1}\|^2_2 \left[ \sum_{\ell=1}^{p_2}\sum_{u=1}^{p_1} \rE \left(\frac{1}{T} \sum_{t=1}^{T-h} \sum_{q=1}^{k_1}\sum_{v=1}^{k_2}e_{t+h,u\ell}F_{t,qv}  I(z_t<r_0)\right)^2 \right]\\
%&&+2 k_2 \|\bR_2\|_2^2  \|\bC_{2}\|^2_2   \left[ \sum_{\ell=1}^{p_2}\sum_{u=1}^{p_1} \rE \left( \frac{1}{T} \sum_{t=1}^{T-h} \sum_{q=1}^{k_1}\sum_{v=1}^{k_2} e_{t+h,u\ell} F_{t,qv}  I(r_0<z_t<r_0+\epsilon)\right)^2 \right]\\
&\leq& 2k_2 a_1^4p_1^{1-\delta_{11}}p_2^{1-\delta_{21}}  \left[ \frac{1}{T^2}\sum_{\ell=1}^{p_2}\sum_{u=1}^{p_1}\sum_{q=1}^{k_1}\sum_{v=1}^{k_2}\sum_{t=1}^{T-h} \rE\left(e_{t+h,u\ell}^2 f_{t,qv}^2  I(z_t<r_0)\right) \right] \nonumber \\
&&+2k_2 a_1^4p_1^{1-\delta_{12}}p_2^{1-\delta_{22}}  \left[ \frac{1}{T^2}\sum_{\ell=1}^{p_2}\sum_{u=1}^{p_1}\sum_{q=1}^{k_1}\sum_{v=1}^{k_2}\sum_{t=1}^{T-h} \rE\left(e_{t+h,u\ell}^2 f_{t,qv}^2  I(r_0<z_t<r_0+\epsilon)\right) \right] \nonumber \\
&\leq&2k_1k_2^2 a_1^4\sigma_e^2\sigma_f^2 p_1^{2-\delta_{11}}p_2^{2-\delta_{21}}T^{-1}+2k_1k_2^2 a_1^4\sigma_e^2\sigma_f^2  \epsilon p_1^{2-\delta_{12}}p_2^{2-\delta_{22}}T^{-1} \nonumber \\
&=&2k_1k_2^2a_1^4\sigma_e^2\sigma_f^2\, (p_1^{2-\delta_{11}}p_2^{2-\delta_{21}}+\epsilon p_1^{2-\delta_{12}}p_2^{2-\delta_{22}})T^{-1},\label{de}
\end{eqnarray}
and similarly we have
\[ \sum_{m=1}^{p_2} \sum_{\ell=1}^{p_2}\rE \|\hat{\bOmega}_{ed,11,m\ell}(h,r_0+\epsilon)\|_2^2\leq 2k_1k_2^2 a_1^4\sigma_e^2 \sigma_f^2 \, (p_1^{2-\delta_{11}}p_2^{2-\delta_{21}}+\epsilon p_1^{2-\delta_{12}}p_2^{2-\delta_{22}})T^{-1}.
\]
For the noise term,
\begin{eqnarray*}
\lefteqn{\sum_{m=1}^{p_2} \sum_{\ell=1}^{p_2} \rE \|\hat{\bOmega}_{e,11,m\ell}(h,r_0+\epsilon)\|_2^2}\\
& =&\sum_{m=1}^{p_2} \sum_{\ell=1}^{p_2} \rE \Big\|\frac{1}{T}\sum_{t=1}^{T-h}\be_{t,m} \be_{t+h,\ell}'I_{t,1}(r_0+\epsilon)I_{t+h,1}(r_0+\epsilon)\Big\|_2^2\\
&\leq &\frac{1}{T^2} \sum_{m=1}^{p_2} \sum_{\ell=1}^{p_2} \sum_{u=1}^{p_1}\sum_{v=1}^{p_1} \sum_{t=1}^{T-h}\rE (e_{t,um}^2e^2_{t+h,v \ell }I_{t,1}(r_0+\epsilon)I_{t+h,1}(r_0+\epsilon))\leq \sigma_e^4p_1^2p_2^2T^{-1}.
\end{eqnarray*}
It follows
\begin{eqnarray*}
\lefteqn{\sum_{m=1}^{p_2} \sum_{\ell=1}^{p_2}\rE \|\hat{\bOmega}_{x,11,m\ell}(h,r_0+\epsilon)-\bOmega_{x,11,m\ell}(h,r_0+\epsilon)\|_2^2}\\
&\leq &  \sum_{m=1}^{p_2} \sum_{\ell=1}^{p_2} \left(4\rE \|\hat{\bOmega}_{d,11,m\ell}(h,r_0+\epsilon)-{\bOmega}_{d,11,m\ell}(h,r_0+\epsilon)\|_2^2+4 \rE \|\hat{\bOmega}_{de,11,m\ell} (h, r_0+\epsilon) \|_2^2 \right. \\
&&\left. + 4\rE  \|\hat{\bOmega}_{ed,11,m\ell} (h, r_0+\epsilon) \|_2^2 +4\rE \|\hat{\bOmega}_{e,11,m\ell}(h,r_0+\epsilon)\|_2^2 \right)\\
&\leq&16(3h+8\alpha)k_1^2k^4_2 a_1^8 \sigma_f^4  (p_1^{2-2\delta_{11}} p_2^{2-2\delta_{21}}+2\tau_1\epsilon p_1^{2-\delta_{11}-\delta_{12}} p_2^{2-2\delta_{21}-\delta_{22}}+ \tau_1^2 \epsilon^2 p_1^{2-2\delta_{12}} p_2^{2-2\delta_{22}})T^{-1}\\
&&+16k_1k_2^2a_1^4\sigma_e^2\sigma_f^2\, (p_1^{2-\delta_{11}}p_2^{2-\delta_{21}}+\epsilon p_1^{2-\delta_{12}}p_2^{2-\delta_{22}})T^{-1}+4\sigma_e^4p_1^2p_2^2T^{-1}\\
&\leq &64(3h+8\alpha)k_1^2k_2^4a_1^8\sigma_0^4\tau_0^2 \delta_{\eta}^2 p_1^2p_2^2T^{-1}+32k_1k_2^2a_1^4\sigma_0^4 \delta_{\eta}p_1^2p_2^2T^{-1}+4\sigma_0^4 p_1^2p_2^2T^{-1}\\
&\leq & 100(3h+8\alpha)k_1^2k_2^4 a_1^8\sigma_0^4 \tau_0^2\delta_{\eta}^2p_1^2p_2^2T^{-1},
\end{eqnarray*}
where $\tau_0=\max\{\tau_1,1\}$, $\delta_{\eta}=\max\{\eta_2-\eta_1,1\}$, and $\sigma_0=\max\{\sigma_f,\sigma_e\}$.

For other cases, we can prove it in a similar way.
\endp

\begin{lemma}\label{Y}
Under Conditions A1-A4 and C1, for $\epsilon\in (\eta_1-r_0,\eta_2-r_0)$, we have
\begin{align*}
\lefteqn{\sum_{m=1}^{p_2} \sum_{\ell=1}^{p_2}\|\bOmega_{x,11,m\ell}(h,r_0+\epsilon)\|_2^2}\\
&\leq& \left\{
\begin{array}{ll}
k_1^2k_2^4 a_1^8 \sigma_f^4p_1^{2-2\delta_{11}}p_2^{2-2\delta_{21}}	&\epsilon \leq 0,\\
4k_1^2 k_2^4 a_1^8 \sigma_f^4\, ( p_1^{2-2\delta_{11}}p_2^{2-2\delta_{21}}+2 \epsilon^2 p_1^{2-\delta_{11}-\delta_{12}}p_2^{2-\delta_{21}-\delta_{22}}+ \epsilon^4 p_1^{2-2\delta_{12}}p_2^{2-\delta_{22}}) &\epsilon >0,
\end{array}
\right.
\end{align*}
\begin{align*}
\lefteqn{\sum_{m=1}^{p_2} \sum_{\ell=1}^{p_2}\|\bOmega_{x,22,m\ell}(h,r_0+\epsilon)\|_2^2}\\
&\leq& \left\{
\begin{array}{ll}
4k_1^2k_2^4 a_1^8\sigma_f^4\, (p_1^{2-2\delta_{12}}p_2^{2-2\delta_{22}}+2\epsilon^2 p_1^{2-\delta_{11}-\delta_{12}}p_2^{2-\delta_{21}-\delta_{22}}+\epsilon^4 p_1^{2-2\delta_{11}}p_2^{2-2\delta_{21}})	&\epsilon \leq 0,\\
k_1^2k_2^4 a_1^8 \sigma_f^4 p_1^{2-2\delta_{12}}p_2^{2-2\delta_{22}}	&\epsilon >0,\\
\end{array}
\right.
\end{align*}
when $i,j \in \{1,2\}$ and $i \neq j$,
\begin{eqnarray*}
\lefteqn{\sum_{m=1}^{p_2} \sum_{\ell=1}^{p_2}\|\bOmega_{x,ij,m\ell}(h,r_0+\epsilon)\|_2^2}\\
&\leq& \left\{
\begin{array}{ll}
2k_1^2k_2^4a_1^8\sigma_f^4(p_1^{2-\delta_{11}-\delta_{12}}p_2^{2-\delta_{21}-\delta_{22}}+\epsilon^2 p_1^{2-2\delta_{11}}p_2^{2-2\delta_{21}})	&\epsilon < 0,\\
2k_1^2k_2^4a_1^8\sigma_f^4 p_1^{2-\delta_{11}-\delta_{12}}p_2^{2-\delta_{21}-\delta_{22}}	&\epsilon=0,\\
2k_1^2k_2^4a_1^8\sigma_f^4(p_1^{2-\delta_{11}-\delta_{12}}p_2^{2-\delta_{21}-\delta_{22}}+\epsilon^2 p_1^{2-2\delta_{12}}p_2^{2-2\delta_{22}})	&\epsilon > 0.
\end{array}
\right.
\end{eqnarray*}
\end{lemma}
\noindent{\it Proof:} By the definition of $\bOmega_{x,11,m\ell}(h,r_0+\epsilon)$ and Lemma \ref{fc}, if $\epsilon>0$, we have
\begin{eqnarray*}
\lefteqn{\sum_{m=1}^{p_2} \sum_{\ell=1}^{p_2}\|\bOmega_{x,11,m\ell}(h,r_0+\epsilon)\|_2^2}\\
&\leq &\sum_{m=1}^{p_2} \sum_{\ell=1}^{p_2} 4\|\bR_1\bOmega_{fc,11,m\ell}(h,-\infty, r_0, -\infty, r_0) \bR_1' \|_2^2 +4 \|\bR_1\bOmega_{fc,12,m\ell}(h,-\infty, r_0,  r_0, r_0+\epsilon) \bR_2' \|_2^2\\
&&+4\|\bR_2\bOmega_{fc,21,m\ell}(h,r_0, r_0+\epsilon,-\infty, r_0)\bR_1'\|_2^2+4\|\bR_2\bOmega_{fc,22,m\ell}(h, r_0,r_0+\epsilon, r_0, r_0+\epsilon) \bR_2' \|_2^2\\
&\leq&4 k_1^2k_2^4 a_1^8 \sigma_f^4 p_1^{2-2\delta_{11}}p_2^{2-2\delta_{21}} +8k_1^2k_2^4a_1^8\sigma_f^4\epsilon^2 p_1^{2-\delta_{11}-\delta_{12}}p_2^{2-\delta_{21}-\delta_{22}}+4k_1^2k_2^4a_1^8 \sigma_f^4\epsilon^4 p_1^{2-2\delta_{12}}p_2^{2-\delta_{22}}\\
&\leq& 4k_1^2 k_2^4 a_1^8 \sigma_f^4\, ( p_1^{2-2\delta_{11}}p_2^{2-2\delta_{21}}+2\epsilon^2 p_1^{2-\delta_{11}-\delta_{12}}p_2^{2-\delta_{21}-\delta_{22}}+ \epsilon^4 p_1^{2-2\delta_{12}}p_2^{2-\delta_{22}}).
\end{eqnarray*}
%\begin{eqnarray*}
%\lefteqn{\sum_{m=1}^{p_2} \sum_{\ell=1}^{p_2}\|\bOmega_{x,12,m\ell}(h,r_0+\epsilon)\|_2^2}\\
%&\leq &\sum_{m=1}^{p_2} \sum_{\ell=1}^{p_2} 2 \|\bR_1\bOmega_{fc,12,m\ell}(h,-\infty, r_0,  r_0+\epsilon, +\infty) \bR_2' \|_2^2+2\|\bR_2\bOmega_{fc,22,m\ell}(h, r_0,r_0+\epsilon,  r_0+\epsilon, +\infty) \bR_2' \|_2^2\\
%&\leq&2k_1^2k_2^4a_1^8\sigma_f^4 p_1^{2-\delta_{11}-\delta_{12}}p_2^{2-\delta_{21}-\delta_{22}}+2k_1^2k_2^4a_1^8 \sigma_f^4\epsilon^2 p_1^{2-2\delta_{12}}p_2^{2-\delta_{22}}.
%\end{eqnarray*}

Other equations can be obtained in a similar fashion.
\endp

\begin{lemma}\label{dist}
Under Conditions A4 and C7, it holds that 
\[
a_2^2 b_2^2p_1^{\beta_1-\delta_{12}}\leq \|\bB_{1,1}' \bR_2\|_2^2 \leq  k_1 a_1^2  b_1^2 p_1^{\beta_1-\delta_{12}},
\]
and
\[
a_2^2 b_2^2 p_1^{\beta_1-\delta_{12}}\leq \|\bB_{1,2}' \bR_1\|_2^2 \leq  k_1 a_1^2 b_1^2  p_1^{\beta_1-\delta_{12}},
\]
where $b_1$ and $b_2$ are positive constants such that $b_2 p^{\beta_1/2-1/2} \leq \cD(\cM(\bQ_{s,1}),\cM (\bQ_{s,2})) \leq b_1 p^{\beta_1/2-1/2}$, for $s=1,2$, and $a_2$ is a positive constant such that $\|\bC_i\|_{\min} \geq a_2 p_2^{1/2-\delta_{2i}/2}$ for $i=1,2$.
\end{lemma}
\noindent{\it Proof.} Note that 
\begin{eqnarray*}
\lefteqn{\tr \left[ \bQ_{1,2}'  \left( \begin{array}{cc}
\bQ_{1,1} 	&\bB_{1,1}
\end{array}
\right)  \left(\begin{array}{c} 
\bQ_{1,1}'\\
\bB_{1,1}'
\end{array} \right)
\bQ_{1,2} \right]}\\
& =& \tr( \bQ_{1,2}'\bQ_{1,1}\bQ_{1,1}' \bQ_{1,2}) +\tr(\bQ_{1,2}' \bB_{1,1} \bB_{1,1}' \bQ_{1,2}) \\
&=&k_1\{1-  \left[{\cal D}({\cal M}(\bQ_{1,1}), {\cal M}(\bQ_{1,2}))\right]^2\} +\tr(\bQ_{1,2}' \bB_{1,1} \bB_{1,1}' \bQ_{1,2}).
\end{eqnarray*}
On the other hand, 
\begin{eqnarray*}
\tr \left[ \bQ_{1,2}'  \left( \begin{array}{cc}
\bQ_{1,1} 	&\bB_{1,1}	
\end{array}
\right)  \left(\begin{array}{c} 
\bQ_{1,1}'\\
\bB_{1,1}'
\end{array} \right)
\bQ_{1,2} \right] =\tr( \bQ_{1,2}'\bQ_{1,2}) =k_1.
\end{eqnarray*}
Hence, $\tr(\bQ_{1,2}' \bB_{1,1} \bB_{1,1}' \bQ_{1,2}) =k_1 [\cD(\cM(\bQ_{1,1}),\cM (\bQ_{1,2}))]^2$. Then Condition C7 indicates that we can find two positive constants $b_1$ and $b_2$ such that  $b_2 p^{\beta_1/2-1/2} \leq \cD(\cM(\bQ_{1,1}),\cM (\bQ_{1,2})) \leq b_1 p^{\beta_1/2-1/2}$. Therefore, $\|\bB_{1,1}' \bQ_{1,2}\|_2^2 \geq \tr(\bQ_{1,2}' \bB_{1,1} \bB_{1,1}' \bQ_{1,2})/k_1 \geq b_2^2p_1^{\beta_1-1}$ and $\|\bB_{1,1}' \bQ_{1,2}\|_2^2 \leq \tr(\bQ_{1,2}' \bB_{1,1} \bB_{1,1}' \bQ_{1,2}) \leq k_1 b_1^2 p_1^{\beta_1-1}$. With Condition A4, we have
\[
a_2^2 b_2^2 p_1^{\beta_1-\delta_{12}}\leq \|\bB_{1,1}' \bR_2\|_2^2 \leq k_1 a_1^2 b_1^2  p_1^{\beta_1-\delta_{12}}.
\]
\endp

\begin{lemma}\label{BY}
Under Conditions A1-A4, C5, and C7 for $\epsilon\in (\eta_1-r_0,\eta_2-r_0)$, we have
\begin{eqnarray*}
\lefteqn{\sum_{m=1}^{p_2} \sum_{\ell=1}^{p_2}\|\bB_{1,1}(\eta_1,\eta_2)'\bOmega_{x,11,m\ell}(h,r_0+\epsilon) \|_2^2}\\
&& \left\{
\begin{array}{ll}
=0	&\epsilon \leq 0,\\
\leq 4k_1^3k_2^4a_1^8 b_1^2\sigma_f^4\, ( \epsilon^2 p_1^{1+\beta_1-\delta_{11}-\delta_{12}}p_2^{2-\delta_{21}-\delta_{22}}+\epsilon^4 p_1^{1+\beta_1-2\delta_{12}}p_2^{2-2\delta_{22}})	&\epsilon >0,
\end{array}
\right.
\end{eqnarray*}
\begin{eqnarray*}
\sum_{m=1}^{p_2} \sum_{\ell=1}^{p_2}\|\bB_{1,1}(\eta_1,\eta_2)'\bOmega_{x,12,m\ell}(h,r_0+\epsilon)\|_2^2 \left\{
\begin{array}{ll}
=0	&\epsilon \leq 0\\
\leq 2k_1^3k_2^4a_1^8b_1^2 \sigma_f^4\, \epsilon^2 p_1^{1+\beta_1-2\delta_{12}}p_2^{2-2\delta_{22}}	&\epsilon >0,
\end{array}
\right.
\end{eqnarray*}
\begin{eqnarray*}
\sum_{m=1}^{p_2} \sum_{\ell=1}^{p_2}\|\bB_{1,2}(\eta_1,\eta_2)'\bOmega_{x,21,m\ell}(h,r_0+\epsilon)\|_2^2 \left\{
\begin{array}{ll}
\leq 2k_1^3k_2^4a_1^8b_1^2 \sigma_f^4\, \epsilon^2 p_1^{1+\beta_1-2\delta_{11}}p_2^{2-2\delta_{21}}	  &\epsilon \leq 0,\\
=0	&\epsilon >0,\\
\end{array}
\right.
\end{eqnarray*}
\begin{eqnarray*}
\lefteqn{\sum_{m=1}^{p_2} \sum_{\ell=1}^{p_2}\| \bB_{1,2}(\eta_1,\eta_2)'\bOmega_{x,22,m\ell}(h,r_0+\epsilon)\|^2_2}\\
&& \left\{
\begin{array}{ll}
\leq 4k_1^3k_2^4a_1^8 b_1^2 \sigma_f^4 \,(\epsilon^2 p_1^{1+\beta_1-\delta_{11}-\delta_{12}}p_2^{2-\delta_{21}-\delta_{22}}+\epsilon^4 p_1^{1+\beta_1-2\delta_{11}}p_2^{2-2\delta_{21}})	&\epsilon \leq 0,\\
=0 &\epsilon >0.\\
\end{array}
\right.
\end{eqnarray*}
\end{lemma}
\noindent{\it Proof:}
If $\epsilon>0$, by the definition, we have
\begin{eqnarray*}
\lefteqn{\bOmega_{x,11,m\ell}(h,r_0+\epsilon)}\\
&=& \bR_1\bOmega_{fc,11,m\ell}(h,-\infty, r_0, -\infty, r_0)\bR_1'+\bR_1\bOmega_{fc,12,m\ell}(h,-\infty, r_0, r_0,+\infty)\bR_2'\\
&&+\bR_2\bOmega_{fc,21,m\ell}(h, r_0, r_0+\epsilon,-\infty, r_0)\bR_1'+\bR_2\bOmega_{fc,22,m\ell}(h,r_0, r_0+\epsilon,r_0,r_0+\epsilon)\bR_2'.
\end{eqnarray*}
If $r_0 \in(\eta_1,\eta_2)$, by the definition, we have $\cM(\bB_{s,i})=\cM(\bB_{s,i}(\eta_1,\eta_2))$ for $s,i=1,2$. Hence, there exists a $(p_i-k_i)\times (p_i-k_i)$ orthogonal matrix $\bGamma_{s,i}$ such that $\bB_{s,i}=\bB_{s,i}(\eta_1,\eta_2)\bGamma_{s,i}$. By Lemmas \ref{fc} and \ref{dist} and Cauchy-Schwarz inequality, 
\begin{eqnarray*}
\lefteqn{\sum_{m=1}^{p_2} \sum_{\ell=1}^{p_2}\|\bB_{1,1}(\eta_1,\eta_2)' \bOmega_{x,11,m\ell}(h,r_0+\epsilon)\|_2^2=\sum_{m=1}^{p_2} \sum_{\ell=1}^{p_2}\|\bB_{1,1}' \bOmega_{x,11,m\ell}(h,r_0+\epsilon)\|_2^2 }\\
&\leq &\sum_{m=1}^{p_2} \sum_{\ell=1}^{p_2} \Big[4\|\bB_{1,1}'\bR_1\bOmega_{fc,11,m\ell}(h,-\infty, r_0, -\infty, r_0)\bR_1' \|_2^2\\
&& +4 \|\bB_{1,1}'\bR_1\bOmega_{fc,12,m\ell}(h,-\infty, r_0,  r_0, r_0+\epsilon)\bR_2' \|_2^2+4\|\bB_{1,1}'\bR_2\bOmega_{fc,21,m\ell}(h,r_0, r_0+\epsilon,-\infty, r_0)\bR_1' \|_2^2\\
&&+4\|\bB_{1,1}' \bR_2\bOmega_{fc,22,m\ell}(h, r_0,r_0+\epsilon, r_0, r_0+\epsilon)\bR_2'\|_2^2 \Big]\\
&\leq &\sum_{m=1}^{p_2} \sum_{\ell=1}^{p_2} \Big[4\|\bB_{1,1}'\bR_2\|_2^2 \cdot \|\bOmega_{fc,21,m\ell}(h,r_0,r_0+\epsilon -\infty, r_0)\|_2^2 \cdot \|\bR_1 \|_2^2\\
&&+4\|\bB_{1,1}'\bR_2\|_2^2 \cdot \|\bOmega_{fc,22,m\ell}(h, r_0,r_0+\epsilon, r_0, r_0+\epsilon)\|_2^2 \cdot \|\bR_2\|_2^2 \Big]\\
&\leq&4 k_1^3k_2^4a_1^8b_1^2 \sigma_f^4( \epsilon^2 p_1^{1+\beta_1-\delta_{11}-\delta_{12}}p_2^{2-\delta_{21}-\delta_{22}}+\epsilon^4 p_1^{1+\beta_1-2\delta_{12}}p_2^{2-\delta_{22}}).
\end{eqnarray*}
\endp

\begin{lemma}\label{min}
Under Conditions A1-A4 and C6-C7,  for any $ \epsilon \in (\eta_1-r_0,0)$, 
\begin{align*}
\lambda_{k_1}\left(\sum_{m=1}^{p_2} \sum_{\ell=1}^{p_2} \bOmega_{fc,11,m\ell}(h_1^*,r_0+\epsilon,r_0,-\infty,r_0+\epsilon)\bOmega_{fc,11,m\ell}(h_1^*,r_0+\epsilon,r_0,-\infty,r_0+\epsilon)'\right) \geq  \pi_1^2a_2^2 \gamma_0^2 \tau_2^2 \epsilon^2 p_2^{2-2\delta_{21}},\\
\lambda_{k_1}\left(\sum_{m=1}^{p_2} \sum_{\ell=1}^{p_2} \bOmega_{fc,12,m\ell}(h_1^*,r_0+\epsilon,r_0, r_0,+\infty)\bOmega_{fc,12,m\ell}(h_1^*,r_0+\epsilon,r_0,r_0,+\infty)'\right) \geq \pi_2^2 a_2^2 \gamma_0^2 \tau_2^2\epsilon^2 p_2^{2-\delta_{21}-\delta_{22}},
\end{align*}
and for any $\epsilon \in (0,\eta_2-r_0)$,
\begin{align*}
\lambda_{k_1}\left(\sum_{m=1}^{p_2} \sum_{\ell=1}^{p_2} \bOmega_{fc,21,m\ell}(h_2^*,r_0, r_0+\epsilon,-\infty,r_0)\bOmega_{fc,21,m\ell}(h_1^*, r_0,r_0+\epsilon,-\infty,r_0)'\right) \geq \pi_1^2 a_2^2 \gamma_0^2 \tau_2^2 \epsilon^2p_2^{2-\delta_{21}-\delta_{22}},\\
\lambda_{k_1}\left(\sum_{m=1}^{p_2} \sum_{\ell=1}^{p_2} \bOmega_{fc,22,m\ell}(h_2^*,r_0,r_0+\epsilon,r_0+\epsilon,+ \infty)\bOmega_{fc,22,m\ell}(h_1^*,r_0,r_0+\epsilon,r_0+\epsilon,+\infty)'\right) \geq  \pi_2^2 a_2^2 \gamma_0^2 \tau_2^2 \epsilon^2 p_2^{2-2\delta_{22}},
\end{align*}
where $\lambda_i(\bH)$ is the $i$-th largest eigenvalue of $\bH$.
\end{lemma}
\noindent{\it Proof:} By definition and properties of Kronecker product, we have
\begin{eqnarray*}
\lefteqn{\bOmega_{fc,ij,m\ell}(h,c_1,c_2,c_3,c_4)}\\
&=&\frac{1}{T}\sum_{t=1}^{T-h} \rE[(\bc_{i,m\cdot}' \otimes \bI_{k_1}) {\rm vec}(\bF_t) {\rm vec}(\bF_{t+h})' (\bc_{\ell, j\cdot} \otimes \bI_{k_1})I_t(h,c_1,c_2,c_3,c_4)]\\
&=&P(c_1<z_t<c_2, c_3<z_{t+h}<c_4)(\bc_{i,m\cdot}' \otimes \bI_{k_1}) \bSigma_{f}(h,c_1,c_2,c_3,c_4)(\bc_{j,\ell \cdot} \otimes \bI_{k_1}).
\end{eqnarray*}
Under Conditions A1-A3, following the proof of Lemma 5 in \cite{wang2019}, we can obtain
\begin{eqnarray*}
\lefteqn{\frac{1}{P(c_1<z_t<c_2, c_3<z_{t+h}<c_4)^2}\cdot \lambda_{k_1}\left(\sum_{m=1}^{p_2} \sum_{\ell=1}^{p_2} \bOmega_{fc,ij,m\ell}(h,c_1,c_2,c_3,c_4)\bOmega_{fc,ij,m\ell}(h,c_1,c_2,c_3,c_4)'\right)}\\
%&=&\lambda_{k_1} \left( \sum_{m=1}^{p_2} \sum_{\ell=1}^{p_2} (\bc_{i,m\cdot}' \otimes \bI_{k_1}) \bSigma_{f}(h,c_1,c_2,c_3,c_4)(\bc_{j,\ell \cdot} \otimes \bI_{k_1})(\bc_{j,\ell \cdot}' \otimes \bI_{k_1}) \bSigma_{f}(h,c_1,c_2,c_3,c_4)'(\bc_{i,m\cdot} \otimes \bI_{k_1})\right)\\
%&\geq & \lambda_{k_1}\left(\sum_{m=1}^{p_2} \sum_{\ell=1}^{p_2} (\bc_{i,m\cdot}' \otimes \bI_{k_1}) \bSigma_{f}(h,c_1,c_2,c_3,c_4)(\bc_{j,\ell \cdot} \bc_{j,\ell \cdot}'\otimes \bI_{k_1}) \bSigma_{f}(h,c_1,c_2,c_3,c_4)'(\bc_{i,m\cdot} \otimes \bI_{k_1})    \right)\\
%&= & \lambda_{k_1}\left(\sum_{m=1}^{p_2}  (\bc_{i,m\cdot}' \otimes \bI_{k_1}) \bSigma_{f}(h,c_1,c_2,c_3,c_4)(\bC_j'\bC_j  \otimes \bI_{k_1}) \bSigma_{f}(h,c_1,c_2,c_3,c_4)'(\bc_{i,m\cdot} \otimes \bI_{k_1})    \right)\\
%&= & \lambda_{k_1}\left(\sum_{m=1}^{p_2}  (\bc_{i,m\cdot}' \otimes \bI_{k_1}) \bSigma_{f}(h,c_1,c_2,c_3,c_4)(\bC_j' \otimes \bI_{k_1}) ( \bC_j \otimes \bI_{k_1}) \bSigma_{f}(h,c_1,c_2,c_3,c_4)'(\bc_{i,m\cdot} \otimes \bI_{k_1})    \right)\\
%&= & \lambda_{k_1}\left(\sum_{m=1}^{p_2} ( \bC_j \otimes \bI_{k_1})\bSigma_{f}(h,c_1,c_2,c_3,c_4)'(\bc_{i,m\cdot} \otimes \bI_{k_1})  (\bc_{i,m\cdot}' \otimes \bI_{k_1}) \bSigma_{f}(h,c_1,c_2,c_3,c_4)(\bC_j' \otimes \bI_{k_1}) \right)\\
&\geq& \lambda_{k_1}\left( ( \bC_j \otimes \bI_{k_1})\bSigma_{f}(h,c_1,c_2,c_3,c_4)'(\bC_i' \bC_i \otimes \bI_{k_1})  \bSigma_{f}(h,c_1,c_2,c_3,c_4)(\bC_j' \otimes \bI_{k_1}) \right).
\end{eqnarray*}
Since $\bC_i'\bC_i$ is a $k_2\times k_2$ symmetric positive definite matrix, we can find a $k_2\times k_2$ positive definite matrix $\bU_i$ such that $\bC_i'\bC_i=\bU_i\bU_i'$ and $\|\bU_i\|_2 \geq  \|\bU_i\|_{\min} \geq a_2 p_2^{1/2-\delta_{2i}/2} $, for $i=1,2$, where $a_2$ is defined in Lemma \ref{dist}. With the property of Kronecker product, it can be seen that $\sigma_1(\bU_i \otimes \bI_{k_1}) \geq \sigma_{(k_1k_2)}(\bU_i \otimes \bI_{k_1})\geq a_2 p_2^{1/2-\delta_{2i}/2}$. By Theorem 9 in \cite{merikoski2004}, Lemma \ref{fc}, and Condition C6, we have $\sigma_{k_1} (\bSigma_f(h_2^*,r_0,r_0+\epsilon,r_0+\epsilon,+\infty)(\bU_i \otimes \bI_{k_1}))\geq a_2 \gamma_0 p_2^{1/2-\delta_{2i}}$. 

Similar to proof of Lemma 5 in \cite{wang2019}, we have
\begin{eqnarray*}
\lefteqn{\frac{1}{p_\epsilon^2}\lambda_{k_1}\left(\sum_{m=1}^{p_2} \sum_{\ell=1}^{p_2} \bOmega_{fc,22,m\ell}(h_2^*,r_0, r_0+\epsilon,r_0+\epsilon,+ \infty)\bOmega_{fc,22,m\ell}(h_2^*,r_0, r_0+\epsilon,r_0+\epsilon,+ \infty)'\right)}\\
%&\geq&\lambda_{k_1}\left(\sum_{m=1}^{p_2} ( \bC_2 \otimes \bI_{k_1})\bSigma_{f}(h_2^*,r_0,r_0+\epsilon,r_0+\epsilon,+ \infty)'(\bU_2\otimes \bI_{k_1}) \right. \\
%&&\cdot (\bU_2' \otimes \bI_{k_1})  \bSigma_{f}(h_2^*,r_0, r_0+\epsilon,r_0+\epsilon,+ \infty)(\bC_2' \otimes \bI_{k_1})  \Big)\\
%&=& \lambda_{k_1}\left(\sum_{m=1}^{p_2}  (\bU_2'\otimes \bI_{k_1})  \bSigma_{f}(h_2^*,r_0, r_0+\epsilon,r_0,+ \infty)(\bC_2'  \bC_2 \otimes \bI_{k_1})\right.\\
%&&\cdot \bSigma_{f}(h_2^*,r_0,r_0+\epsilon,r_0,+ \infty)'(\bU_2\otimes \bI_{k_1}) \Big)\\
%&\geq&\lambda_{k_1}\left(\sum_{m=1}^{p_2}  (\bU_2'\otimes \bI_{k_1})  \bSigma_{f}(h_2^*,r_0,r_0+\epsilon,r_0,+ \infty)(\bU_2 \otimes \bI_{k_1})\right.\\
%&&\cdot (\bU_2' \otimes \bI_{k_1})\bSigma_{f}(h_2^*,r_0,r_0+\epsilon,r_0,+ \infty)'(\bU_2\otimes \bI_{k_1})  \Big)\\
&\geq&\left[\sigma_{k_1}\left((\bU_2' \otimes \bI_{k_1})\bSigma_f(h_2^*,r_0,r_0+\epsilon,r_0,+ \infty) (\bU_2 \otimes \bI_{k_1})\right) \right]^2\geq a_2^2 \gamma_0^2 p_2^{2-2\delta_{22}},
\end{eqnarray*}
where $p_\epsilon=P(r_0<z_t<r_0+\epsilon, z_{t+h}>r_0+\epsilon)$. The conclusion follows,
\begin{eqnarray*}
\lefteqn{\lambda_{k_1}\left(\sum_{m=1}^{p_2} \sum_{\ell=1}^{p_2} \bOmega_{fc,22,m\ell}(h_2^*,r_0, r_0+\epsilon,r_0+\epsilon,+ \infty)\bOmega_{fc,22,m\ell}(h_2^*,r_0, r_0+\epsilon,r_0+\epsilon,+ \infty)'\right)}\\
&\geq & P(r_0<z_t<r_0+\epsilon, z_{t+h}>r_0+\epsilon)^2a_2^2\gamma_0^2 p_2^{2-2\delta_{22}} \geq \pi_2^2 a_2^2 \gamma_0^2 \tau_2^2 \epsilon^2 p_2^{2-2\delta_{22}}.
\end{eqnarray*}
\endp

\noindent{\bf Proof of Theorem 1.} When $r_1 \leq r_0$ and $r_2 \geq r_0$ and $r_1$ and $r_2$ are fixed, under Conditions A4 and B1, similar to Lemmas \ref{F}, and \ref{fc}-\ref{Y}, we can show that
\begin{eqnarray*}
\sum_{m=1}^{p_2}\sum_{\ell=1}^{p_2} \|\bOmega_{x,11,m\ell}(h,r_1,r_2)\|^2_2=\|\bR_1 \bOmega_{fc,11,m\ell}(-\infty,r_1,-\infty,r_1)\bR_1'\|_2\leq k_1^2 k_2^4 a_1^8 \sigma_f^4 p_1^{2-2\delta_{11}}p_2^{2-2\delta_{21}}.\\
\sum_{m=1}^{p_2}\sum_{\ell=1}^{p_2} \|\bOmega_{x,12,m\ell}(h,r_1,r_2)\|^2_2=\|\bR_1 \bOmega_{fc,12,m\ell}(-\infty,r_1,r_2,+\infty)\bR_2'\|_2=
k_1^2 k_2^4 a_1^8 \sigma_f^4 p_1^{2-\delta_{11}-\delta_{12}}p_2^{2-\delta_{21}-\delta_{22}},
\end{eqnarray*}
and
\begin{eqnarray*}
\sum_{m=1}^{p_2}\sum_{\ell=1}^{p_2}{\rE} \|\hat{\bOmega}_{x,1j,m\ell}(h,r_1,r_2)-\bOmega_{x,1j,m\ell}(h,r_1,r_2)\|_2^2\leq 100(3h+8\alpha)k_1^2k_2^4 a_1^8\sigma_0^4  p_1^2p_2^2 T^{-1}, \mbox{ for } i=1,2.
\end{eqnarray*}
Since since $p_1^{\delta_{11}/2+\delta_{12}/2}p_2^{\delta_{21}/2+\delta_{22}/2}T^{-1/2}=o(1)$, by Cauchy-Schwarz inequality, it follows
\begin{eqnarray*}
\lefteqn{\rE \|\hat{\bM}_{1,1}(r_1,r_2)-\bM_{1,1}(r_1,r_2)\|_2}\\
&\leq &\sum_{h=1}^{h_0} \sum_{j=1}^2\Bigg[ \sum_{m=1}^{p_2}\sum_{\ell=1}^{p_2}\rE \|\hat{\bOmega}_{x,1j,m\ell}(h,r_1,r_2)-\bOmega_{x,1j,m\ell}(h,r_1,r_2)\|_2^2\\
&+& \sum_{m=1}^{p_2}\sum_{\ell=1}^{p_2} \|\bOmega_{x,1j,m\ell}\|_2 \cdot \rE \|\hat{\bOmega}_{x,1j,m\ell}(h,r_1,r_2)-\bOmega_{x,1j,m\ell}(h,r_1,r_2)\|_2\Bigg]\\
&\leq &\sum_{h=1}^{h_0} \sum_{j=1}^2\Bigg[ \sum_{m=1}^{p_2}\sum_{\ell=1}^{p_2} \rE \|\hat{\bOmega}_{x,1j,m\ell}(h,r_1,r_2)-\bOmega_{x,ij,m\ell}(h,r_1,r_2)\|_2^2\\
&+& \left(\sum_{m=1}^{p_2}\sum_{\ell=1}^{p_2} \|\bOmega_{x,1j,m\ell}(h,r_1,r_2)\|_2^2\right)^{1/2}\left(\sum_{m=1}^{p_2}\sum_{\ell=1}^{p_2} \rE \|\hat{\bOmega}_{x,1j,m\ell}(h,r_1,r_2)-\bOmega_{x,1j,m\ell}(h,r_1,r_2)\|_2^2\right)^{1/2}\Bigg]\\
&\leq&200(3h_0+8\alpha)h_0k_1^2k_2^4a_1^8\sigma_0^4 p_1^2 p_2^2 T^{-1}\\
&&+20\sqrt{(3h_0+8\alpha)}h_0k_1^2k_2^4a_1^8\sigma_0^4 p_1^{2-\delta_{11}/2-\delta_{1\min}/2}p_2^{2-\delta_{21}/2-\delta_{2\min}/2}T^{-1/2})\\
&\leq & 220(3h_0+8\alpha)h_0k_1^2 k_2^4 a_1^8\sigma_0^4 p_1^{2-\delta_{11}/2-\delta_{1\min}/2}p_2^{2-\delta_{21}/2-\delta_{2\min}/2}T^{-1/2}.
\end{eqnarray*}
Similar to Lemma \ref{min}, with Condition B2 we can prove that there exists a positive constant $C$ such that $\sum_{m=1}^{p_2} \sum_{\ell=1}^{p_2}\| \bOmega_{fc,1j,m\ell}(h_{1j},r_1,r_2)\|_{\min}^2 \geq C p_1^{2-\delta_{11}-\delta_{1j}}$ for $j=1,2$. By Theorem 9 in \cite{merikoski2004}, we have
\begin{eqnarray}
\lefteqn{\|\bM_{1,1}(r_1,r_2)\|_{\min} \geq \sum_{h=1}^{h_0}\sum_{j=1}^2\sum_{m=1}^{p_2} \sum_{\ell=1}^{p_2} \|\bR_1 \bOmega_{fc,1j,m\ell}(h,r_1,r_2) \bR_j'\|_{\min}^2} \nonumber \\
&\geq&\sum_{h=1}^{h_0}\sum_{j=1}^2 \|\bR_1\|_{2}^2 \|\bR_j\|_{2}^2 \left(\sum_{m=1}^{p_2} \sum_{\ell=1}^{p_2}\| \bOmega_{fc,1j,m\ell}(h,r_1,r_2)\|_{\min}^2 \right) \nonumber \\
&\geq&Ca_2^2 p_1^{2-\delta_{11}-\delta_{1\min}}p_2^{2-\delta_{21}-\delta_{2\min}}. \label{Mr1r2}
\end{eqnarray}
Since $r_1<r_0<r_2$, we know that $\cM(\bQ_{1,1}(r_1,r_2))=\cM(\bQ_{1,1})$. Following the proof for Theorem 1 in \cite{lam2011} and Theorem 3 in \cite{liu2016}, we have $\cD(\cM(\hat{\bQ}_{1,1}(r_1,r_2)),\cM(\bQ_{1,1}))=O_p(p_1^{\delta_{11}/2+\delta_{1\min}/2}p_2^{\delta_{21}/2+\delta_{2\min}/2}T^{-1/2}$).
\endp

\begin{lemma}\label{Gmin}
Under Conditions A1-A4 and C1-C7, for $\epsilon \in (\eta_1-r_0,\eta_2-r_0)$, with true $k_0$, when $p_1$ and $p_2$ are large enough, we have $G(r_0)=0$, and 
\begin{eqnarray*}
G(r_0+\epsilon) \geq \left\{
\begin{array}{ll}
2\pi_0^2 a_2^4 b_2^2\gamma_0^2 \tau_2^2 \epsilon^2 d_{\max}^2 p_1^{2-\delta_{11}-\delta_{1\min}}p_2^{2-\delta_{21}-\delta_{2\min}}&\epsilon<0,\\
2\pi_0^2 a_2^4b_2^2\gamma_0^2 \tau_2^2 \epsilon^2 d_{\max}^2 p_1^{2-\delta_{12}-\delta_{1\min}}p_2^{2-\delta_{22}-\delta_{2\min}}&\epsilon>0,
\end{array}
\right.
\end{eqnarray*}
where $\pi_0=\min\{\pi_1/2,\pi_2/2\}$.
\end{lemma}
\noindent{\it Proof:} 
When $\epsilon>0$, if $p_1^{\delta_{11}-\delta_{12}}p_2^{\delta_{21}-\delta_{22}}=o(1)$, when $p_1$ and $p_2$ are large enough by Theorem 9 in \cite{merikoski2004} and Lemmas \ref{fc}, \ref{dist}, and \ref{min}, we have
\begin{eqnarray*}
\lefteqn{\|\bB_{1,1}'\bM_{1,1}(r_0+\epsilon) \bB_{1,1}\|_2 \geq \Big\| \sum_{m=1}^{p_2} \sum_{\ell=1}^{p_2}\bB_{1,1}' \bOmega_{x,11,m\ell}(h_2^*,r_0+\epsilon)\bOmega_{x,11,m\ell}(h_2^*,r_0+\epsilon)' \bB_{1,1} \Big\|_2}\\
& =& \Big\| \sum_{m=1}^{p_2} \sum_{\ell=1}^{p_2}\bB_{1,1}'\bR_2\left( \bOmega_{fc,21,m\ell}(h_2^*,r_0,r_0+\epsilon, -\infty,r_0)\bR_1' +\bOmega_{fc,22,m\ell}(h_2^*,r_0,r_0+\epsilon, r_0,r_0+\epsilon)\bR_2' \right) \\
&& \cdot \left(\bR_1\bOmega_{fc,21,m\ell}(h_2^*,r_0,r_0+\epsilon, -\infty,r_0)'  +\bOmega_{fc,22,m\ell}(h_2^*,r_0,r_0+\epsilon, r_0,r_0+\epsilon)'\bR_2 \right) \bR_2'\bB_{1,1} \Big\|_2\\
&\geq& \frac{1}{2}\|\bB_{1,1}' \bR_2\|_2^2 \cdot  \sum_{m=1}^{p_2} \sum_{\ell=1}^{p_2} \| \bOmega_{fc,21,m\ell}(h_2^*,r_0,r_0+\epsilon, -\infty,r_0) \|_{\min}^2 \|\bR_1\|_2^2\\
&&-\|\bB_{1,1}' \bR_2\|_2^2 \cdot  \sum_{m=1}^{p_2} \sum_{\ell=1}^{p_2} \| \bOmega_{fc,22,m\ell}(h_2^*,r_0,r_0+\epsilon,r_0, r_0+\epsilon) \|_{2}^2 \|\bR_2\|_2^2\\
 &\geq&\pi_1^2 a_2^4 b_2^2 \gamma_0^2 \tau_2^2\epsilon^2 p_1^{1+\beta_1-\delta_{11}-\delta_{12}}p_2^{2-\delta_{21}-\delta_{22}}/2-  k_1^2k_2^4 a_2^4 b_2^2\sigma_f^4\epsilon^4 p_1^{1+\beta_1-2\delta_{12}}p_2^{2-2\delta_{22}}. %\\
 %&\geq & \pi_1 a_2^4 b_2^2 \gamma_0^2 \tau_2^2\epsilon^2 p_1^{1+\beta_1-\delta_{11}-\delta_{12}}p_2^{2-\delta_{21}-\delta_{22}}/2.
\end{eqnarray*}
If $p_1^{\delta_{11}-\delta_{12}}p_2^{\delta_{21}-\delta_{22}}\geq C$ as $p_1$, $p_2$, and $T$ go to infinity,  then we have
\begin{eqnarray*}
\lefteqn{\|\bB_{1,1}'\bM_{1,1}(r_0+\epsilon) \bB_{1,1}\|_2 \geq \Big\| \sum_{m=1}^{p_2} \sum_{\ell=1}^{p_2}\bB_{1,1}' \bOmega_{x,12,m\ell}(h,r_0+\epsilon)\bOmega_{x,12,m\ell}(h,r_0+\epsilon)' \bB_{1,1} \Big\|_2}\\
& =& \Big\| \sum_{m=1}^{p_2} \sum_{\ell=1}^{p_2}\bB_{1,1}'\bR_2 \bOmega_{fc,22,m\ell}(h_2^*,r_0,r_0+\epsilon, r_0+\epsilon, +\infty)\bR_2' \bR_2\bOmega_{fc,22,m\ell}(h_2^*,r_0,r_0+\epsilon, r_0+\epsilon, +\infty) \bB_{1,1}\Big\|_2  \\
&\geq &\pi_2^2a_2^4 b_2^2 \gamma_0^2\tau_2^2 \epsilon^2 p_1^{1+\beta_1-2\delta_{12}}p_2^{2-2\delta_{22}}.
\end{eqnarray*}
In sum, when $p_1$ and $p_2$ are large enough we have 
\[
\|\bB_{1,1}'\bM_{1,1}(r_0+\epsilon) \bB_{1,1}\|_2\geq  \pi_0^2a_2^4 b_2^2\gamma_0^2 \tau_2^2 \epsilon^2 p_1^{1+\beta_1-\delta_{12}-\delta_{1\min}}p_2^{2-\delta_{22}-\delta_{2\min}},
\]
where $\pi_0=\min\{\pi_1/2,\pi_2/2\}$. We can also show that 
\[
\|\bB_{2,1}'\bM_{2,1}(r_0+\epsilon) \bB_{2,1}\|_2 \geq \pi_0^2 a_2^4 b_2^2\gamma_0^2 \tau_2^2 \epsilon^2 p_1^{2-\delta_{12}-\delta_{1\min}}p_2^{1+\beta_2-\delta_{22}-\delta_{2\min}}.
\]
It indicates that 
\[
G(r_0+\epsilon)\geq 2\pi_0^2 a_2^4 b_2^2\gamma_0^2 \tau_2^2 \epsilon^2 d_{\max}^2p_1^{2-\delta_{12}-\delta_{1\min}}p_2^{2-\delta_{22}-\delta_{2\min}},\mbox{ if } \epsilon>0.
\]
It can be shown that $G(r_0+\epsilon)\geq 2\pi_0^2 a_2^4 b_2^2 \gamma_0^2 \tau_2^2 \epsilon^2 d_{\max}^2  p_1^{2-\delta_{11}-\delta_{1\min}}p_2^{2-\delta_{21}-\delta_{2\min}}$ when $\epsilon<0$ and $G(r_0)=0$ by definition and Lemmas \ref{BY} and \ref{min}.
\endp

\begin{lemma}\label{B2}
Under Conditions A1-A4 and C1-C3, if $p_1^{\delta_{11}/2+\delta_{12}/2}p_2^{\delta_{21}/2+\delta_{22}/2} T^{-1/2}=o(1)$, with true $k_0$, as $p_1,p_2, T \to \infty$, it holds that
\[
\rE \|\hat{\bB}_{1,i}(\eta_1, \eta_2)-\bB_{1,i}(\eta_1,\eta_2)\|_2^2 \leq C p_1^{\delta_{1i}+\delta_{1\min}}p_2^{\delta_{2i}+\delta_{2\min}}T^{-1}, \mbox { for } i=1,2,
\]
where $C$ is a generic uniformly positive constant which only depends on the parameters.
\end{lemma}

\noindent{\it Proof:} Let $Y_{t}=f_{t,qv}f_{t+h,m\ell}\cdot I_t(h,c_1,c_2,c_3,c_4)-\rE[f_{t,qv}f_{t+h,m\ell}\cdot I_t(h,c_1,c_2,c_3,c_4)]$, and $c_1,c_2,c_3,c_4$ are real numbers in $[\eta_1,\eta_2]$, $-\infty$, or $+\infty$. Condition A2 and Jensen's inequality indicate that $\rE(Y_t^2)<\sigma_f^4$, $\rE (Y_t^4)<\sigma_f^8$, and $\rE Y_t^{2\gamma}<\sigma_f^{4\gamma}$. Thus, by Cauchy-Schwartz inequality, Proposition 2.5 in \cite{fan2003} and Lemma \ref{F},
\begin{eqnarray*}
\lefteqn{\frac{1}{T^4} \rE\left( \sum_{t=1}^{T-h} Y_{t}^4\right)}\\
&\leq&\frac{1}{T^4}\sum_{t=1}^{T-h} \rE(Y_{t}^4)+\frac{2\binom 4 1}{T^4}\sum_{t_1< t_2} \rE(Y_{t_1}^3Y_{t_2})+\frac{2\binom 4 2}{T^4}\sum_{t_1< t_2} \rE(Y_{t_1}^2Y_{t_2}^2)+ \frac{\binom 4 2 \binom 2 1}{T^4}\sum_{\substack{t_1\neq t_2,t_2\neq t_3\\ t_1 \neq t_3}} \rE (Y_{t_1}^2Y_{t_2}Y_{t_3})\\
&&+\frac{4!}{T^4} \sum_{t_1<t_2<t_3<t_4}\rE (Y_{t_1}Y_{t_2}Y_{t_3}Y_{t_4})\\
&<&\frac{\sigma_f^8}{T^3}+\frac{10\sigma_f^8}{T^2} + \frac{12\sigma_f^8}{T}+\frac{24}{T^4}\sum_{\substack{t_1<t_2<t_3<t_4 \\ t_3-t_2 \leq h}} \rE (Y_{t_1}Y_{t_2}Y_{t_3}Y_{t_4})+\frac{24}{T^4} \sum_{\substack{t_1<t_2<t_3<t_4 \\ t_3-t_2>h}}\rE (Y_{t_1}Y_{t_2}Y_{t_3}Y_{t_4})\\
&<& \frac{23 \sigma_f^8}{T}+\frac{24h(T-2h)^3 \sigma_f^8}{T^4}+\frac{24}{T^4}\sum_{\substack{t_1<t_2<t_3<t_4\\ t_3-t_2> h}} \cov(Y_{t_1}Y_{t_2}, Y_{t_3}Y_{t_4})+\rE(Y_{t_1}Y_{t_2})\rE(Y_{t_3}Y_{t_4})\\
&<&\frac{47h\sigma_f^8}{T}+\frac{48\sigma_f^8}{T^2}\sum_{u=1}^{T-2h}\alpha(u)^{1-2/\gamma}+\frac{192}{T^4} \left( \sum_{t_1=1}^{T-h} \sum_{t_2=1}^{T-h} |\cov(Y_{t_1},Y_{t-2})|\right) \cdot \left( \sum_{t_3=1}^{T-h} \sum_{t_4=1}^{T-h} |\cov(Y_{t_3},Y_{t-4})|\right)\\
&<&\frac{(47h+48\alpha)\sigma_f^8}{T}+\frac{3\sigma_f^8}{T^2}\left(\sum_{u=1}^{T-2h} \alpha(u)^{1-2/\gamma} \right)^2\\
&\leq& \frac{(47h+48\alpha+192\alpha^2)\sigma_f^8}{T}.
\end{eqnarray*}
Together with Lemma \ref{f}, we have
\begin{eqnarray}
\lefteqn{\rE \left(\sum_{m=1}^{p_2}\sum_{\ell=1}^{p_2} \| \hat{\bOmega}_{fc,ij,m\ell}(h,c_1,c_2,c_3,c_4)-\bOmega_{fc,ij,m\ell}(h,c_1,c_2,c_3,c_4)\|_2^2 \right)^2} \nonumber \\
&\leq& \left( \sum_{m=1}^{p_2}  \|\bc_{i,m\cdot}\|_2^2\right)^2 \cdot \left( \sum_{\ell=1}^{p_2} \|\bc_{j,\ell\cdot}\|_2^2\right)^2 \nonumber\\
&& \cdot \rE \Big\| \frac{1}{T} \sum_{t=1}^{T-h} [ \bF_{t+h} \otimes \bF_t I_t(h,c_1,c_2,c_3,c_4) - \rE(  \bF_{t+h} \otimes \bF_t I_t(h,c_1,c_2,c_3,c_4)) ] \Big\|_F^4\nonumber \\
&\leq&  \| \bC_i \|_F^4 \| \bC_j\|_F^4  \cdot \frac{\rho_{c_1,c_2}\rho_{c_3,c_4} k_1^2k_2^2(47h+48\alpha+192\alpha^2)\sigma_f^8}{T} \nonumber \\
&\leq& \frac{\rho_{c_1,c_2}\rho_{c_3,c_4}a_0^8k_1^2k_2^6 (47h+48\alpha+192\alpha^2)\sigma_f^8p_2^{4-2\delta_{2i}-2\delta_{2j}}}{T}.\label{d2}
\end{eqnarray}
Define
\begin{align*}
\hat{\bOmega}_{d,ij,m\ell}(h,r_1,r_2)&=\frac{1}{T}\sum_{t=1}^{T-h}\bd_{t,m}'\bd_{t+h,\ell}I_{t,i}(r_i)I_{t+h,j}(r_j).\\
\hat{\bOmega}_{de,ij,m\ell}(h,r_1,r_2)&=\frac{1}{T}\sum_{t=1}^{T-h}\bd_{t,m}'\be_{t+h,\ell}I_{t,i}(r_i)I_{t+h,j}(r_j).\\
\hat{\bOmega}_{ed,ij,m\ell}(h,r_1,r_2)&=\frac{1}{T}\sum_{t=1}^{T-h}\be_{t,m}'\bd_{t+h,\ell}I_{t,i}(r_i)I_{t+h,j}(r_j).
\end{align*}
(\ref{d2}) shows 
\[
\rE\left( \sum_{m=1}^{p_2}\sum_{\ell=1}^{p_2}\|\hat{\bOmega}_{d,ij,m\ell}(h,r_1,r_2)\|_2^2\right)^2 \leq C  p_1^{4-2\delta_{1i}-2\delta_{1j}}p_2^{4-2\delta_{2i}-2\delta_{2j}}T^{-1}.
\]
For the interaction of the common component and noise, with (\ref{de}) we can show in a similar way that
\begin{eqnarray*}
\lefteqn{\rE\left(\sum_{m=1}^{p_2}\sum_{\ell=1}^{p_2} \|\hat{\bOmega}_{de,ij,m \ell}(h,r_1,r_2)\|_2^2\right)^2}\\
&\leq& k_2^2 a_0^8p_1^{2-\delta_{1i}-\delta_{1j}}p_2^{2-\delta_{2i}-\delta_{2j}} \rE \left(\frac{1}{T^2}\sum_{\ell=1}^{p_2} \sum_{u=1}^{p_1}\sum_{q=1}^{k_1}\sum_{v=1}^{k_2}\sum_{t=1}^{T-h} e_{t+h,u\ell}^2 F_{t,qv}^2 I_{t,i}(r_i)I_{t+h,j}(r_j) \right)^2\\
&\leq& k_2^2 a_0^8\sigma_0^8 p_1^{4-\delta_{1i}-\delta_{1j}}p_2^{4-\delta_{2i}-\delta_{2j}} T^{-2},
\end{eqnarray*}
and 
\begin{eqnarray*}
\rE\left(\sum_{m=1}^{p_2}\sum_{\ell=1}^{p_2} \|\hat{\bOmega}_{ed,ij,m \ell}(h,r_1,r_2)\|_2^2\right)^2
\leq k_2^2 a_0^8\sigma_0^8 p_1^{4-\delta_{1i}-\delta_{1j}}p_2^{4-\delta_{2i}-\delta_{2j}} T^{-2}.
\end{eqnarray*}
For the noise term,
\begin{eqnarray*}
\lefteqn{\rE \left(\sum_{m=1}^{p_2}\sum_{\ell=1}^{p_2} \|\hat{\bOmega}_{e,11,m\ell}(h,r_1,r_2)\|_2^2 \right)^2}\\
&\leq&\frac{1}{T^4} \rE \left(\sum_{m=1}^{p_2}\sum_{\ell=1}^{p_2} \sum_{u=1}^{p_1} \sum_{v=1}^{p_1} \sum_{t=1}^{T-h}e_{t,um}^2e_{t+h,v\ell}^2 I_{t,i}(r_i)I_{t+h,j}(r_j) \right)^2\leq \sigma_0^8 p_1^4p_2^4T^{-2}.
\end{eqnarray*}
%By Lemmas \ref{F} and \ref{f}, we can find the upper bound for $\sum_{m=1}^{p_2}\sum_{\ell=1}^{p_2} \|\bOmega_{fc,ij,m\ell}(h,c_1,c_2,c_3,c_4)\|_2^4$.
%\begin{eqnarray}
%\lefteqn{\sum_{m=1}^{p_2}\sum_{\ell=1}^{p_2} \|\bOmega_{fc,ij,m\ell}(h,c_1,c_2,c_3,c_4)\|_2^4} \nonumber \\
%&\leq &\left(\sum_{m=1}^{p_2} \|\bc_{i,m \cdot}\|_2^4 \right) \left( \sum_{m=1}^{p_2} \|\bc_{i,\ell \cdot}\|_2^4\right) \Big\| \frac{1}{T} \sum_{t=1}^{T-h} \rE (\bF_{t+h}\otimes \bF_{t}I_t(h,c_1,c_2,c_3,c_4)\Big\|_F^4 \nonumber \\
%&\leq & \rho_{c_1,c_2}^4\rho_{c_3,c_4}^4 C_0^8 \sigma_f^8 p_2^{4-2\delta_{2i}-2\delta_{2j}}.
%\end{eqnarray}
Similar to the proof of Lemma \ref{hatX}, it can be shown that there exists a positive constant $C$ such that
\begin{eqnarray*}
\lefteqn{\rE \left(\sum_{m=1}^{p_2}\sum_{\ell=1}^{p_2}\|\hat{\bOmega}_{x,ij,m\ell}(h,r_1,r_2) -\bOmega_{x,ij,m\ell}(h,r_1,r_2)\|_2^2 \right)^2}\\
& \leq &C_1p_1^{4-2\delta_{1i}-2\delta_{1j}}p_2^{4-2\delta_{2i}-2\delta_{2j}}T^{-1}+ C_2 p_1^4p_2^4 T^{-2}, 
\end{eqnarray*}
for $i,j=1,2$, where $C_1$ and $C_2$ do not depend on $p_1$, $p_2$, $T$ or $\epsilon$.\\
If $p_1^{\delta_{11}/2+\delta_{12}/2}p_2^{\delta_{21}/2+\delta_{22}/2}T^{-1/2}=o(1)$, we have
\begin{eqnarray*}
\lefteqn{\rE \|\bM_{1,1}(r_1,r_2)-\bM_{1,1}(r_1,r_2)\|_2^2}\\
&\leq & 4h_0\sum_{h=1}^{h_0} \sum_{j=1}^2 \rE \left( \sum_{m=1}^{p_2} \sum_{\ell=1}^{p_2} \|   \hat{\bOmega}_{x,1j,m\ell}(h,r_1,r_2) - \bOmega_{x,1j,m\ell}(h,r_1,r_2\|_2^2\right)^2 \\
&&  + 4h_0\sum_{h=1}^{h_0} \sum_{j=1}^2  \rE \left( \sum_{m=1}^{p_2} \sum_{\ell=1}^{p_2} 2 \|\bOmega_{x,1j,m\ell}(h,r_1,r_2) \|_2 \|   \hat{\bOmega}_{x,1j,m\ell}(h,r_1,r_2) - \bOmega_{x,1j,m\ell}(h,r_1,r_2\|_2  \right)^2\\
&\leq &4h_0\sum_{h=1}^{h_0} \sum_{j=1}^2 \left[ \rE \left( \sum_{m=1}^{p_2} \sum_{\ell=1}^{p_2} \|   \hat{\bOmega}_{x,1j,m\ell}(h,r_1,r_2) - \bOmega_{x,1j,m\ell}(h,r_1,r_2\|_2^2\right)^2 \right. \\
&&+ \left. \left( \sum_{m=1}^{p_2} \sum_{\ell=1}^{p_2}  \|\bOmega_{x,1j,m\ell}(h,r_1,r_2) \|_2^2 \right) \rE \left( \sum_{m=1}^{p_2} \sum_{\ell=1}^{p_2}  2\| \hat{\bOmega}_{x,1j,m\ell}(h,r_1,r_2) - \bOmega_{x,1j,m\ell}(h,r_1,r_2\|_2^2\right) \right]\\
&\leq&Cp_1^{4-\delta_{11}-\delta_{1\min}} p_2^{4-\delta_{21}-\delta_{2\min}} T^{-1}.
\end{eqnarray*}
Together with (\ref{Mr1r2}), following the proof for Theorem 1 in \cite{lam2011} and Theorem 3 in \cite{liu2016}, we can reach the conclusion.
\endp

\begin{lemma}\label{Ghat}
Under Condition A1-A4 and C1-C7, with true $k_0$, when $\eta_1-r_0<\epsilon <0$,
\begin{eqnarray*}
\lefteqn{\rE |\hat{G}(r_0+\epsilon)-G(r_0+\epsilon)|}\\
&\leq&C_1p_1^2p_2^2T^{-1}+C_2\epsilon d_{\max}p_1^{2-\delta_{11}/2-\delta_{1\min}/2}p_2^{2-\delta_{21}/2-\delta_{2\min}/2}T^{-1/2}\\
&&+C_3\epsilon^2 d_{\max} p_1^{2-\delta_{11}/2+\delta_{12}/2-\delta_{1\min}/2}p_2^{2-\delta_{21}/2+\delta_{22}/2-\delta_{2\min}/2}T^{-1/2},
\end{eqnarray*}when $0<\epsilon <\eta_2-r_0$,
\begin{eqnarray*}
\lefteqn{\rE|\hat{G}(r_0+\epsilon)-G(r_0+\epsilon)|}\\
&\leq&C_1p_1^2p_2^2T^{-1}+C_2\epsilon d_{\max} p_1^{2-\delta_{12}/2-\delta_{1\min}/2}p_2^{2-\delta_{22}/2-\delta_{2\min}/2}T^{-1/2}\\
&&+C_3\epsilon^2 d_{\max}p_1^{2+\delta_{11}/2-\delta_{12}-\delta_{1\min}/2}p_2^{2+\delta_{21}/2-\delta_{22}-\delta_{2\min}/2}T^{-1/2},
\end{eqnarray*}
and
\[
\rE |\hat{G}(r_0)-G(r_0)|\leq C_1p^2_1p_2^2T^{-1}.
\]
%where $C_1$,$C_2$, and $C_3$ are uniformly positive constants which only depend on the parameters.
\end{lemma}
\noindent{\it Proof:} Since $r_0 \in(\eta_1,\eta_2)$, by the definition, $\cM(\bB_{s,i})=\cM(\bB_{s,i}(\eta_1,\eta_2))$ for $s,i=1,2$. Hence, there exists a $(p_s-k_s)\times (p_s-k_s)$ orthogonal matrix $\bGamma_{s,i}$ such that $\bB_{s,i}=\bB_{s,i}(\eta_1,\eta_2)\bGamma_{s,i}$. Then we have
\begin{eqnarray}\label{hatG}
\lefteqn{\Bigg| \sum_{i=1}^2\left( \|\hat{\bB}_{1,i}(\eta_1,\eta_2)' \hat{\bM}_{1,i}(r)\hat{\bB}_{1,i}(\eta_1,\eta_2)\|_2- \|\bB_{1,i}'\bM_{1,i}(r)\bB_{1,i} \|_2\right)\Bigg|}\nonumber\\
&=&\Bigg| \sum_{i=1}^2\left(\|\hat{\bB}_{1,i}(\eta_1,\eta_2)' \hat{\bM}_{1,i}(r)\hat{\bB}_{1,i}(\eta_1,\eta_2)\|_2- \|\bB_{1,i}(\eta_1,\eta_2)'\bM_{1,i}(r)\bB_{1,i}(\eta_1,\eta_2)\|_2\right) \Bigg| \nonumber\\
&\leq & \sum_{h=1}^{h_0}\sum_{i=1}^2\sum_{j=1}^2 \sum_{m=1}^{p_2}\sum_{\ell=1}^{p_2}  \Big( \| \hat{\bB}_{1,i}(\eta_1,\eta_2)'\hat{\bOmega}_{x,ij,m\ell}(h,r)- {\bB}_{s,i}(\eta_1,\eta_2)'  {\bOmega}_{x,ij,m\ell}(h,r)\|_2^2\nonumber\\
&& + 2 \| \bB_{1,i}(\eta_1,\eta_2)' {\bOmega}_{x,ij,m\ell}(h,r)[ \hat{\bOmega}_{x,ij,m\ell}(h,r)'\hat{\bB}_{1,i}(\eta_1,\eta_2)-{\bOmega}_{x,ij,m\ell}(h,r)'{\bB}_{s,i}(\eta_1,\eta_2)] \big\|_2\Big) \nonumber\\
&\leq& \sum_{h=1}^{h_0} \sum_{i=1}^2  \sum_{j=1}^2 \Big[ \sum_{m=1}^{p_2}\sum_{\ell=1}^{p_2} \Big( \|  \hat{\bB}_{1,i}(\eta_1,\eta_2)\|_2 \| \hat{\bOmega}_{x,ij,m\ell}(h,r)-{\bOmega}_{x,ij,m\ell}(h,r)\|_2 \nonumber \\
&&+\| \hat{\bB}_{1,i}(\eta_1,\eta_2)- {\bB}_{1,i}(\eta_1,\eta_2)\|_2  \|{\bOmega}_{x,ij,m\ell}(h,r)\|_2\Big)^2 +2 \sum_{m=1}^{p_2}\sum_{\ell=1}^{p_2}\|{\bB}_{1,i}(\eta_1,\eta_2){\bOmega}_{x,ij,m\ell}(h,r)\|_2  \nonumber\\
&&\cdot  \| \hat{\bOmega}_{x,ij,m\ell}(h,r)-{\bOmega}_{x,ij,m\ell}(h,r)\|_2 \|\hat{\bB}_{1,i}(\eta_1,\eta_2)\|_2 + 2 \sum_{m=1}^{p_2}\sum_{\ell=1}^{p_2}\|{\bB}_{1,i}(\eta_1,\eta_2)\nonumber\\
&&\cdot {\bOmega}_{x,ij,m\ell}(h,r) {\bOmega}_{x,ij,m\ell}(h,r)'\|_2\| \hat{\bB}_{1,i}(\eta_1,\eta_2)-{\bB}_{1,i}(\eta_1,\eta_2)\|_2    \Big] \nonumber\\
&=&  \sum_{i=1}^2  \sum_{j=1}^2 [L_{i,j,1}(r)+L_{i,j,2}(r)+L_{i,j,3}(r)]. 
\end{eqnarray}

When $\epsilon>0$, by Cauchy-Schwarz inequality and Lemmas \ref{hatX}-\ref{BY} and \ref{B2},
\begin{eqnarray*}
\lefteqn{\rE (L_{1,1,1}(r_0+\epsilon))}\\
&\leq & 2\sum_{h=1}^{h_0} \sum_{m=1}^{p_2}\sum_{\ell=1}^{p_2}  \rE \| \hat{\bOmega}_{x,11,m\ell}(h,r_0+\epsilon)-{\bOmega}_{x,11,m\ell}(h,r_0+\epsilon)\|_2^2\\
&&+2  \rE \| \hat{\bB}_{1,1}(\eta_1,\eta_2)- {\bB}_{1,1}(\eta_1,\eta_2)\|_2^2  \left( \sum_{m=1}^{p_2}\sum_{\ell=1}^{p_2}\|{\bOmega}_{x,11,m\ell}(h,r_0+\epsilon)\|_2^2\right) \\
&\leq&O(p_1^2p_2^2T^{-1})+O(p_1^{2-\delta_{11}+\delta_{1\min}}p_2^{2-\delta_{21}+\delta_{2\min}}T^{-1})+O(\epsilon^2 p_1^{2-\delta_{12}+\delta_{1\min}}p_2^{2-\delta_{22}+\delta_{2\min}}T^{-1})\\
&&+ O(\epsilon^4 p_1^{2+\delta_{11}-2\delta_{12}+\delta_{1\min}}p_2^{2+\delta_{21}-2\delta_{22}+\delta_{2\min}}T^{-1} )\\
&=&O(p_1^2p_2^2T^{-1})+ O(\epsilon^4 p_1^{2+\delta_{11}-2\delta_{12}+\delta_{1\min}}p_2^{2+\delta_{21}-2\delta_{22}+\delta_{2\min}}T^{-1} ),
\end{eqnarray*}
\begin{eqnarray*}
\lefteqn{\rE(L_{1,1,2}(r_0+\epsilon))}\\
&\leq& 2\sum_{h=1}^{h_0} \left( \sum_{m=1}^{p_2}\sum_{\ell=1}^{p_2}\|{\bB}_{1,1}(\eta_1,\eta_2){\bOmega}_{x,11,m\ell}(h,r_0+\epsilon)\|_2^2  \right)^{1/2} \\
&&\cdot \left(  \sum_{m=1}^{p_2}\sum_{\ell=1}^{p_2} \rE \| \hat{\bOmega}_{x,11,m\ell}(h,r_0+\epsilon)-{\bOmega}_{x,11,m\ell}(h,r_0+\epsilon)\|_2^2 \right)^{1/2}\\
&= &O(\epsilon p_1^{3/2+\beta_1/2-\delta_{11}/2-\delta_{12}/2}p_2^{2-\delta_{21}/2-\delta_{22}/2}T^{-1/2})+O(\epsilon^2 p_1^{3/2+\beta_1/2-\delta_{12}}p_2^{2-\delta_{22}}T^{-1/2}).
\end{eqnarray*}
and
\begin{eqnarray*}
\lefteqn{\rE(L_{1,1,3}(r_0+\epsilon))}\\
&\leq& 2\sum_{h=1}^{h_0} \left( \sum_{m=1}^{p_2}\sum_{\ell=1}^{p_2}\|{\bB}_{1,1}(\eta_1,\eta_2){\bOmega}_{x,11,m\ell}(h,r_0+\epsilon)\|_2^2  \right)^{1/2} \\
&&\cdot \left(  \sum_{m=1}^{p_2}\sum_{\ell=1}^{p_2}  \|{\bOmega}_{x,11,m\ell}(h,r_0+\epsilon)\|_2^2 \right)^{1/2} \rE \| \hat{\bB}_{1,i}(\eta_1,\eta_2)-{\bB}_{1,i}(\eta_1,\eta_2)\|_2  \\
&=&O(\epsilon p_1^{3/2+\beta_1/2-\delta_{11}-\delta_{12}/2+\delta_{1\min}/2}p_2^{2-\delta_{21}-\delta_{22}/2+\delta_{2\min}/2}T^{-1/2})\\
&&+O(\epsilon^2 p_1^{3/2+\beta_1/2-\delta_{11}/2-\delta_{12}+\delta_{1\min}/2}p_2^{2-\delta_{21}/2-\delta_{22}+\delta_{2\min}/2}T^{-1/2})\\
&&+O(\epsilon^3 p_1^{3/2+\beta_1/2-3\delta_{12}/2+\delta_{1\min}/2}p_2^{2-3\delta_{22}/2+\delta_{2\min}/2}T^{-1/2})\\
&&+O(\epsilon^4 p_1^{3/2+\beta_1/2+\delta_{11}/2-2\delta_{12}+\delta_{1\min}/2}p_2^{2+\delta_{21}/2-2\delta_{22}+\delta_{2\min}/2}T^{-1/2}).
\end{eqnarray*}
\begin{eqnarray*}
\rE(L_{1,2,1}(r_0+\epsilon))&\leq &O(p_1^2p_2^2T^{-1}+ \epsilon^2 p_1^{2+\delta_{11}-2\delta_{12}+\delta_{1\min}}p_2^{2+\delta_{21}-2\delta_{22}+\delta_{2\min}}T^{-1}),\\
\rE(L_{1,2,2}(r_0+\epsilon))&\leq&O(\epsilon p_1^{3/2+\beta_1/2-\delta_{12}}p_2^{2-\delta_{22}}T^{-1/2}),\\
\rE(L_{1,2,3}(r_0+\epsilon))&\leq &O(\epsilon p_1^{3/2+\beta_1/2-3\delta_{12}/2+\delta_{1\min}/2}p_2^{2-3\delta_{22}/2+\delta_{2\min}/2}T^{-1/2})\\
&&+O(\epsilon^2 p_1^{3/2+\beta_1/2+\delta_{11}/2-2\delta_{12}+\delta_{1\min}/2}p_2^{2+\delta_{21}/2-2\delta_{22}+\delta_{1\min}/2}T^{-1/2}),\\
\rE(L_{2,1,1}(r_0+\epsilon))&\leq &O(p_1^2p_2^2T^{-1}),\quad L_{2,1,2}(r_0+\epsilon)=0, \quad L_{2,1,3}(r_0+\epsilon)=0,\\
\rE(L_{2,2,1}(r_0+\epsilon))&\leq&O(p_1^2p_2^2T^{-1}),\quad L_{2,2,2}(r_0+\epsilon)=0, \quad L_{2,2,3}(r_0+\epsilon)=0.
\end{eqnarray*}
It follows from (\ref{hatG}),
\begin{eqnarray*}
\lefteqn{\rE \Big|\sum_{i=1}^2\left( \|\hat{\bB}_{1,i}(\eta_1,\eta_2)' \hat{\bM}_{1,i}(r)\hat{\bB}_{1,i}(\eta_1,\eta_2)\|_2- \|\bB_{1,i}'\bM_{1,i}(r)\bB_{1,i} \|_2\right)\Big|}\\
&\leq&O(p_1^2p_2^2T^{-1})+O(\epsilon p_1^{3/2+\beta_1/2-\delta_{12}/2-\delta_{1\min}/2}p_2^{2-\delta_{22}/2-\delta_{2\min}/2}T^{-1/2})\\
&&+O(\epsilon^2 p_1^{3/2+\beta_1/2+\delta_{11}/2-\delta_{12}-\delta_{1\min}/2}p_2^{2+\delta_{21}/2-\delta_{22}-\delta_{2\min}/2}T^{-1/2}).
\end{eqnarray*}
Similarly we can establish the rate of convergence for $\sum_{i=1}^2( \|\hat{\bB}_{2,i}(\eta_1,\eta_2)' \hat{\bM}_{2,i}(r)\hat{\bB}_{2,i}(\eta_1,\eta_2)\|_2- \|\bB_{2,i}'\bM_{2,i}(r)\bB_{2,i} \|_2)$. Then when $\epsilon >0$, we have
\begin{eqnarray*}
\lefteqn{|\hat{G}(r_0+\epsilon)-G(r_0+\epsilon)|}\\
&\leq & \Bigg| \sum_{s=1}^2\sum_{i=1}^2\left( \|\hat{\bB}_{s,i}(\eta_1,\eta_2)' \hat{\bM}_{s,i}(r)\hat{\bB}_{s,i}(\eta_1,\eta_2)\|_2- \|\bB_{s,i}'\bM_{s,i}(r)\bB_{s,i} \|_2\right) \Bigg|\\
&=&O(p_1^2p_2^2T^{-1})+O(\epsilon d_{\max} p_1^{2-\delta_{12}/2-\delta_{1\min}/2}p_2^{2-\delta_{22}/2-\delta_{2\min}/2}T^{-1/2})\\
&&+O(\epsilon^2 d_{\max} p_1^{2+\delta_{11}/2-\delta_{12}-\delta_{1\min}/2}p_2^{2+\delta_{21}/2-\delta_{22}-\delta_{2\min}/2}T^{-1/2}).
\end{eqnarray*}
where $d_{\max}=\max\{ p_1^{\beta_1/2-1/2}, p_2^{\beta_2/2-1/2}\}$.
\endp

\noindent{\bf Proof of Theorem 2.} Let $g_0=\pi_0^2 a_2^4b_2^2\gamma_0^2 \tau_2^2d_{\max}^{2}p_1^{2-\delta_{12}-\delta_{1\min}}p_2^{2-\delta_{22}-\delta_{2\min}}$. Following the proof of Theorem 2 in \cite{liu2019}, we can reach the conclusion.
\endp

\noindent{\bf Proof of Theorem 3.} When $\epsilon>0$, if $p_1^{\delta_{11}/2+\delta_{12}/2}p_2^{\delta_{21}/2+\delta_{22}/2}T^{-1/2}d_{\max}^{-1}=o(1)$, by Cauchy-Schwarz inequality and Lemmas \ref{hatX}-\ref{Y}, we have 
\begin{eqnarray}\label{Mhat}
\lefteqn{\|\hat{\bM}_{1,1}(r_0+\epsilon)-\bM_{1,1}(r_0+\epsilon)\|_2}\nonumber\\
&\leq & \sum_{h=1}^{h_0}\sum_{j=1}^2 \sum_{m=1}^{p_2} \sum_{\ell=1}^{p_2}\| \hat{\bOmega}_{x,1j,m\ell}(h,r_0+\epsilon) -\bOmega_{x,1j,m\ell}(h,r_0+\epsilon) \|_2^2\nonumber \\
&&+2 \sum_{h=1}^{h_0} \sum_{j=1}^2 \sum_{m=1}^{p_2} \sum_{\ell=1}^{p_2} \|\bOmega_{x,1j,m\ell}(h,r_0+\epsilon)\|_2 \cdot \|\hat{\bOmega}_{x,1j,m\ell}(h,r_0+\epsilon) -\bOmega_{x,1j,m\ell}(h,r_0+\epsilon)\|_2\nonumber \\
&\leq& \sum_{h=1}^{h_0}\sum_{j=1}^2 \sum_{m=1}^{p_2} \sum_{\ell=1}^{p_2}\| \hat{\bOmega}_{x,1j,m\ell}(h,r_0+\epsilon) -\bOmega_{x,1j,m\ell}(h,r_0+\epsilon) \|_2^2\nonumber \\
&&+2 \sum_{h=1}^{h_0}\sum_{j=1}^2 \left(  \sum_{m=1}^{p_2} \sum_{\ell=1}^{p_2} \|\bOmega_{x,1j,m\ell}(h,r_0+\epsilon)\|_2^2\right)^{1/2} \cdot \left( \|\hat{\bOmega}_{x,1j,m\ell}(h,r_0+\epsilon) -\bOmega_{x,1j,m\ell}(h,r_0+\epsilon)\|_2^2 \right)^{1/2}\nonumber \\
&=&O_p(p_1^{2-\delta_{11}/2-\delta_{1\min}/2}p_2^{2-\delta_{21}/2-\delta_{2\min}/2}T^{-1/2})+O_p(\epsilon p_1^{2-\delta_{12}/2-\delta_{1\min}/2}p_2^{2-\delta_{22}/2-\delta_{2\min}/2}T^{-1/2}).
\end{eqnarray}
Under Condition A4, it follows from Lemma \ref{fc},
\begin{eqnarray*}
\lefteqn{\sum_{m=1}^{p_2}\sum_{\ell=1}^{p_2}\|\bOmega_{x,11,m\ell}(h,r_0+\epsilon)-\bOmega_{x,11,m\ell}(h,r_0)\|_2^2}\\
&\leq & \sum_{m=1}^{p_2}\sum_{\ell=1}^{p_2}\|\bR_1 \bOmega_{fc,12,m\ell}(h, -\infty, r_0, r_0, r_0+\epsilon) \bR_2'\|_2^2+  \sum_{m=1}^{p_2}\sum_{\ell=1}^{p_2}\|\bR_2 \bOmega_{fc,21,m\ell}(h, r_0, r_0+\epsilon, -\infty,r_0) \bR_1'\|_2^2\\
&& + \sum_{m=1}^{p_2}\sum_{\ell=1}^{p_2}\|\bR_2 \bOmega_{fc,22,m\ell}(h, r_0, r_0+\epsilon, r_0,r_0+\epsilon) \bR_2'\|_2^2\\
&=&O_p(\epsilon^2 p_1^{2-\delta_{11}-\delta_{12}} p_2^{2-\delta_{21}-\delta_{22}})+O_p(\epsilon^4 p_1^{2-2\delta_{12}} p_2^{2-2\delta_{22}}),
\end{eqnarray*}
and
\begin{eqnarray*}
\lefteqn{\sum_{m=1}^{p_2}\sum_{\ell=1}^{p_2}\|\bOmega_{x,12,m\ell}(h,r_0+\epsilon)-\bOmega_{x,12,m\ell}(h,r_0)\|_2^2}\\
&\leq &\sum_{m=1}^{p_2}\sum_{\ell=1}^{p_2}\|\bR_1 \bOmega_{fc,12,m\ell}(h, -\infty, r_0, r_0+\epsilon, +\infty) \bR_2'\|_2^2\\
&&+ \sum_{m=1}^{p_2}\sum_{\ell=1}^{p_2}\|\bR_2 \bOmega_{fc,22,m\ell}(h, r_0, r_0+\epsilon, r_0+\epsilon, +\infty) \bR_2'\|_2^2\\  
&=&O_p(\epsilon^2 p_1^{2-2\delta_{12}} p_2^{2-2\delta_{22}}).%+O_p(\epsilon^2 p_1^{2-2\delta_{12}} p_2^{2-2\delta_{22}}).
\end{eqnarray*}
Hence,
\begin{eqnarray*}
\lefteqn{\|\bM_{1,1}(r_0+\epsilon)-\bM_{1,1}\|_2}\\
&\leq &\sum_{h=1}^{h_0}\sum_{j=1}^2 \sum_{m=1}^{p_2} \sum_{\ell=1}^{p_2} \|\bOmega_{x,1j,m\ell}(h,r_0+\epsilon) \bOmega_{x,1j,m\ell}(h,r_0+\epsilon)'-\bOmega_{x,1j,m\ell}(h,r_0) \bOmega_{x,1j,m\ell}(h,r_0)'\|_2\\
&\leq&\sum_{h=1}^{h_0}\sum_{j=1}^2 \Bigg( \sum_{m=1}^{p_2} \sum_{\ell=1}^{p_2}   \|\bOmega_{x,1j,m\ell}(h,r_0+\epsilon) -\bOmega_{x,1j,m\ell}(h,r_0)\|_2^2 \\
&&+2  \sum_{m=1}^{p_2} \sum_{\ell=1}^{p_2}    \|\bOmega_{x,1j,m\ell}(h,r_0)\|_2 \cdot\|\bOmega_{x,1j,m\ell}(h,r_0+\epsilon) -\bOmega_{1j,m\ell}(h,r_0)\|_2  \Bigg)\\
&\leq&\sum_{h=1}^{h_0}\sum_{j=1}^2 \Bigg[ \sum_{m=1}^{p_2} \sum_{\ell=1}^{p_2}   \|\bOmega_{x,1j,m\ell}(h,r_0+\epsilon) -\bOmega_{x,1j,m\ell}(h,r_0)\|_2^2 \\
&&+2 \left( \sum_{m=1}^{p_2} \sum_{\ell=1}^{p_2}    \|\bOmega_{x,1j,m\ell}(h,r_0)\|_2^2\right)^{1/2} \left( \sum_{m=1}^{p_2} \sum_{\ell=1}^{p_2} \|\bOmega_{x,1j,m\ell}(h,r_0+\epsilon) -\bOmega_{x,1j,m\ell}(h,r_0)\|_2^2  \right)^{1/2} \Bigg]\\
&=&O(\epsilon^2 p_1^{2-\delta_{12}-\delta_{1\min}} p_2^{2-\delta_{22}-\delta_{2\min}})+O(\epsilon p_1^{2-\delta_{11}/2-\delta_{12}/2-\delta_{1\min}} p_2^{2-\delta_{21}/2-\delta_{22}/2-\delta_{2\min}}).
\end{eqnarray*}
If $p_1^{\delta_{11}/2+\delta_{12}/2}p_2^{\delta_{21}/2+\delta_{22}/2}T^{-1/2}d_{\max}^{-1}=o(1)$, together with (\ref{Mhat}), we have
\begin{eqnarray*}
\lefteqn{\|\hat{\bM}_{1,1}(r_0+\epsilon) -\bM_{1,1}\|_2}\\
&\leq& \|\hat{\bM}_{1,1}(r_0+\epsilon)-\bM_{1,1}(r_0+\epsilon)\|_2 +\|{\bM}_{1,1}(r_0+\epsilon)-\bM_{1,1}\|_2\\
&=&O_p(p_1^{2-\delta_{11}/2-\delta_{1\min}/2}p_2^{2-\delta_{21}/2-\delta_{2\min}/2}T^{-1/2})+O_p(\epsilon p_1^{2-\delta_{11}/2-\delta_{12}/2-\delta_{1\min}} p_2^{2-\delta_{21}/2-\delta_{22}/2-\delta_{2\min}})\\
&&+O_p(\epsilon^2 p_1^{2-\delta_{12}-\delta_{1\min}} p_2^{2-\delta_{22}-\delta_{2\min}}).
\end{eqnarray*}
Theorem 2 states that if $\hat{r}>r_0$, $|\hat{r}-r_0|=O_p(p_1^{\delta_{12}/2+\delta_{1\min}/2}p_2^{\delta_{22}/2+\delta_{2\min}/2}T^{-1/2}d_{\max}^{-1})$. Therefore,
\begin{eqnarray}\label{Mhat2}
\|\hat{\bM}_{1,1}(\hat{r})-\bM_{1,1}\|_2=O_p(p_1^{2-\delta_{11}/2-\delta_{1\min}/2}p_2^{2-\delta_{21}/2-\delta_{2\min}/2}T^{-1/2}d_{\max}^{-1}), \mbox{ if } \hat{r}>r_0.
\end{eqnarray}
On the other hand, under Condition C4 following the proof of Lemma \ref{min}, we can obtain 
\begin{eqnarray*}
\lambda_{k_1}\left( \sum_{m=1}^{p_2} \sum_{\ell=1}^{p_2} \bOmega_{fc,11,m\ell}(h_{11},-\infty,r_0,-\infty,r_0)\bOmega_{fc,11,m\ell}(h_{11},-\infty,r_0,-\infty,r_0)'   \right)\geq C p_2^{2-2\delta_{21}},\\
\lambda_{k_1}\left( \sum_{m=1}^{p_2} \sum_{\ell=1}^{p_2} \bOmega_{fc,12,m\ell}(h_{11},-\infty,r_0,r_0,+\infty)\bOmega_{fc,12,m\ell}(h_{11},-\infty,r_0,r_0,+\infty)'   \right)\geq C p_2^{2-\delta_{21}-\delta_{22}}.
\end{eqnarray*}
If $p_1^{\delta_{11}-\delta_{12}}p_2^{\delta_{21}-\delta_{22}}=o(1)$, by Theorems 6 and 9 in \cite{merikoski2004}, we can show that
\begin{eqnarray*}
\lefteqn{\|\bM_{1,1}\|_{\min}\geq \sum_{m=1}^{p_2} \sum_{\ell=1}^{p_2} \|\bOmega_{x,11,m\ell}(h_{11},r_0)\|_{\min}^2}\\
&\geq& \sum_{m=1}^{p_2} \sum_{\ell=1}^{p_2}  \|\bR_1\bOmega_{fc,11,m\ell}(h_{11},-\infty,r_0,-\infty,r_0)\bR_1'\|_{\min}^2\\
&\geq & \|\bR_1\|_{\min}^4 \cdot  \lambda_{k_1}\left( \sum_{m=1}^{p_2} \sum_{\ell=1}^{p_2} \bOmega_{fc,11,m\ell}(h_{11},-\infty,r_0,-\infty,r_0)\bOmega_{fc,11,m\ell}(h_{11},-\infty,r_0,-\infty,r_0)'   \right) \\
&\geq& Cp_1^{2-2\delta_{11}}p_2^{2-2\delta_{21}}.
\end{eqnarray*}
If $p_1^{\delta_{11}-\delta_{12}}p_2^{\delta_{21}-\delta_{22}}>C$ as $p_1$ and $p_2$ go to infinity, 
\begin{eqnarray*}
\lefteqn{\|\bM_{1,1}\|_{\min}\geq \sum_{m=1}^{p_2} \sum_{\ell=1}^{p_2} \|\bOmega_{x,12,m\ell}(h_1,r_0)\|_{\min}^2}\\
&\geq& \sum_{m=1}^{p_2} \sum_{\ell=1}^{p_2}  \|\bR_1\bOmega_{fc,12,m\ell}(h_{12},-\infty,r_0,r_0,+\infty)\bR_2'\|_{\min}^2\\
&\geq & \|\bR_1\|_{\min}^2 \|\bR_2\|_{\min}^2 \cdot  \lambda_{k_1}\left( \sum_{m=1}^{p_2} \sum_{\ell=1}^{p_2} \bOmega_{fc,12,m\ell}(h_{12},-\infty,r_0,r_0,+\infty)\bOmega_{fc,12,m\ell}(h_{12},-\infty,r_0,r_0,+\infty)'   \right) \\
&\geq& C p_1^{2-\delta_{11}-\delta_{12}}p_2^{2-\delta_{21}-\delta_{22}}.
\end{eqnarray*}
Hence, $\|\bM_{1,1}\|_{\min} \geq C p_1^{2-\delta_{11}-\delta_{1\min}}p_2^{2-\delta_{21}-\delta_{2\min}}$. Similar to the proof of Theorems 1 in \cite{lam2011} and Theorem 3 in \cite{liu2016}, with (\ref{Mhat2}), the conclusions can be reached.
\endp

\begin{lemma}\label{extra}
For $\hat{k}_1 >k_1$, let $\bB_{1,i}^*$ be a $p_s \times (p_1-\hat{k}_1)$ orthogonal matrix such that $\cM(\bB_{1,i}^*) \in \cM(\bB_{1,i})$ for $i=1,2$. Under Conditions A1-A4 and C1-C8, for any $\bB_{1,i}^*$, it holds that
\begin{eqnarray*}
\lefteqn{\sum_{m=1}^{p_2} \sum_{\ell=1}^{p_2}\|{\bB_{1,1}^*(\eta_1,\eta_2)}'\bOmega_{x,11,m\ell}(h,r_0+\epsilon) \|_2}\\
&& \left\{
\begin{array}{ll}
=0	&\epsilon \leq 0,\\
\leq O(\epsilon^2 p_1^{1+\beta_1-\delta_{11}-\delta_{12}}p_2^{2-\delta_{21}-\delta_{22}})+O(\epsilon^4 p_1^{1+\beta_1-2\delta_{12}}p_2^{2-2\delta_{22}})	&\epsilon >0,
\end{array}
\right.
\end{eqnarray*}
\begin{eqnarray*}
\sum_{m=1}^{p_2} \sum_{\ell=1}^{p_2}\|{\bB_{1,1}^*(\eta_1,\eta_2)}'\bOmega_{x,12,m\ell}(h,r_0+\epsilon)\|_2^2 \left\{
\begin{array}{ll}
=0	&\epsilon \leq 0,\\
\leq O(\epsilon^2 p_1^{1+\beta_1-2\delta_{12}}p_2^{2-2\delta_{22}})	&\epsilon >0,
\end{array}
\right.
\end{eqnarray*}
\begin{eqnarray*}
\sum_{m=1}^{p_2} \sum_{\ell=1}^{p_2}\|{\bB_{1,2}^*(\eta_1,\eta_2)}'\bOmega_{x,21,m\ell}(h,r_0+\epsilon)\|_2^2 \left\{
\begin{array}{ll}
\leq O(\epsilon^2 p_1^{1+\beta_1-2\delta_{11}}p_2^{2-2\delta_{21}}) &\epsilon \leq 0,\\
=0	&\epsilon >0,\\
\end{array}
\right.
\end{eqnarray*}
\begin{eqnarray*}
\lefteqn{\sum_{m=1}^{p_2} \sum_{\ell=1}^{p_2}\|{ \bB_{1,2}^*(\eta_1,\eta_2)}'\bOmega_{x,22,m\ell}(h,r_0+\epsilon)\|^2_2}\\
&& \left\{
\begin{array}{ll}
\leq O(\epsilon^2 p_1^{1+\beta_1-\delta_{11}-\delta_{12}}p_2^{2-\delta_{21}-\delta_{22}})+O(\epsilon^4 p_1^{1+\beta_1-2\delta_{11}}p_2^{2-2\delta_{21}})	&\epsilon \leq 0,\\
=0&\epsilon >0.\\
\end{array}
\right.
\end{eqnarray*}
\end{lemma}
\noindent{\it Proof:} Similar to proof of Lemma 7 in \cite{liu2019}.
\endp

\noindent{\bf Proof of Corollary 1.} Similar to proof of and Corollary 1 in \cite{liu2016}.

\noindent{\bf Proof of Theorem 4.} With Lemma \ref{extra}, following the proof of Theorem 4 in \cite{liu2019}, we can reach the conclusion.

\noindent{\bf Proof of Theorem 5.} Similar to proof of Theorem 3.

%\section{More Tables and Plots} \label{appendix:more_tableplot}

%\begin{figure}[H]
%	\centering
%	\includegraphics[width=0.7\textwidth, keepaspectratio]{loading_spdist_exmp2_free}
%	\caption{Boxplots of space distance for row and column loading matrices $\hat{\bR}_1$, $\hat{\bR}_2$, $\hat{\bC}_1$, and $\hat{\bC}_2$ under different combinations of factor strength and thresholding effect. Note that the y-axis are under different scales. }. 
%	\label{fig:loading_spdist_exmp2_free} 
%\end{figure}

\section{Multinational Macroeconomic Indexes Dataset} \label{appendix:cross_country_macro_dataset}

Table \ref{table:macro_index_data} lists the short name of each series, its mnemonic (the series label used in the OECD database), the transformation applied to the series, and a brief data description. All series are from the OECD Database. In the transformation column, $\Delta$ denote the first difference, $\Delta\ln$ denote the first difference of the logarithm. GP denotes the measure of growth rate last period.

\begin{table}[H]
\centering 
\resizebox{\textwidth}{!}{%
\begin{tabular}{lccp{10cm}}
\hline 
Short name & Mnemonic & Tran & description\tabularnewline
\hline 
CPI: Food & CPGDFD & $\Delta^2 \ln$ & Consumer Price Index: Food, seasonally adjusted \tabularnewline
\hline 
CPI: Ener & CPGREN & $\Delta^2 \ln$ & Consumer Price Index: Energy, seasonally adjusted \tabularnewline
\hline 
CPI: Tot & CPALTT01 & $\Delta^2 \ln$ & Consumer Price Index:  Total, seasonally adjusted \tabularnewline
\hline 
IR: Long & IRLT & $\Delta$ & Interest Rates: Long-term gov bond yields\tabularnewline
\hline 
IR: 3-Mon & IR3TIB & $\Delta$ & Interest Rates: 3-month Interbank rates and yields\tabularnewline
\hline 
P: TIEC & PRINTO01 & $\Delta \ln$ & Production: Total industry excl
construction \tabularnewline
\hline 
P: TM  & PRMNTO01 & $\Delta \ln$ & Production: Total manufacturing \tabularnewline
\hline 
GDP  & LQRSGPOR & $\Delta \ln$ & GDP: Original (Index 2010 = 1.00, seasonally adjusted) \tabularnewline
\hline 
IT: Ex & XTEXVA01 & $\Delta \ln$ & International Trade: Total Exports Value (goods) \tabularnewline
\hline 
IT: Im & XTIMVA01 & $\Delta \ln$ & International Trade: Total Imports Value (goods) \tabularnewline
\hline 
\end{tabular}
} %
\caption{Data transformations, and variable definitions}
\label{table:macro_index_data}
\end{table}

\begin{table}[H]
\centering
\scalebox{1}{
\begin{tabular}{lc||lc}
\hline
\multicolumn{1}{c}{Country} & ISO ALPHA-3 Code & \multicolumn{1}{c}{Country} & ISO ALPHA-3 Code \\ \hline
United States of America & USA & United Kingdom & GBR \\
Canada & CAN & Finland & FIN \\
New Zealand & NZL & Sweden & SWE \\
Australia & AUS & France & FRA \\
Norway & NOR & Netherlands & NLD \\
Ireland & IRL & Austria & AUT \\
Denmark & DNK & Germany & DEU \\ \hline
\end{tabular} }
\caption{Countries and ISO Alpha-3 Codes in Macroeconomic Indices Application}
\label{table:oecd_country_sel}
\end{table}

\end{appendices}

\end{document}